\documentclass[twocolumn]{aastex6}

\usepackage{graphicx}
\usepackage{array}
\usepackage{amssymb}
\usepackage{natbib}
\usepackage{float}
\usepackage{color}
\usepackage[utf8]{inputenc}  
\usepackage{amsmath}  
\usepackage{comment}
\usepackage{longtable,tabu}
\usepackage{url}
\usepackage{pbox}
\usepackage[caption2]{ccaption}
\usepackage[figuresright]{rotating}
\usepackage{chngcntr}
\usepackage{footnotebackref}
\usepackage{threeparttablex} 
\usepackage{booktabs}        

\begin{document}


\title{Spectral properties of quasars from Sloan Digital Sky Survey data release 14: The catalog}



\author{Suvendu Rakshit$^{1}$, C. S. Stalin$^{2}$ and Jari Kotilainen$^{1,3}$. }

\affil{$^{1}$Finnish Centre for Astronomy with ESO (FINCA), FI-20014 University of Turku, Finland}
\affil{$^{2}$Indian Institute of Astrophysics, Block II, Koramangala, Bangalore-560034, India}
\affil{$^{3}$Tuorla Observatory, Department of Physics and Astronomy, FI-20014 University of Turku, Finland}
\email{suvenduat@gmail.com}
\shorttitle{Properties of SDSS DR14 quasars}
\shortauthors{Rakshit et al.}

\begin{abstract}
    
We present measurements of the spectral properties for a total of 526,265 quasars, out of which 63\% have continuum S/N$>3$ pixel$^{-1}$, selected from the fourteenth data release of the Sloan Digital Sky Survey (SDSS-DR14) quasar catalog. We performed a careful and homogeneous analysis of the SDSS spectra of these sources, to estimate the continuum and line properties of several emission lines such as H${\alpha}$, H${\beta}$, H${\gamma}$, Mg \textsc{ii}, C \textsc{iii]}, C \textsc{iv} and Ly${\alpha}$. From the derived emission line parameters, we estimated single-epoch virial black hole masses ($M_{\mathrm{BH}}$) for the sample using H${\beta}$, Mg \textsc{ii} and C \textsc{iv} emission lines. The sample covers a wide range in  bolometric luminosity ($\log L_{\mathrm{bol}}$; erg s$^{-1}$) between 44.4 and 47.3 and $\log M_{\mathrm{BH}}$ between 7.1 and 9.9 $M_{\odot}$. Using the ratio of $L_{\mathrm{bol}}$ to the Eddington luminosity as a measure of the accretion rate, the logarithm of the accretion rate is found to be in the range between $-$2.06 and 0.43. We performed several correlation analyses between different emission line parameters and found them to match with that known earlier using smaller samples. We noticed that strong Fe \textsc{ii} sources with large Balmer line width, and highly accreting sources with large $M_{\mathrm{BH}}$ are rare in our sample. We make available online an extended and complete catalog that contains various spectral properties of 526,265 quasars derived in this work along with other properties culled from the  SDSS-DR14 quasar catalog.
\end{abstract}





\section{Introduction}

Quasars, a class of active galactic nuclei (AGN), are powered by accretion of 
matter onto a super massive black hole surrounded by an accretion disk 
\citep[e.g.][]{1993ARA&A..31..473A}.  The availability of a large number of quasars 
with measured line and continuum properties is of paramount importance in a 
wide variety of astrophysical research such as galaxy evolution, 
black hole growth, etc. For example, the mass of the black holes ($M_{\mathrm{BH}}$) 
in AGN is found to be strongly correlated with host galaxies velocity 
dispersion suggesting the co-evolution of the black hole and host galaxy 
\citep[e.g.,][]{2013ARA&A..51..511K}.  Thus, measuring $M_{\mathrm{BH}}$ 
for a large sample of quasars is required to study the growth and evolution 
of black hole across cosmic time. A direct method to measure $M_{\mathrm{BH}}$ in 
quasars over a large range of redshifts is via the technique of reverberation 
mapping \citep{1982ApJ...255..419B,1993PASP..105..247P} and such studies show a strong correlation between the 
quasar monochromatic luminosity ($L$) at 5100 \AA  ~~and  
the size ($R$) of the broad line region \citep[BLR; e.g.,][]{2000ApJ...533..631K,2009ApJ...697..160B,
2013ApJ...767..149B}. Since reverberation mapping requires 
long-term monitoring campaign, which is difficult for high redshift and high luminosity objects, 
the size-luminosity ($R-L$) relation has been used to estimate $M_{\mathrm{BH}}$ from the single-epoch 
spectrum for which monochromatic luminosity and emission line width measurements are available 
\citep[e.g.,][]{2002ApJ...579..530W,2011ApJS..194...45S}. The values of $M_{\mathrm{BH}}$ estimated from 
single-epoch spectrum are mostly consistent with the reverberation mapping 
$M_{\mathrm{BH}}$ estimates within a factor of few (e.g., \citealt{1999ApJ...526..579W,2002ApJ...571..733V,2002MNRAS.337..109M,2017ApJ...851...21G} but also see \citealt{2006ApJ...641..689V}, \citealt{2013BASI...41...61S} and \citealt{2014SSRv..183..253P} for merits and caveats of single-epoch $M_{\mathrm{BH}}$).

Also, statistical studies of quasars will help
in a better understanding of quasar properties \citep{1995PASP..107..803U,1989AJ.....98.1195K} 
such as the quasar luminosity function \citep{2006AJ....131.2766R}, black hole 
mass function, which shows a peak at $z\sim2$ \citep{2009ApJ...699..800V,2010ApJ...719.1315K},
and the Eddington ratio distribution, which peaks at $L_{\mathrm{bol}}/L_{\mathrm{Edd}} \sim 0.05$ 
\citep{2010ApJ...719.1315K}, where, $L_{\mathrm{bol}}$ is the bolometric luminosity and
$L_{\mathrm{Edd}} = 1.3 \times 10^{38}$ ($M_{\mathrm{BH}}$/$M_{\odot}$) erg s$^{-1}$ is the Eddington luminosity. Several correlations between continuum and emission line 
properties in quasars are available, e.g., the anti-correlation between line 
equivalent width (EW) and continuum luminosity \citep{1977ApJ...214..679B}, 
especially strong in C \textsc{iv} and Mg \textsc{ii} lines \citep{2011ApJS..194...45S}, 
correlations between continuum luminosity and line widths and luminosities, 
etc. \citep[e.g.,][]{1992ApJS...80..109B,2005ApJ...630..122G,2011ApJS..194...45S,2017ApJS..229...39R,2017MNRAS.472.4051C}.
Also, studies of the emission lines from AGN will help in enhancing our understanding of the physical conditions of
the gas close to the central regions of AGN \citep{1989agna.book.....O}.

All of the above studies require large samples of quasars. Since the discovery of quasars 
about more than half a century ago \citep{1963Natur.197.1040S}, the number of quasars that are known 
has increased gradually. A significant increase in the number of quasars 
happened in the last two decades with the bulk of the contribution 
coming from the Sloan Digital Sky Survey \citep[SDSS;][]{2000AJ....120.1579Y}. 
In addition to SDSS, other surveys too have contributed
to the increase in the  number of quasars such as the 2dF quasar redshift  
survey \citep[2QZ;][]{2004MNRAS.349.1397C},
the bright quasar survey \citep{1983ApJ...269..352S} and the large bright quasar survey (LBQS; \citealt{1995AJ....109.1498H}).
Also, the number of quasars is expected to 
increase manifold in the future from the next generation large optical imaging survey using the 
Large Synoptic Survey Telescope \citep[LSST, now known as Vera C. Rubin Observatory;][]{2019ApJ...873..111I,2014IAUS..304...11I}.

Among the many available quasar surveys, SDSS has provided us with
the largest homogeneous sample of quasars with optical spectra. Each SDSS quasar survey had different science goals. For example SDSS DR7 quasar catalog \citep{2010AJ....139.2360S} consists of 105,783 spectroscopically confirmed quasars from SDSS-I/II survey \citep{2000AJ....120.1579Y} whose aim was to study quasar luminosity function \citep[e.g.][]{2006AJ....131.2766R} and clustering properties \citep[e.g.][]{2006AJ....131....1H,2007AJ....133.2222S}.  The survey also led to the discovery of many high redshift $z\sim6$ quasars \citep[e.g.,][]{2006AJ....131.1203F,2008AJ....135.1057J}, and broad absorption line quasars \citep[e.g.,][]{2003AJ....125.1711R,2006ApJS..165....1T,2008ApJ...675..985G}. The SDSS-III/BOSS survey \citep{2011AJ....142...72E,2013AJ....145...10D} was intended to discover a large sample of quasars with Lyman-$\alpha$ forest, which fall in the redshift range of $2.15-3.5$ to constrain the Baryon Acoustic Oscillation (BAO) scale. This survey lead to the discovery of 270,000 quasars, mostly at $z>2$, which helped to provide strong cosmological constraints at z$\sim2.5$ through the auto-correlation of Lyman-$\alpha$ forest \citep[e.g.][]{2017A&A...603A..12B} and cross-correlation of quasars and Lyman-$\alpha$ forest \citep[e.g.,][]{2017A&A...608A.130D}. 

The SDSS-IV has multiple goals, SDSS-IV/eBOSS \citep[see][]{2016AJ....151...44D} is dedicated to measure percent-level angular diameter distance $d_{A}$(z) and Hubble parameter H(z) using 250,000 new spectroscopically confirmed luminous red galaxies, 195,000 new emission line galaxies, 500,000 spectroscopically confirmed quasars and 60,000 new Lyman-$\alpha$ forest quasar measurements at redshifts $z > 2.1$. The time-domain Spectroscopic Survey (TDSS) of SDSS-IV was designed to study the spectroscopic variability of quasars, and the Spectroscopic Identification of eROSITA Sources (SPIDERS) program was designed to investigate X-ray sources in SDSS-IV.  \citet{2018A&A...613A..51P} recently compiled a quasar catalog from SDSS-IV including all previously spectroscopically selected quasars from SDSS I, II and III surveys. This catalog consists of 526,356 quasars over 9376 degree$^2$ region of the sky from SDSS with 144,046 newly discovered quasars from SDSS-IV. The catalog of \citet{2018A&A...613A..51P} is, therefore, a unique and the largest list of spectroscopically confirmed quasars selected homogeneously and covering a large part of the northern sky. Once all the spectral properties of the quasars in \citet{2018A&A...613A..51P} are available, the catalog can serve as the largest quasar database useful to address a wide variety of astrophysical problems and/or revisit the correlations already known between various quasar properties.

 The spectral properties of SDSS DR7 quasars have been studied by \citet[][hereafter S11]{2011ApJS..194...45S}, consisting of about 100,000 quasars. \citet[][hereafter C17]{2017MNRAS.472.4051C} studied spectral properties of about 70,000 quasars at $z<2$ from SDSS-DR10 \citep{2014ApJS..211...17A}, which contains the first data release from SDSS-III. The latest SDSS-DR14 quasar catalog of \citet{2018A&A...613A..51P} not only increases the number of quasars by a factor of 5 compared to SDSS DR7, it also covers about 1.5 mag fainter sources ($i$-band absolute magnitude M$_{i} (z=2)<-20.5$) than SDSS DR7 (M$_{i} (z=2)<-22$). As the DR14 quasar catalog includes much fainter quasars, this opens up the possibility of the exploration of the properties of quasars over a large range in luminosity. Though the catalog contains the X-ray, UV, optical, IR and radio imaging properties
  of the quasars wherever available, it lacks spectral information of the sources. About 332,000 ($\sim$ 63\%) sources in DR14 catalog have the continuum S/N$>3$ pixel$^{-1}$.
 This is a factor of 3 larger than the entire sample of S11
 catalog. Thus, DR14 catalog with spectral information will be useful for the astronomical community not only
 for statistical studies of quasars but also to discover and investigate peculiar objects. 
 
We, therefore, carried out detailed spectral modeling of all the quasars 
cataloged in \citet{2018A&A...613A..51P} and provide a new catalog of continuum 
and emission line properties of 526,265 quasars along with $M_{\mathrm{BH}}$ and Eddington 
ratio. This paper is structured as follows. Our data and spectral analysis 
procedures are described in section \ref{sec:analysis}. We compare our 
measurements with the previous works in section 
\ref{sec:results}. In section \ref{sec:SNR}, we discuss the impact of the S/N of the spectra on the derived spectral quantities. We discuss some applications of the catalog in section 
\ref{sec:discussion} with a summary in section \ref{sec:summary}. In Appendix \ref{sec:flag} we define and describe the quality of our spectral measurements and in Appendix \ref{sec:catalog} we present the spectral catalog.
A cosmology with $H_0 = 70\, \mathrm{km\,s^{-1}\, Mpc^{-1}}$, 
$\Omega_\mathrm{m} = 0.3$, and $\Omega_{\lambda} = 0.7$ is assumed throughout.


\section{Data and spectral analysis}\label{sec:analysis}
We started with the SDSS DR14 quasar catalog (version ``DR14Q$\_$v4$\_$4'') by
\citet[][hereafter DR14Q]{2018A&A...613A..51P}, which includes all the 
spectroscopically confirmed quasars observed during any SDSS data release, 
consisting of 526,356 quasars based on $i$-band absolute magnitude M$_{i}(z=2) 
<-20.5$ and having at least one emission line with full width at half maximum 
(FWHM) larger than 500 km$s^{-1}$ or having interesting/complex absorption 
features. It was constructed from SDSS-DR14 \citep{2018ApJS..235...42A} and 
a major part of the newly discovered quasars in DR14Q are from the extended 
Baryon Oscillation Spectroscopic Survey (eBOSS) of SDSS IV 
\citep{2015ApJS..221...27M}. A detailed description of DR14Q 
can be found in \citet{2018A&A...613A..51P}. 

To measure the spectral information of the quasars in DR14Q, we first downloaded 
all the processed and calibrated\footnote{\url{https://www.sdss.org/dr14/spectro/pipeline/}} spectra from the SDSS database\footnote{\url{https://www.sdss.org/dr14/}}. 
We then analysed each spectrum 
using the publicly available multi-component spectral fitting code \textsc{PyQSOFit}\footnote{\url{https://github.com/legolason/PyQSOFit}} 
developed by \citet{2018ascl.soft09008G}. A detailed description of the code and 
its applications can be found in \citet{2019MNRAS.482.3288G} and 
\citet{2019ApJS..241...34S}. First, we corrected each spectrum for Galactic 
extinction using the \citet{1998ApJ...500..525S} map and a Milky Way extinction 
law of \citet{1999PASP..111...63F} with $R_V=3.1$. We then transformed the 
observed spectrum to the rest frame wavelength\footnote{$\lambda_{\mathrm{rest}}=\lambda_{\mathrm{obs}}/(1+z)$, flux $f_{\mathrm{rest}}=f_{\mathrm{obs}}\times(1+z)$ and error in flux $ef_\mathrm{rest}=ef_\mathrm{obs}\times(1+z)$} using the redshift ($z$) value provided in DR14Q.
Finally, we performed  multi-component spectral fittings to each spectrum.

 \begin{table}
 \caption{Emission line fitting parameters. The columns are as follows (1) name of the line complex, (2) wavelength range (in \AA) of the line fitting window, (3) name of the emission line and (4) number of Gaussian used in the fitting.}
	\begin{center}
	\hspace*{-1.1cm}
 	\resizebox{1.2\linewidth}{!}{%
     \begin{tabular}{ l l l l}\hline \hline 
      Complex name    &  wavelength range & emission line name  & Number of Gaussian \\
      				  & (\AA)             &                     &                     \\
      (1)             & (2)               & (3)                 & (4)               \\\hline 
       H$\alpha$      & 6400-6800         & H$\alpha$ broad      & 3                 \\ 
                      &                   & H$\alpha$ narrow     & 1                 \\
                      &                   & [N \textsc{ii}]6549  & 1                 \\
                      &                   & [N \textsc{ii}]6585  & 1                 \\
                      &                   & [S \textsc{ii}]6718  & 1                 \\
                      &                   & [S \textsc{ii}]6732  & 1                 \\
       H$\beta$       & 4640-5100         & H$\beta$ broad       & 3                 \\
                      &                   & H$\beta$ narrow      & 1                 \\
                      &                   & [O \textsc{iii}]4959 core & 1            \\
                      &                   & [O \textsc{iii}]4959 wing & 1            \\
                      &                   & [O \textsc{iii}]5007 core & 1            \\
                      &                   & [O \textsc{iii}]5007 wing & 1            \\
       H$\gamma$      & 4250-4440         & H$\gamma$ broad           & 1           \\
                      &                   & H$\gamma$ narrow          & 1           \\
                      &                   & [O \textsc{iii}]4364      & 1           \\
       Mg \textsc{ii} & 2700-2900         & Mg \textsc{ii} broad      & 2           \\
       				  &                   & Mg \textsc{ii} narrow     & 1           \\
       C \textsc{iii]} & 1850-1970        & C \textsc{iii]}     & 2                 \\
       C \textsc{iv}   & 1500-1600        & C \textsc{iv}       & 3                 \\
       Ly$\alpha$     & 1150-1290         & Ly$\alpha$          & 3                 \\
       				  &      			  & N \textsc{v} 1240   & 1                 \\ \hline\hline
        \end{tabular} } 
        \label{Table:line_complex}
        \end{center}
    \end{table}

   \begin{figure}
  \resizebox{9cm}{8cm}{\includegraphics{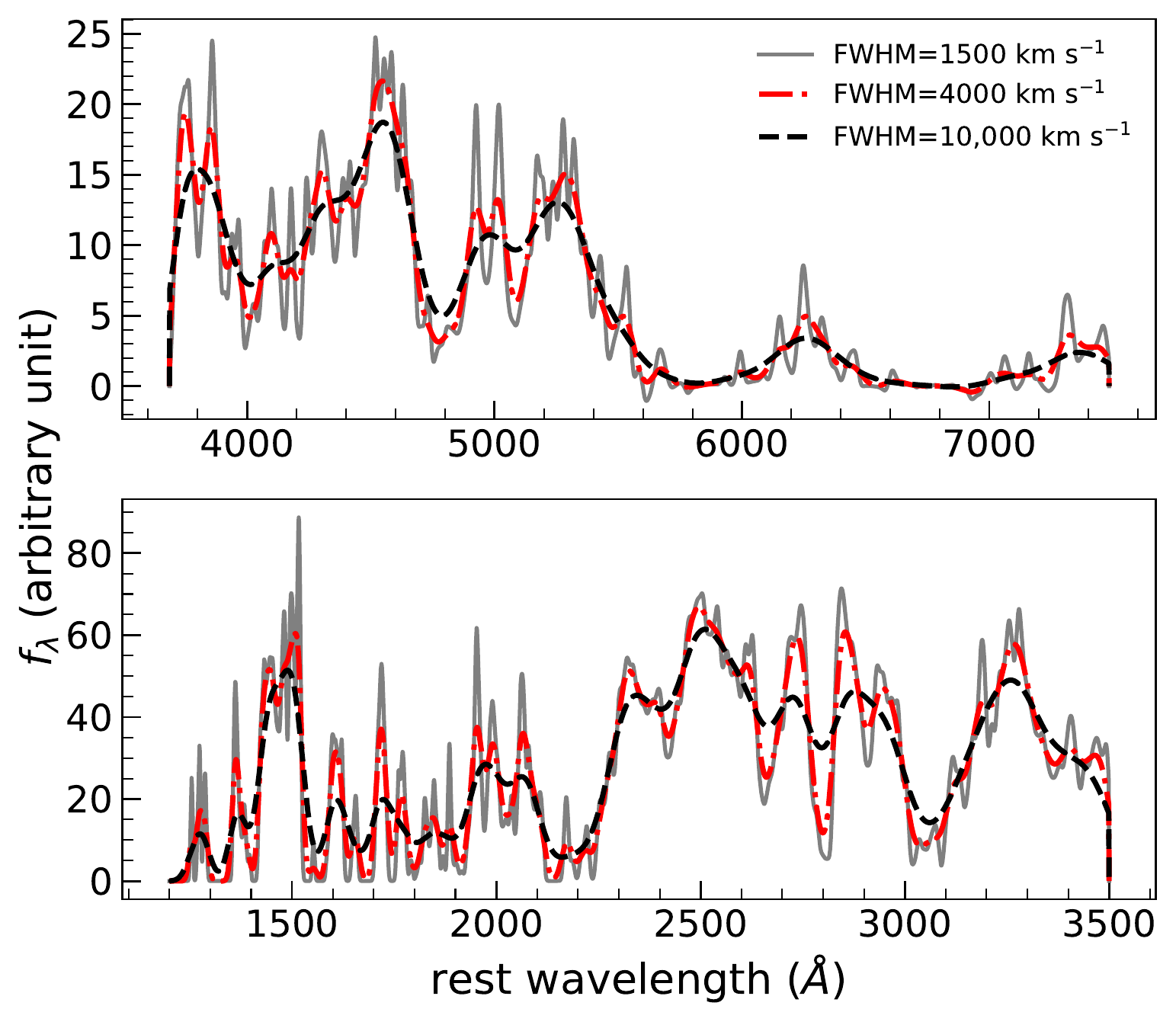}} 
  \caption{Examples of iron template broadened by the convolution of a Gaussian of different FWHM (parameter $b1$ in equation \ref{eq:feii}). The upper panel is the optical Fe II from \citet{1992ApJS...80..109B} and lower panel is the UV Fe II built by \citet{2019ApJS..241...34S} from the templates of \citet{2001ApJS..134....1V}, \citet{2006ApJ...650...57T} and 
  \citet{2007ApJ...662..131S}.}\label{Fig:iron_temp} 
  \end{figure}

    \begin{figure}
    \centering
    \resizebox{8cm}{6cm}{\includegraphics{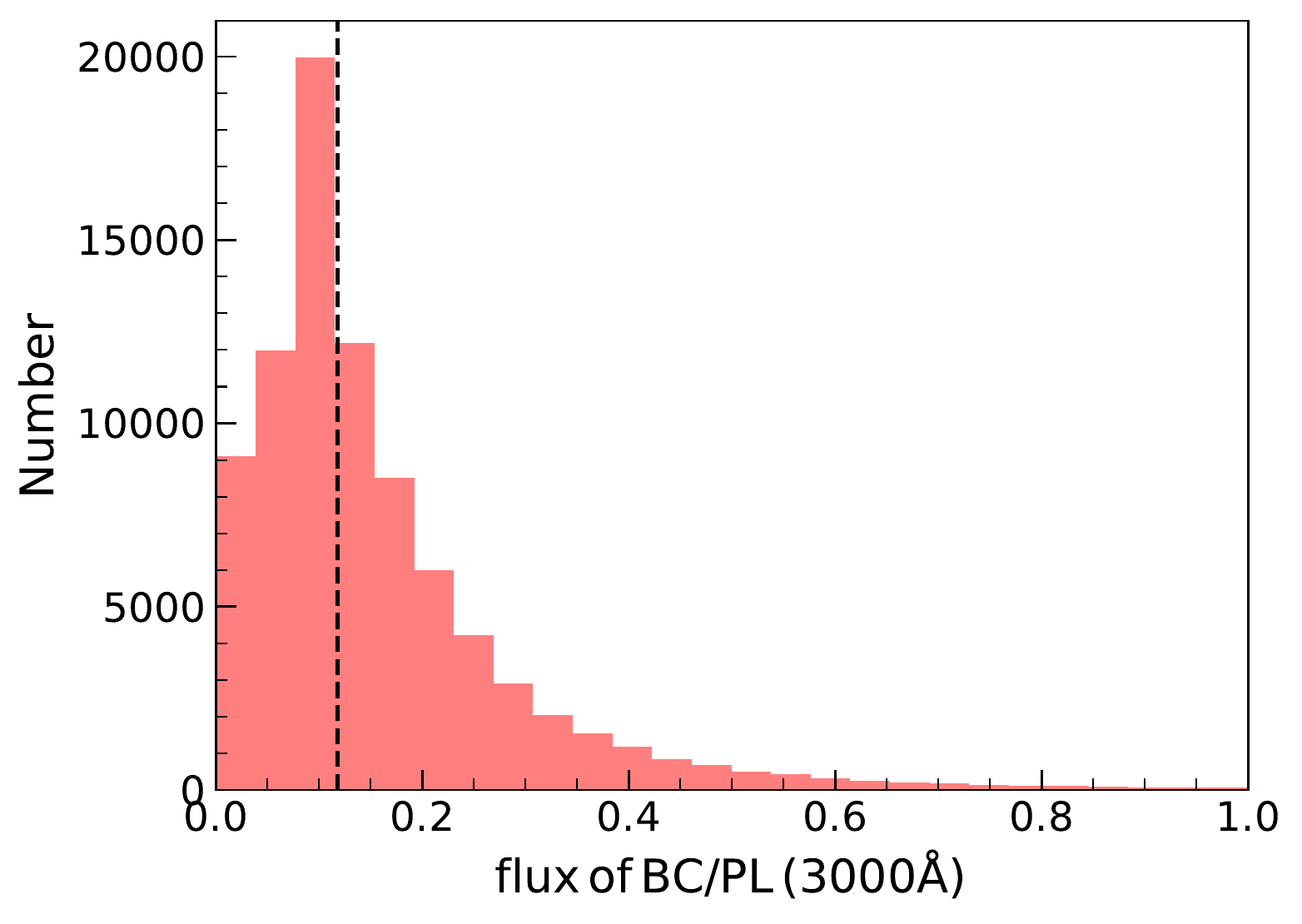}}
    \caption{The distribution of flux ratio of Balmer continuum to power-law at 3000 \AA \, for $z<1.1$. The dashed line represents the median of the distribution.}\label{Fig:fbc} 
    \end{figure}

 \begin{figure*}
 \centering
 \resizebox{18cm}{10cm}{\includegraphics{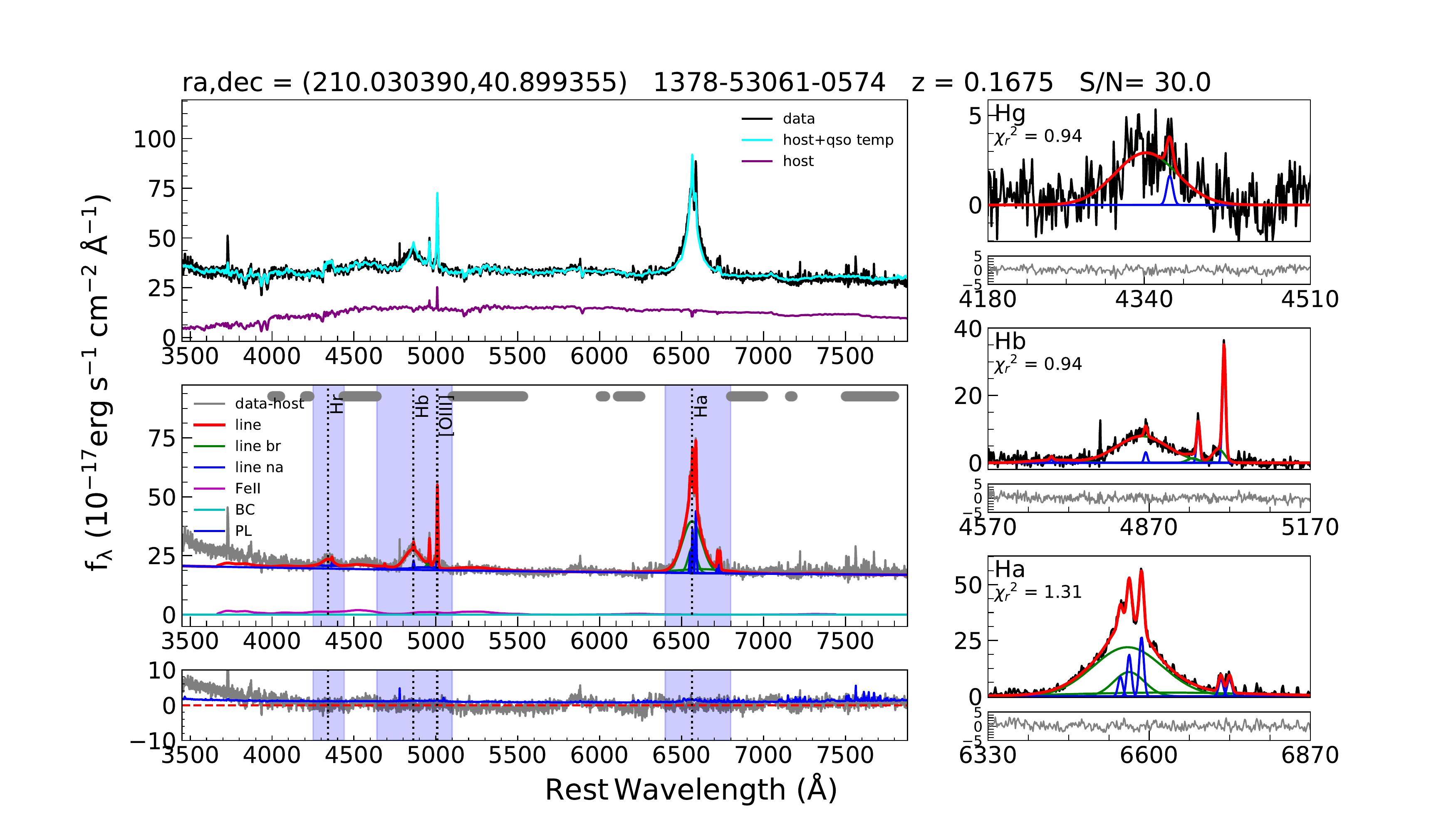}} 
 \resizebox{18cm}{10cm}{\includegraphics{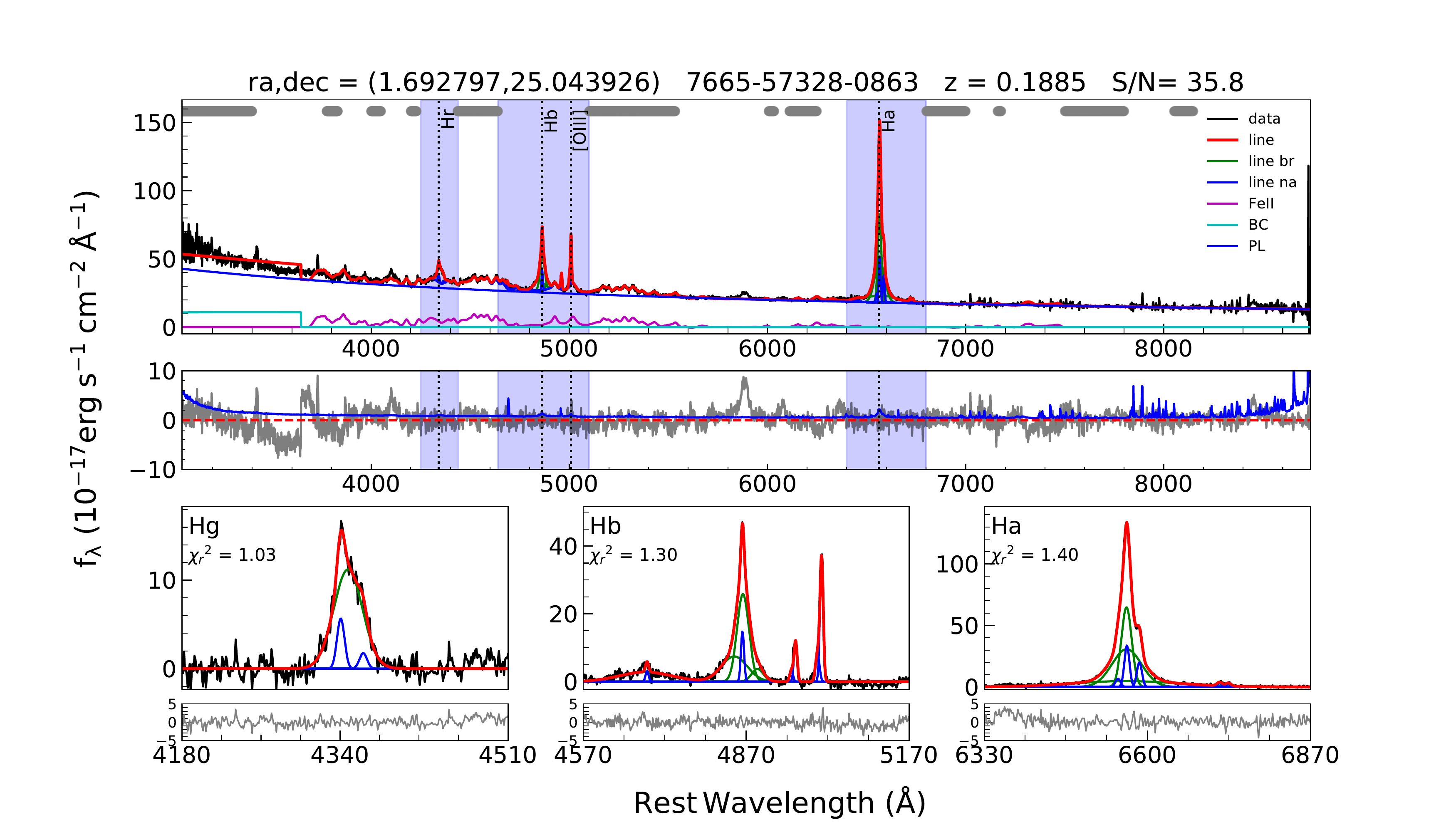}} 
  \caption{Examples of spectral decomposition. Top: an example of a quasar spectrum  
with significant host galaxy contribution. Upper subplot: SDSS data (black), the best fit 
quasar+host galaxy template (cyan) and the decomposed host galaxy (purple). Middle subplot: The host subtracted spectrum (gray), the power-law (blue), 
Fe II (magenta), broad line (green), narrow line (blue) and the total best fit 
model (red), which is a sum of $f_{\mathrm{cont}}$+line are shown. The wavelength windows used to fit AGN continuum is also shown at the top (bar). Bottom subplot: The residual (gray) in the unit of error spectrum i.e., (data-model)/error, is shown along with error spectrum (blue). Bottom: An example of a quasar spectrum  without significant host galaxy contribution. Upper subplot: the power-law (blue), 
Fe \textsc{ii} (magenta), broad line (green), narrow line (blue) and the total best fit 
model (red), which is a sum of $f_{\mathrm{cont}}$+line are shown in each panel. Middle plot: residual (gray) in the unit of error spectrum (blue). A zoomed version of individual line complex is also shown along with the residual in the unit of error spectrum. The shaded region shows the wavelength window at which each line complex was fitted (see Table \ref{Table:line_complex}) and the reduced $\chi^2$ obtained for each of the line complex. The median continuum S/N per pixel is also given.}\label{Fig:fit_hb} 
  \end{figure*}
  \begin{figure*}
 \resizebox{18cm}{10cm}{\includegraphics{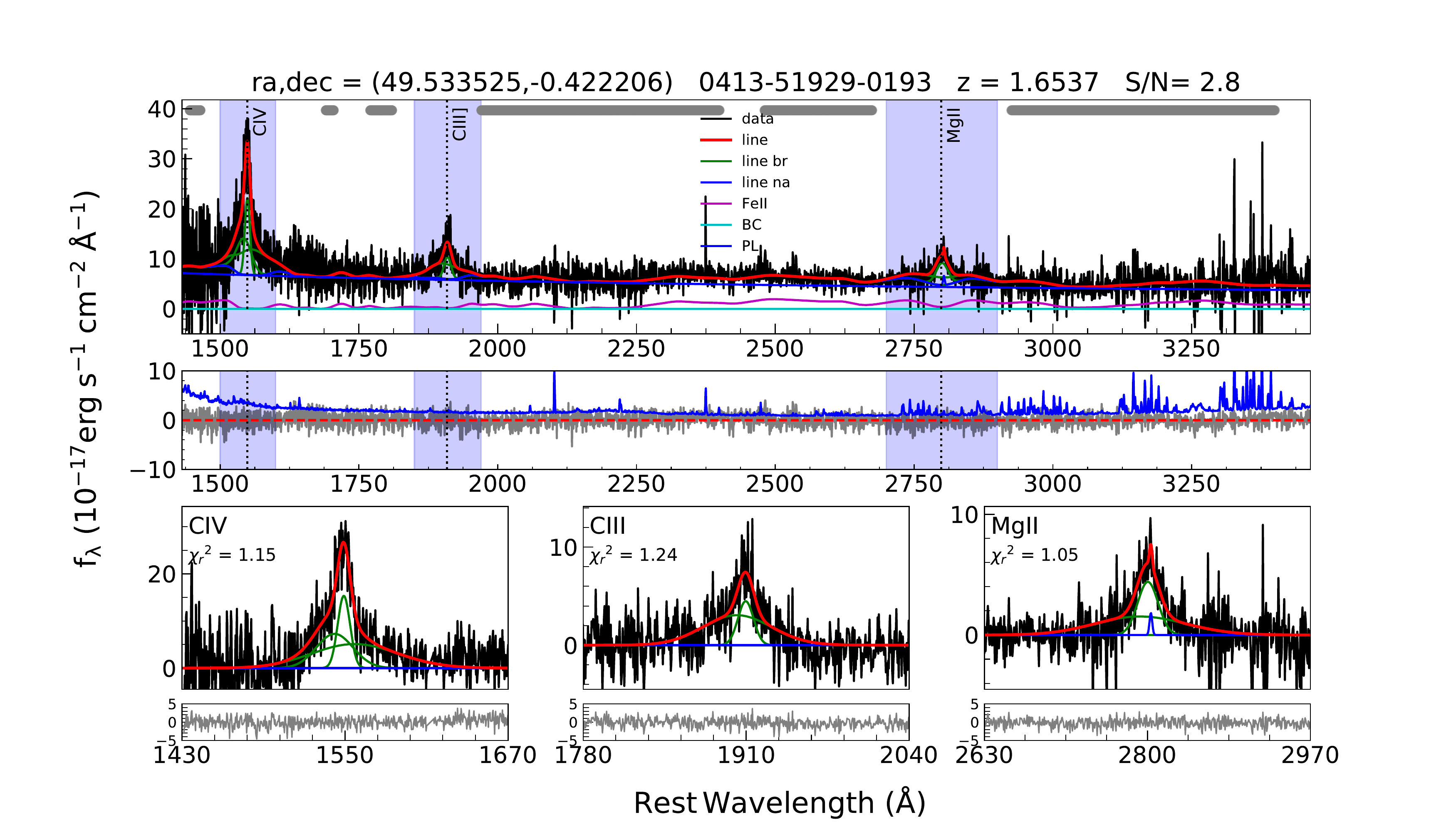}}  
 \resizebox{18cm}{10cm}{\includegraphics{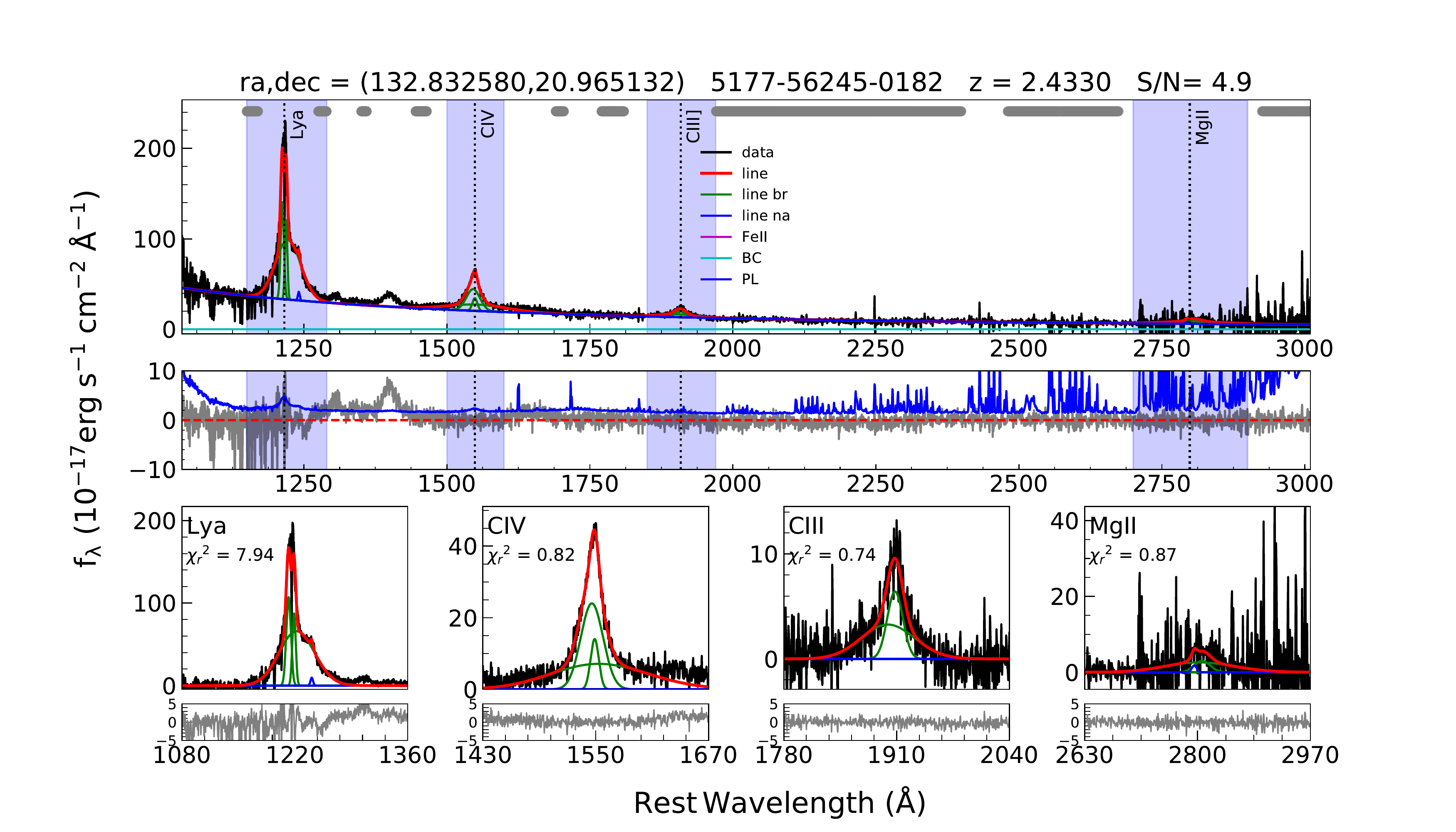}}  
 \caption{Example of spectral decomposition for poor continuum S/N spectra. Labels are same as in Figure \ref{Fig:fit_hb}.}\label{Fig:fit_mgii} 
 \end{figure*}

\subsection{Continuum components}\label{sec:continuum_sub}
The light from stars in the host-galaxy of a quasar can contribute  
to the observed quasar's spectrum, particularly significant for low-z quasars 
($z<0.8$). Thus, to extract intrinsic AGN properties, the host galaxy contribution 
to each spectrum must be removed. We, therefore, carried out host galaxy-quasar decomposition to 
the spectra for $z<0.8$ quasars based on the principal component analysis \citep[PCA;][]{2004AJ....128..585Y,2004AJ....128.2603Y} implemented in \textsc{PyQSOFit} code. The PCA method has been used in several previous studies \citep{2006AJ....131...84V,2008AJ....135..928S,2015ApJ...805...96S} to decompose host galaxy and quasar contribution assuming that the observed composite spectrum is a combination of two independent sets of eigenspectra taken from pure galaxy \citep{2004AJ....128..585Y} and pure quasar \citep{2004AJ....128.2603Y} samples. The first three galaxy eigenspectra contain 98\% of the galaxy sample information while the first ten quasar eigenspectra contain 92\% of the quasar sample information. \citet{2006AJ....131...84V} performed PCA on 11,000 SDSS quasars. They also studied the reliability of spectral decomposition with signal-to-noise ratio (S/N) of the spectrum, host galaxy fraction, galaxy class, etc. The host galaxy decomposition is considered to be successful if the host galaxy fraction in the wavelength range of 4160-4210\AA \, is larger than 10\%. This method also has been applied by \citet{2008AJ....135..928S,2015ApJ...805...96S} to decompose the host galaxy from the SDSS spectra. Here too, we applied PCA to decompose the host-galaxy contribution using 5 PCA 
components for galaxies that can reproduce about 98\%  of the galaxy sample and 20 PCA 
components for quasars that can reproduce about 96\% of the quasar sample and the global model (independent of redshift and luminosity). We then 
subtracted the host contribution, if present, from each spectrum.

Using the host-galaxy subtracted spectrum, we modeled the entire continuum,   
masking the prominent emission lines as 
\begin{equation}
f_{\mathrm{cont}} = f_{\mathrm{pl}} + f_{\mathrm{BC}} + f_{\mathrm{Fe II}}
\end{equation}
where the power-law continuum ($f_{\mathrm{pl}}$) is 
\begin{equation}
f_{\mathrm{pl}} = \beta \big(\lambda/\lambda_0 \big)^{\alpha},
\end{equation}
 with a reference wavelength $\lambda_0=3000\AA$. The parameters $\alpha$ and $\beta$ are the power-law slope and normalization parameter, respectively.

  The Fe \textsc{ii} model ($f_{\mathrm{Fe \, \textsc{ii}}}$) is
  \begin{equation}
  f_{\mathrm{Fe \, \textsc{ii}}} = b_0 \, F_{\mathrm{Fe \, \textsc{ii}}} (\lambda, b1, b2),
  \label{eq:feii}
  \end{equation}
where the parameters $b_0$, $b1$, $b2$ are the normalization, the Gaussian FWHM used to convolve the Fe II template, and the wavelength shift applied to the Fe II template, respectively, to fit the data. Both the UV and optical Fe II 
emission were modeled. In \textsc{PyQSOFit}, the UV Fe II template is a modified template built by \citet{2019ApJS..241...34S} with constant velocity dispersion of 103.6 km s$^{-1}$ from the templates of \citet{2001ApJS..134....1V}, \citet{2006ApJ...650...57T} and 
\citet{2007ApJ...662..131S}. For the wavelength range of $1000-2200$\AA\, 
the template is from \citet{2001ApJS..134....1V}, for $2200-3090$\AA\, the template is from \citet{2007ApJ...662..131S} which extrapolates the Fe II flux 
underneath the Mg II line and for $3090-3500$\AA\, the template is from \citet{2006ApJ...650...57T}. The optical Fe \textsc{ii} template (3686$-$7484 \AA \,) is 
based on \citet{1992ApJS...80..109B}. To model the Fe II emission for each of our spectra, we first convolved the template with the parameter $b1$ which is constrained in the range of 1200-10,000 km s$^{-1}$ with an initial guess of 3000 km s$^{-1}$. A small wavelength shift ($b2$), constrained to be within 1\% of the template wavelength was also applied to fit the data. Then the parameters $b_0$, $b1$, and $b2$ were varied within the range mentioned above until the best fit model which represents the data was found. In Figure \ref{Fig:iron_temp}, we show examples of template broadening for different values of Gaussian FWHM ($b1$).

The Balmer continuum \citep[$f_{\mathrm{BC}}$;][]{1982ApJ...255...25G,2002ApJ...564..581D} is defined as 
 \begin{equation}
  f_{\mathrm{BC}} = F_{\mathrm{BE}} B_{\lambda}(T_\mathrm{e}) \big(1- e^{-\tau_{\lambda} (\lambda/\lambda_{\mathrm{BE}})^3}  \big)
  \label{eq:BC}
  \end{equation}
  where $F_{\mathrm{BE}}$ is the normalized flux density, $\tau_{\lambda}$ is the optical depth at the Balmer edge of wavelength $\lambda_{\mathrm{BE}}=3646$\AA, $B_{\lambda}(T_\mathrm{e})$ is the Planck function at the electron temperature $T_e$. As many low-z objects do not have enough spectral coverage to fit a Balmer component, it is fitted whenever the continuum window has at least 100 pixels below $\lambda_{\mathrm{BE}}$. We used $F_{\mathrm{BE}}$ as a free parameter keeping  $T_e=15,000$ K, and $\tau_{\lambda}=1$ fixed to avoid degeneracy between the parameters following \citet{2002ApJ...564..581D} and C17. Moreover, for sources with $z>1.1$, the Balmer component resembles a simple power-law. Thus, following previous studies (e.g., C17), we further allowed $F_{\mathrm{BE}}$ to vary between 0 and $0.1\times F_{3675}$. Here $F_{3675}$ is the flux density at $\lambda=3675$\AA, where the Fe II contribution is insignificant. The upper limit is also justified from the distribution of flux ratio of Balmer to power-law continuum at 3000\AA \, (see Figure \ref{Fig:fbc}) for $z<1.1$ sources, which has a median of $\sim$0.1.

We noticed that in a few low-z ($z<0.3$) objects the blue part of the spectrum between $3000-4000$ \AA \ is much steeper and the entire continuum cannot be well-fitted with a power-law and Balmer component. For those objects\footnote{These objects have MIN\_WAVE=4000 in the catalog Table \ref{Table:catalog}.}, we limited the spectral fitting range to above 4000\AA, thereby excluding the steep rise towards the UV.

Many high-z spectra were affected by broad and narrow absorption lines that could bias the line 
fitting results, hence,  we used `rej\_abs = True' option in \textsc{PyQSOFit} to reduce this bias. The code first performed continuum modeling (`tmp\_cont') of the spectrum and removed the 3$\sigma$ (where $\sigma$ is the flux uncertainty) outliers below the continuum (i.e., pixels where flux $<$ tmp\_cont $-$3 $\times$ flux uncertainty) for wavelength $< 3500$\AA\, and then performed a second iteration of continuum model fit to the 10 pixels box-car smoothed spectrum excluding the outliers. Such a method is found to be useful to reduce the impact of absorption features as noted in \citet{2011ApJS..194...45S,2019ApJS..241...34S}.

\subsection{Emission line components}\label{sec:line}     
The best fit continuum model was subtracted from each spectrum leading to only 
the line spectrum. Individual line complex was fitted separately, while all the emission lines within a line complex were fitted together. The full list of emission lines 
and the number of Gaussian components used for the individual line are given in 
Table \ref{Table:line_complex}. Broad emission line profiles in many objects can be very complex (e.g., double-peaked, flat top, asymmetric) and can not be well represented by a single Gaussian. Moreover, the line width estimated by a single Gaussian model is systematically larger by 0.1 dex compared to the multiple Gaussian model \citep{2008ApJ...680..169S,2011ApJS..194...45S}. Thus, following previous studies, we used multiple Gaussians to model the broad emission line profiles \citep[e.g.,][]{2005ApJ...630..122G,2011ApJS..194...45S}. During the fitting, the velocity and width of all the narrow components in H$\beta$ and H$\alpha$ complex were tied together 
with an added constraint that the maximum allowed FWHM of narrow components is 900 km s$^{-1}$, while the broad components have FWHM $>900$ km s$^{-1}$. The FWHM criterion was adopted to separate Type 1 AGN from the Type 2 AGN following previous studies \citep[e.g.,][]{2009ApJ...707.1334W,2017MNRAS.472.4051C,2019ApJ...882....4W,2019A&A...625A.123C}. The velocity offsets of the broad and narrow components were restricted to $\pm$3000 km s$^{-1}$ and $\pm$1000 km s$^{-1}$, respectively. Furthermore, the flux ratios of [O \textsc{iii}] and [N \textsc{ii}] doublets were fixed to their 
theoretical values, i.e., $F(5007)/F(4959) =3$ and $F(6585)/F(6549) =3$. Note 
that, we did not use any narrow component to model C \textsc{iii]} and C \textsc{iv} emission lines and the line FWHM and flux were determined from the whole line because of 
ambiguity in the presence of narrow components in these lines 
\citep[also see][]{2011ApJS..194...45S,2019ApJS..241...34S}. 

A few examples of the spectral decomposition are shown 
in Figures \ref{Fig:fit_hb} and \ref{Fig:fit_mgii} for spectra of different qualities. The median continuum S/N (estimated from the rest-frame spectrum in the regions around 5100\AA, 4210\AA, 3000\AA, 2245\AA, and 1350\AA\, depending on the spectral coverage) is also noted in the Figure. Note that due to a large number of quasars, visual inspection of all the spectral fittings was not possible. Thus, only random checks of a few thousand spectra in various redshift and S/N bins were made. All the spectral fitting plots and individual model components are made publicly available for the users. We also provide various quality flags on the spectral quantities to access the reliability of the measurements. The good quality measurements are given a quality flag=0. Any measurements with quality flag $>0$, may not be reliable either due to poor S/N or bad spectral decomposition. Therefore, sources with flag $>0$ should be used cautiously.  \textit{The criteria for fulfillment of each quality flag is defined and described
in Appendix \ref{sec:flag} including detailed statistics on each quality flag.}

\subsection{Spectral quantities}\label{sec:param}   
We measured the continuum (slope, luminosity) and emission line (line peak, FWHM, 
EW, luminosity, etc.) properties from the best fit model\footnote{For 91 out of the 526,356 quasars in DR14Q, there is insufficient or no valid data points in the spectrum to perform spectral decomposition, and therefore, these objects were excluded in this work.}. Various studies \citep[e.g.,][]{2006A&A...456...75C,2011ApJS..194...42R} suggested that line dispersion ($\sigma_{\mathrm{line}}$) i.e., the second moment of the line \citep[see][]{2004ApJ...613..682P} is a better measure of emission line width compared to FWHM. However, FWHM is less affected by the noise in the line wings and treatments of line blending (e.g., H$\beta$ blended with Fe II, He II$\lambda$4686 and [O III]) than the ${\sigma}_{\mathrm{line}}$. On the other hand, ${\sigma}_{\mathrm{line}}$ is less sensitive to the treatments of narrow component removal and peculiar line profiles. Instead of line dispersion, FWHM is preferred because of its easiness of the measurement and repeatability, especially in poor quality spectra where the line wings are difficult to constrain and ${\sigma}_{\mathrm{line}}$ can not be measured reliably. Despite that, following the prescription of \citet{2019ApJ...882....4W}, we also measured ${\sigma}_{\mathrm{line}}$ for all the broad emission lines and included them in the catalog.

The uncertainty in each of the spectral quantities was estimated using Monte Carlo approach \citep[e.g.,][]{2011ApJS..194...45S,2019ApJS..241...34S,2018ApJ...865....5R}. We created a mock spectrum by adding to the original spectrum at each pixel a Gaussian random deviate with zero mean and $\sigma$ given by the flux uncertainty at that pixel. We then performed the same spectral fitting on the mock spectrum as was done for the original spectrum and estimated all the spectral quantities from the mock spectrum. We created 50 such mock spectra for each object allowing us to obtain the distribution of each spectral quantity. Finally, for each spectral quantity, semi-amplitude of the range enclosing the 16th and 84th percentiles of the distribution was taken as the uncertainty of that quantity. Therefore, all the uncertainties of the spectral quantities reported in this work were calculated using the Monte Carlo approach.
   
We calculated $L_{\mathrm{bol}}$ from the monochromatic luminosity using the 
bolometric correction factor given in S11 as adapted from the analysis in \citet{2006ApJS..166..470R}
   \[
        L_{\mathrm{bol}}= 
    \begin{cases}
        9.26\times L_{5100}, & \text{if } z< 0.8\\
        5.15\times L_{3000}, & \text{if } 0.8\leq z<1.9\\       
        3.81\times L_{1350}, & \text{if } z\geq 1.9
    \end{cases}
    \]
Note that the above correction factors are derived from the mean spectral energy 
distribution of AGN and using a single value could lead to 50\% uncertainty in 
$L_{\mathrm{bol}}$ measurements \citep{2006ApJS..166..470R}. Estimating bolometric luminosity for individual source requires multi-band data from radio to X-ray to build spectral energy 
distribution, which is not available for most of the quasars. However, the bolometric correction factor allows us to estimate the bolometric luminosity from their monochromatic luminosity albeit with large uncertainty.

 \begin{figure}
 \centering
 \resizebox{9cm}{9cm}{\includegraphics{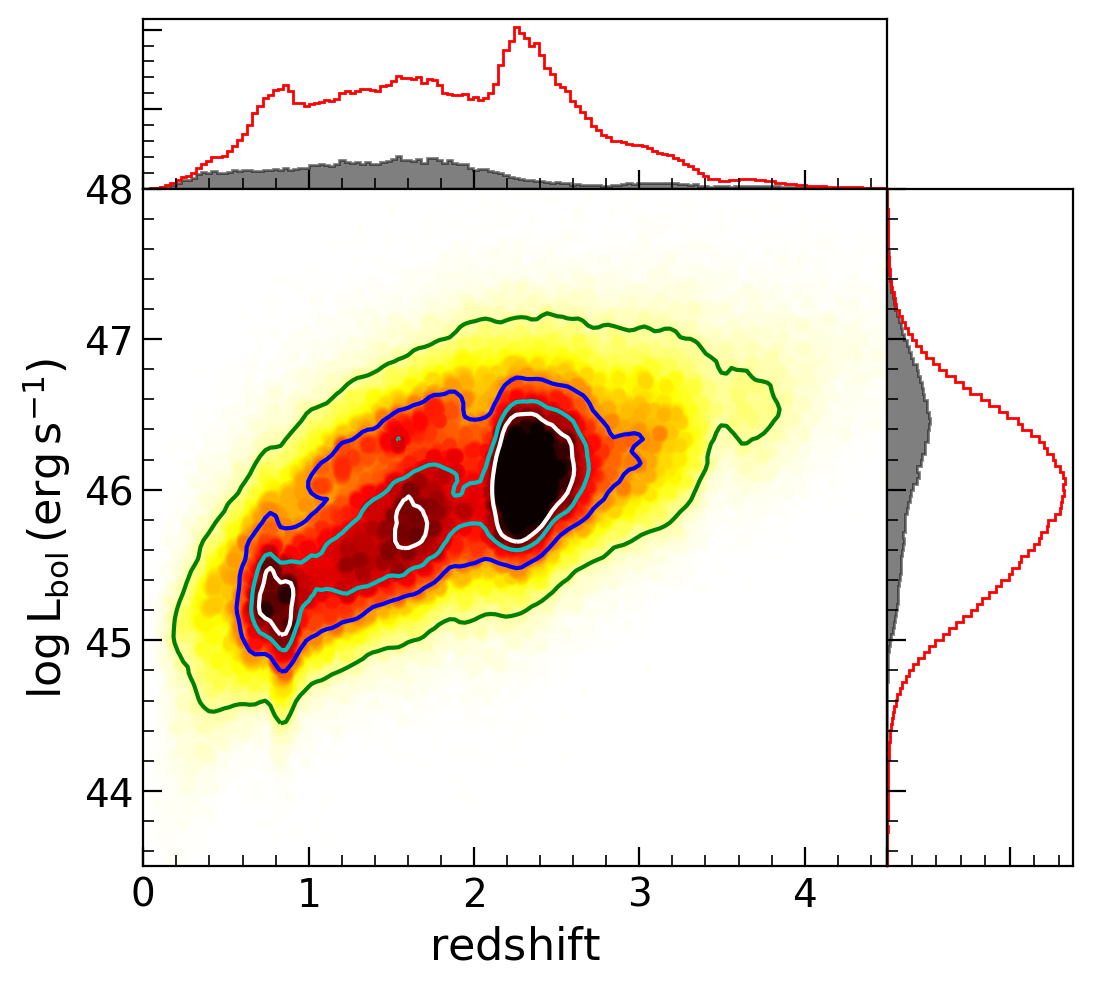}}
 \caption{Bolometric luminosity vs. redshift. The 20, 40, 68 and 95 percentile 
density contours along with density maps are shown. The 
redshift and bolometric luminosity distributions  are also shown for DR14 quasars sample (empty histogram, this work) along with DR7 quasars (filled histogram, S11). A large number of high redshift and low luminous quasars were targeted by DR14 compared to DR7.}\label{Fig:Lbol_z} 
 \end{figure}

 In Figure \ref{Fig:Lbol_z}, we plot $L_{\mathrm{bol}}$ against redshift for 
DR14 quasars. The DR7 quasars from S11 are also shown. As mentioned in 
\citet{2018A&A...613A..51P}, the peak of the redshift distribution at $z\sim2.5$ 
is due to the quasars observed by SDSS-III to access Ly$\alpha$ forest while 
the peaks at $z\sim0.8$ and 1.6 are due to the known degeneracy in color-redshift relation of the quasar 
target selection \citep[see also][]{2012ApJS..199....3R}. For example, a large number of quasars at $z\sim0.8$ have Mg II line at the same wavelength as Ly$\alpha$ at $z\sim3.1$ providing the same broad-band color. Similarly, quasars at $z\sim1.6$ have C IV line at the same wavelength as Ly$\alpha$ at $z\sim2.3$. The bolometric luminosity 
has a median of $\log (L_{\mathrm{bol}})=45.94^{+0.54}_{-0.59} \, \mathrm{erg\, s^{-1}}$ 
with a range of $44.43-47.32$ erg s$^{-1}$ ($3\sigma$ around the median). The errors given in the median bolometric luminosity do not include the uncertainties in the bolometric correction factor. A large fraction of low-luminosity quasars is included in DR14 compared to DR7. For 
example, the fraction of quasars with $\log L_{\mathrm{bol}}<46 \, \mathrm{erg\, s^{-1}}$ in DR14 is about 54\% compared to 27\% in DR7.

We included the commonly used BALnicity Index \citep[BI;][]{1991ApJ...373...23W} 
and its uncertainty from the SDSS DR14 quasars catalog of 
\citet{2018A&A...613A..51P} to flag the broad absorption line quasars (BAL-QSOs) 
in this work. Due to a large number of quasars, \citet{2018A&A...613A..51P} performed a fully automated 
detection of BAL for all $z>1.57$ quasars focusing on C \textsc{iv}  
absorption troughs. A total of 21,876 quasars with C \textsc{iv} absorption troughs 
wider than 2000 km s$^{-1}$ are present in this work. We also included the BAL Flag of SDSS DR7 quasars from S11 who culled the BAL flag from the study of \citet{2009ApJ...692..758G} SDSS DR5 BALQSO catalog and visually inspected post-DR5 BALQSO with redshift $z > 1.45$.

All the parameters and their uncertainties derived in this work are compiled into a catalog (``dr14q\_spec\_prop.fits''), which is 
described in section \ref{sec:catalog} and Table \ref{Table:catalog}, containing 274 columns. We also provide an extended catalog (``dr14q\_spec\_prop\_ext.fits'') where we appended all other 
information from \cite{2018A&A...613A..51P}, which include multi-band imaging 
properties, thereby leading to 380 columns in our extended catalog. Both the catalogs along with other supplementary materials (e.g., best-fit model components and spectral decomposition plots for all objects) are available online\footnote{\url{https://www.utu.fi/sdssdr14/}}.

\begin{figure}
\centering
\resizebox{9cm}{12cm}{\includegraphics{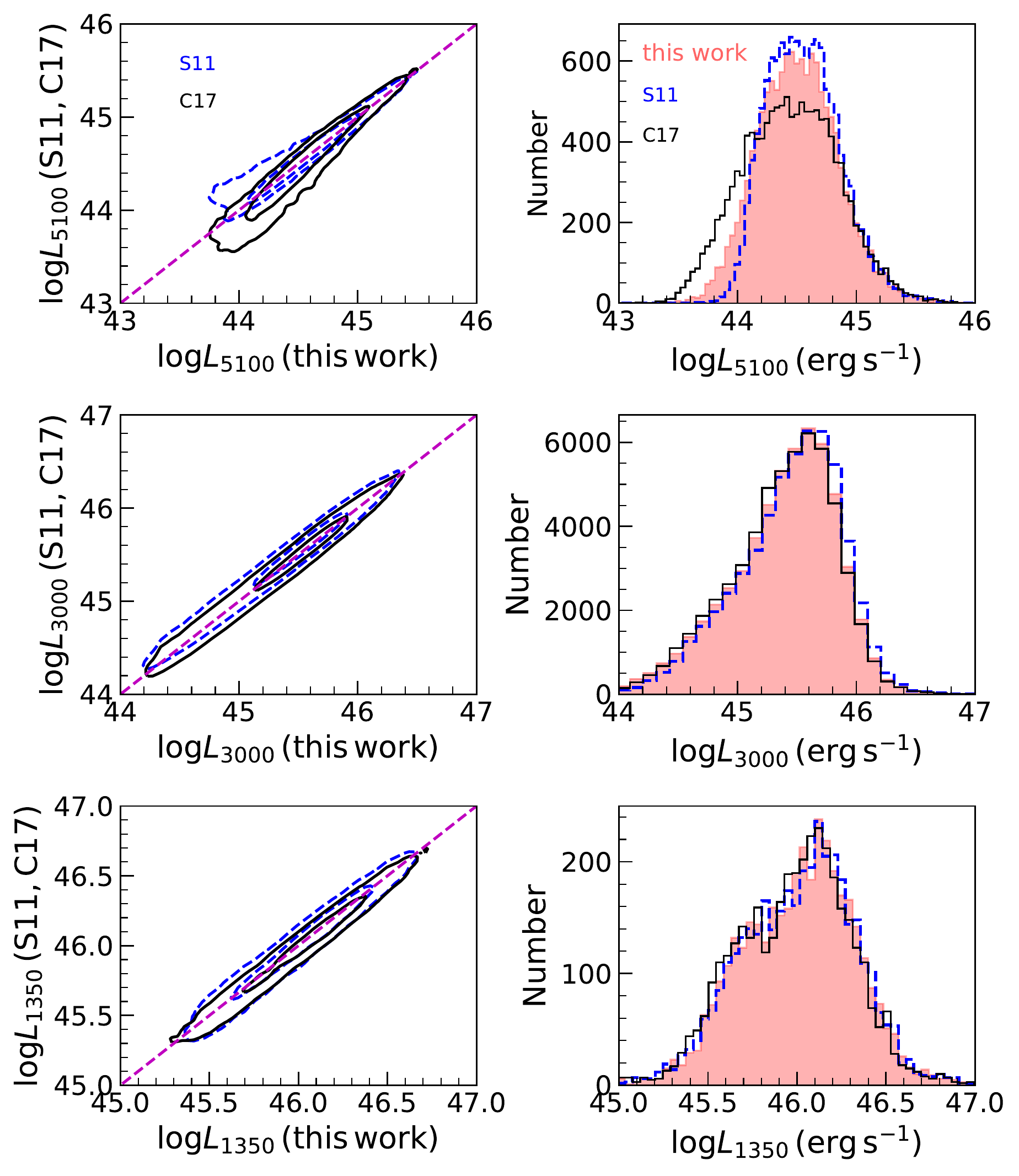}}
\caption{Comparison of continuum luminosity measurements between our work and S11 and C17 for all the common quasars. The inner and outer contours are the $1\sigma$ and $2\sigma$ density contours. The one-to-one line is also shown.}\label{Fig:lum_cont} 
\end{figure}

\begin{figure}
\centering
\resizebox{9cm}{12cm}{\includegraphics{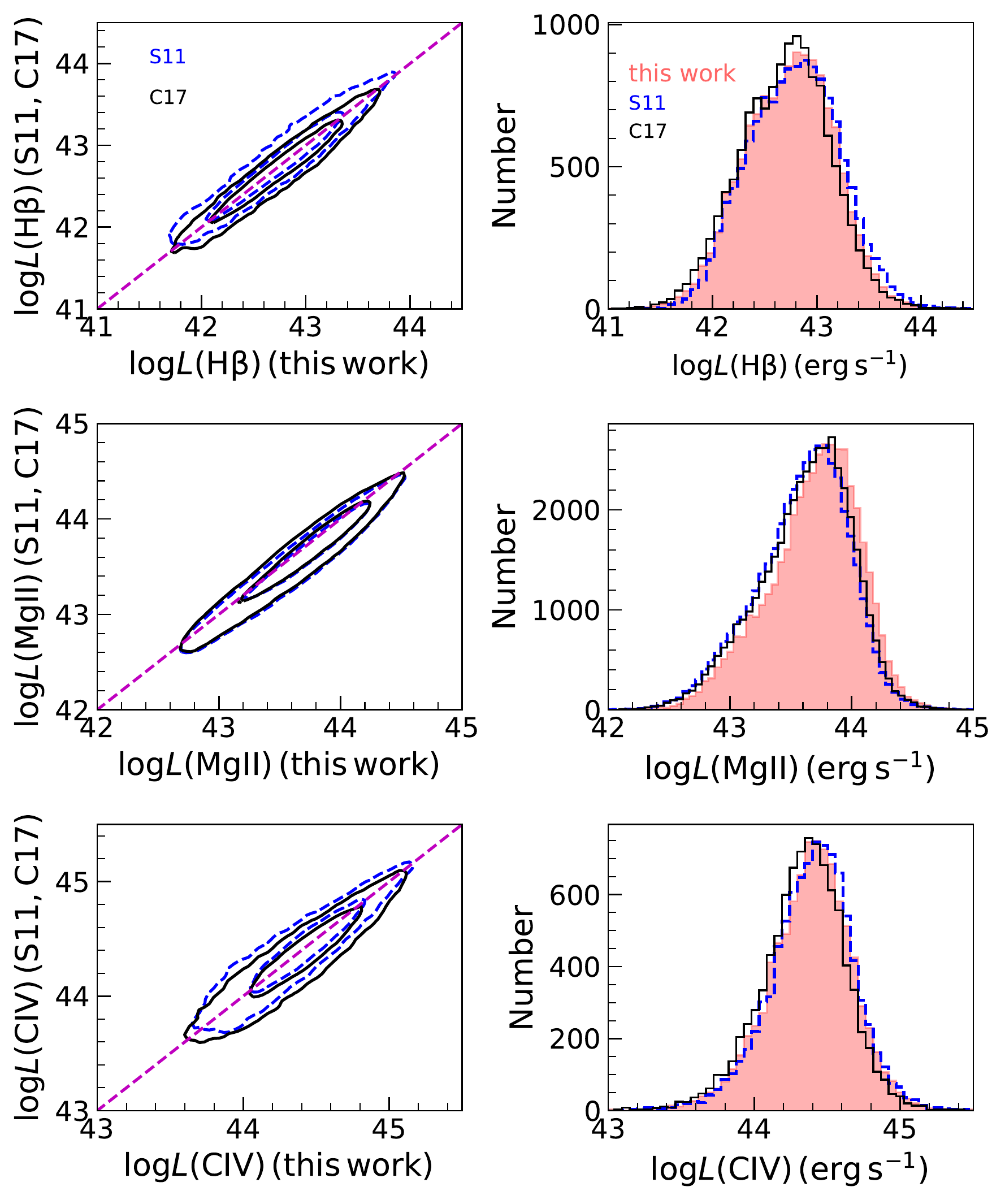}}
\caption{Same as in Figure \ref{Fig:lum_cont}  but for line luminosities.}\label{Fig:lum_line} 
\end{figure}

\begin{figure}
\centering
\resizebox{9cm}{12cm}{\includegraphics{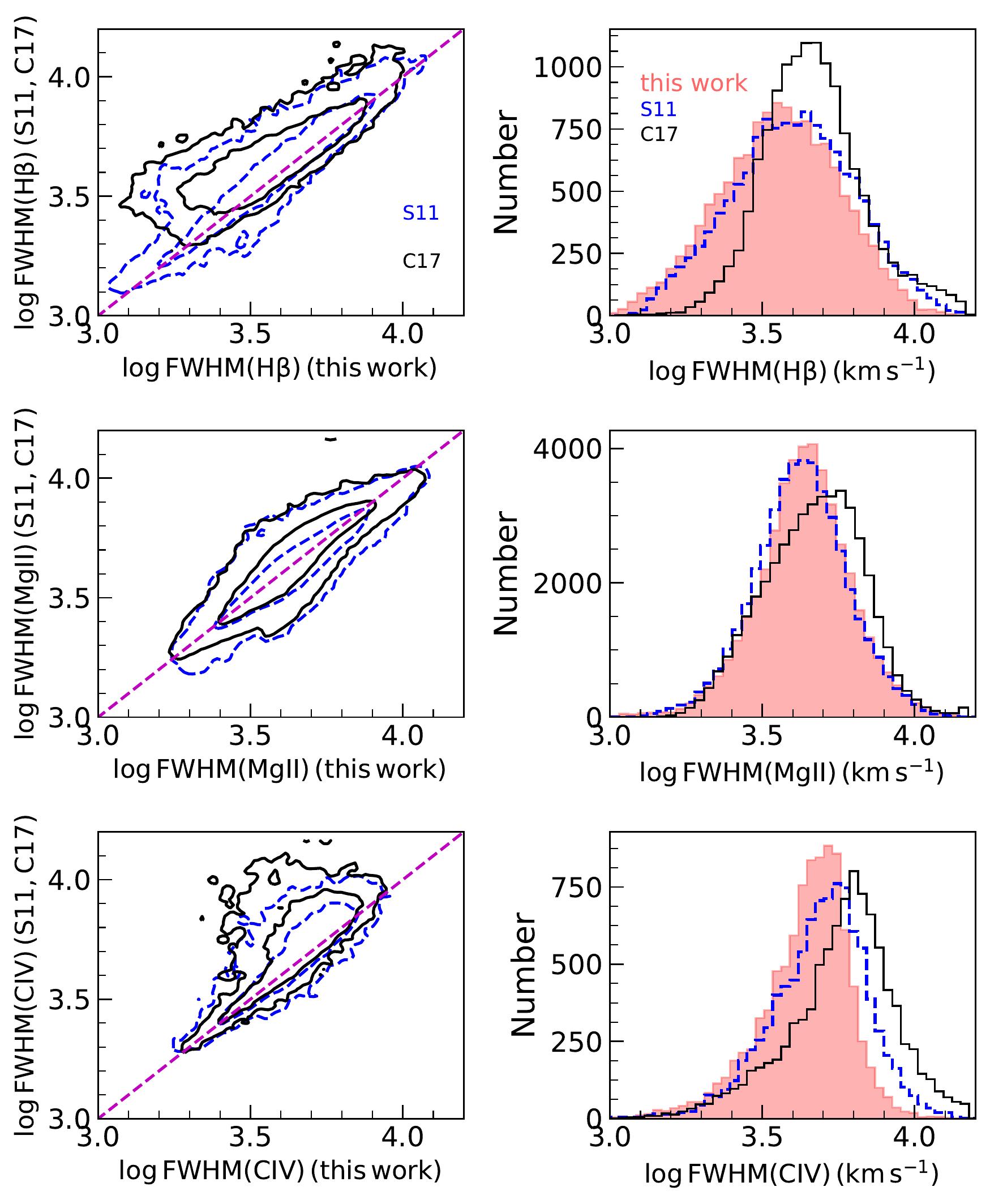}}
\caption{Same as in Figure \ref{Fig:lum_cont} but for line FWHM.}\label{Fig:fwhm} 
\end{figure}

\section{Comparison with previous studies}\label{sec:results}

We compared our measurements with S11 and C17 catalogs. The former catalog is 
based on all DR7 quasars up to $z=5$ with 105,783 entries while the later is based on DR10 
quasars up to $z=2$ with 71,261 entries (catalog version ``qsfit$\_$1.2.4''). Although different methods were used 
by both the authors than the \textsc{PyQSOFit} code used in our analysis, a 
comparison can be made. We refer the readers to \citet{2017MNRAS.472.4051C} for 
a discussion on S11 and C17 spectral analysis methods. Here we summarize the main differences 
\begin{enumerate}
\item S11 didn't decompose host galaxy contamination. C17 used an elliptical galaxy template to represent the host galaxy contamination. We subtracted host galaxy contribution using the PCA method with 5 PCA components for galaxy (see section \ref{sec:continuum_sub}).
\item S11 modeled local AGN continuum (using a power-law) including Fe II emission then fitted the emission lines of the continuum and Fe II subtracted spectrum. C17 fitted continuum (power-law + Balmer continuum) of the whole spectrum and at the final fitting step, they fitted all components (continuum + galaxy + iron + emission lines) simultaneously. We first removed the host galaxy contribution if present and then fitted AGN continuum (power-law + Balmer continuum) and Fe II template of the whole host subtracted spectrum. Finally, we fitted the emission lines of the continuum subtracted spectrum.
\item Both S11 and C17 used the UV Fe II template from \citet{2001ApJS..134....1V}, which is limited to $3090\AA$, while the UV Fe II template used in this work has an extended coverage up to $3500\AA$.
\item S11 fitted broad lines with up to three Gaussian while C17 started their modeling of broad lines with single Gaussian (`known' line) and added more Gaussian if `unknown' emission lines are present close to the known lines. We fitted most of the broad emission lines using multiple Gaussians (see Table \ref{Table:line_complex}).  
\end{enumerate}

 We cross-matched our 
catalog with S11 and C17 using \textsc{TOPCAT}\footnote{http://www.star.bris.ac.uk/~mbt/topcat/} \citep{2005ASPC..347...29T} and 
took only the common entries (71,163) for comparison. However, different catalogs may include spectra of different quality for the same target as repeated observations have been performed by SDSS. Quasars also show spectral variability, which can affect the measurements included in different catalogs. Therefore, we cross-matched sources having the same SDSS plate-mjd-fiber in all three catalogs and found 65,170 matches. Furthermore, we only included measurements having a quality flag of 0 in both C17 and our work.

In Figure \ref{Fig:lum_cont}, we compare our continuum luminosity measurements 
with S11 and C17 where our measurements are plotted along the x-axis in the left 
panels. We also plot the distribution of measurements for all three catalogs 
in the right panels.  In general, we find excellent agreement between the 
measurements. The mean and standard deviation of the difference between this 
work and S11 (C17) is  $0.001\pm0.052$ ($0.023\pm0.031$) dex for $L_{1350}$ (3827 sources), 
$-0.054\pm0.055$ ($0.010\pm0.052$) dex for $L_{3000}$ (56,577 sources) and $-0.045\pm0.104$ 
($0.067\pm0.122$) dex for $L_{5100}$ (12,967 sources). We notice a larger difference in the 
estimates of $L_{5100}$ compared to other luminosities between S11, C17 and our work but mainly for low-luminosity quasars. Our estimates of $L_{5100}$ lie in between S11 and C17. We attribute this difference due
to differences in the host galaxy subtraction procedures. For example, S11 did not perform host galaxy decomposition, thus, their measurements are contaminated 
by the host galaxy contribution.  On the other hand, C17 used a 
single 5 Gyr old elliptical galaxy template to subtract the host galaxy contribution, while, 
we used the PCA method implemented in \textsc{PyQSOFit} to subtract the host galaxy 
(see section \ref{sec:continuum_sub}). Although the PCA host decomposition method allowed us to systematically decompose stellar contribution from a large number of spectra, it is a simplistic approach and in principle one can use other host galaxy decomposition methods \citep[e.g.,][]{2015ApJ...811...91M,2018ApJ...865....5R} using different stellar templates \citep[e.g.,][]{2003MNRAS.344.1000B,2004ApJS..152..251V} to decompose the stellar contribution from the spectra of quasars.

\begin{figure}
\centering
\resizebox{9cm}{8cm}{\includegraphics{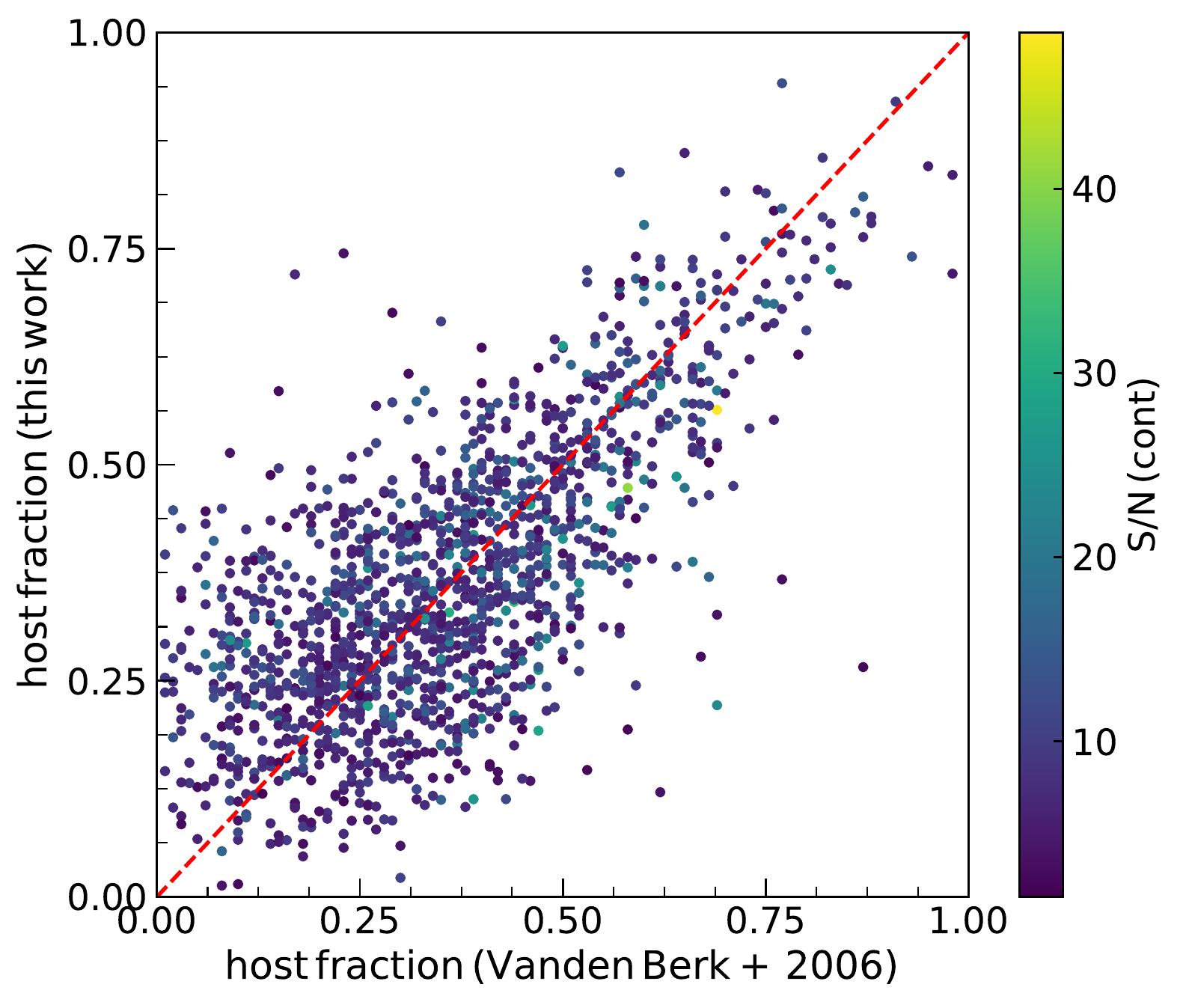}}
\caption{Host galaxy fraction to that of the total flux in the wavelength range of 4160-4210\AA \, in this work (y-axis) is compared to that of \citet{2006AJ....131...84V} color codded by continuum S/N. The one-to-one line is also plotted (dashed-line).}\label{Fig:host} 
\end{figure}

   \begin{figure}
       \resizebox{9cm}{13cm}{\includegraphics{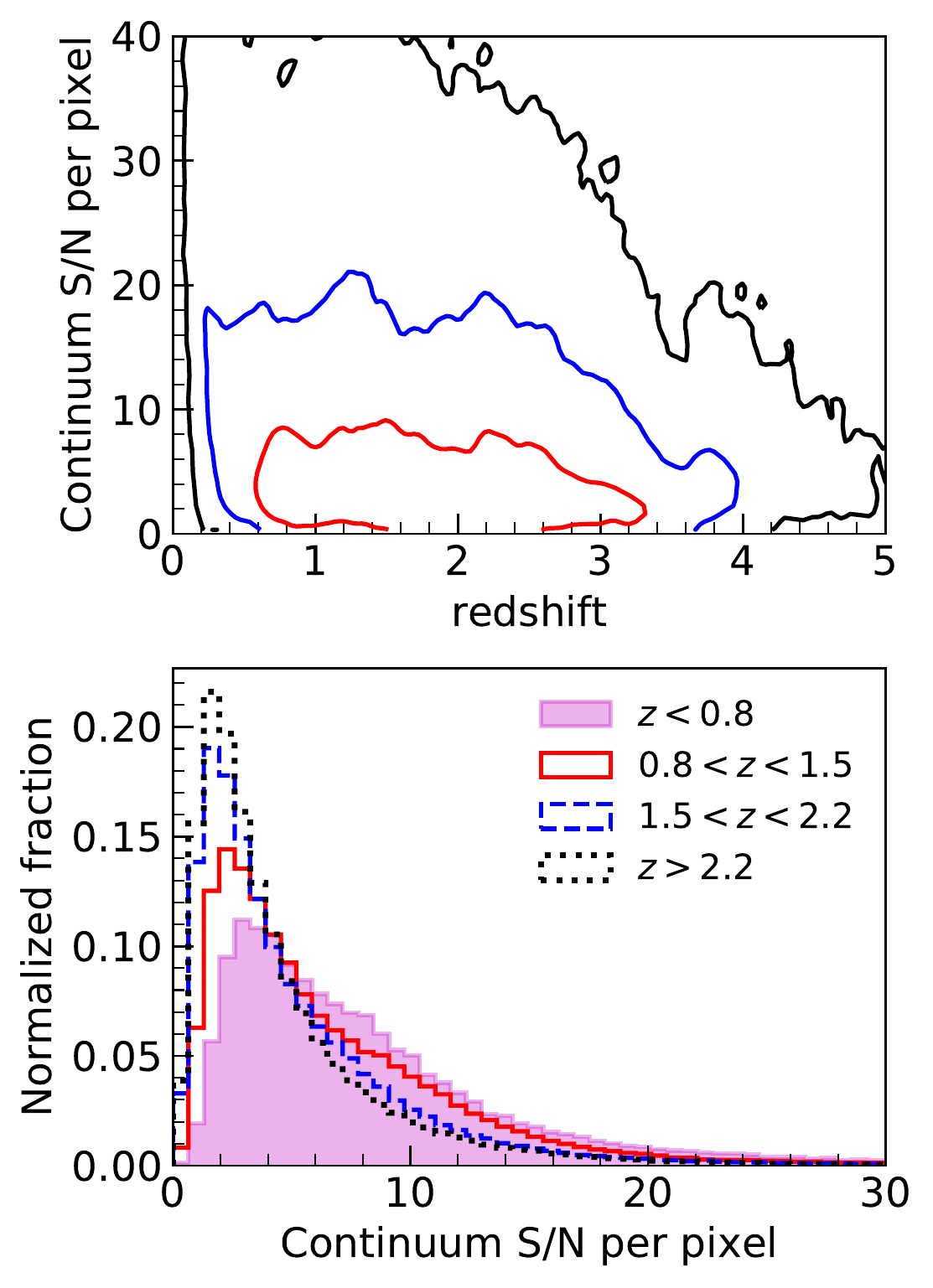}} 
       \caption{Top: The median continuum S/N per pixel is plotted against redshift. The 1$\sigma$ (red), 2$\sigma$ (blue) and 3$\sigma$ (black) density contours are shown. Bottom: Normalized distribution (i.e. the area under each histogram is unity) of continuum S/N for different redshift range. The x-axis is limited to 30.}\label{Fig:cont_snr} 
       \end{figure}

In Figure \ref{Fig:lum_line}, we compare the H$\beta$ (top), Mg \textsc{ii} (middle) and C \textsc{iv} (bottom) line luminosity measurements between all three catalogs. In all cases, we found strong agreement. The mean and standard deviation of H$\beta$ (13,177 sources) line luminosity between this work and S11 (C17) are $-0.045\pm0.111$ ($0.044\pm0.109$) dex, while the same for Mg \textsc{ii} (45,048 sources) and C \textsc{iv} (9,384 sources) line luminosities are $0.094\pm0.083$ ($0.067\pm0.088$) dex and $-0.014\pm0.106$ ($0.043\pm 0.106$) dex, respectively. All the line luminosity plots show a strong correlation with the Spearman rank correlation coefficient $r_s>0.95$ for both H$\beta$ and Mg \textsc{ii} lines, while 0.94 (0.93) for C \textsc{iv} line luminosity between this work and SII (C17). We note that compared to H$\beta$ and CIV line luminosities, Mg II line luminosity shows a larger offset. Our estimated Mg II luminosity is slightly larger compared to S11 and C17. This could be due to the use of different UV Fe II templates. For example, \cite{2019ApJ...874...22S} found that \citet{2006ApJ...650...57T} template provides an average 0.13 dex higher Mg II flux and 0.10 dex lower UV Fe II flux compared to \citet{2001ApJS..134....1V} template.

The emission line widths in different catalogs are less strongly correlated (Figure \ref{Fig:fwhm}) having $r_s=0.85$ (0.75) for H$\beta$, 0.82 (0.82) for Mg \textsc{ii} and 0.72 (0.45) for C \textsc{iv} between this work and S11 (C17) indicating the complexity in the measurement of FWHM. The mean and standard deviation between this work and S11 (C17) is $-0.045\pm0.113$ ($-0.111\pm0.140$) dex for H$\beta$, $0.011\pm0.097$ ($-0.031\pm0.100$) dex for Mg \textsc{ii} and $-0.048\pm0.119$ ($-0.118\pm0.173$) dex for C \textsc{iv} line width measurement. We note that on average our FWHM measurements are more consistent with S11 compared to C17. Although a slight discrepancy between different catalogs is found, measurements are in general agreement with S11 and C17. The discrepancy between different catalogs is due to the use of different spectral decomposition methods as mentioned above.

We compared our estimated host fraction with that of \citet{2006AJ....131...84V} where the host fraction, the ratio of host flux to the total flux, is estimated in the wavelength range of 4160-4210\AA. To avoid any difference due to the spectral quality between the two catalogs, we only considered objects having the same spectra in both the works (SDSS plate-mjd-fiber). There are 1486 sources, which have host contribution $>0$ in both the works. We plotted them in Figure \ref{Fig:host} (color-codded by continuum S/N). Our results are consistent with them having a median ratio (our to their) of $1.01^{+0.58}_{-0.31}$. Therefore, our stellar fraction measurements are consistent with that of \citet{2006AJ....131...84V}.

\begin{figure*}
\centering
\resizebox{5cm}{9cm}{\includegraphics{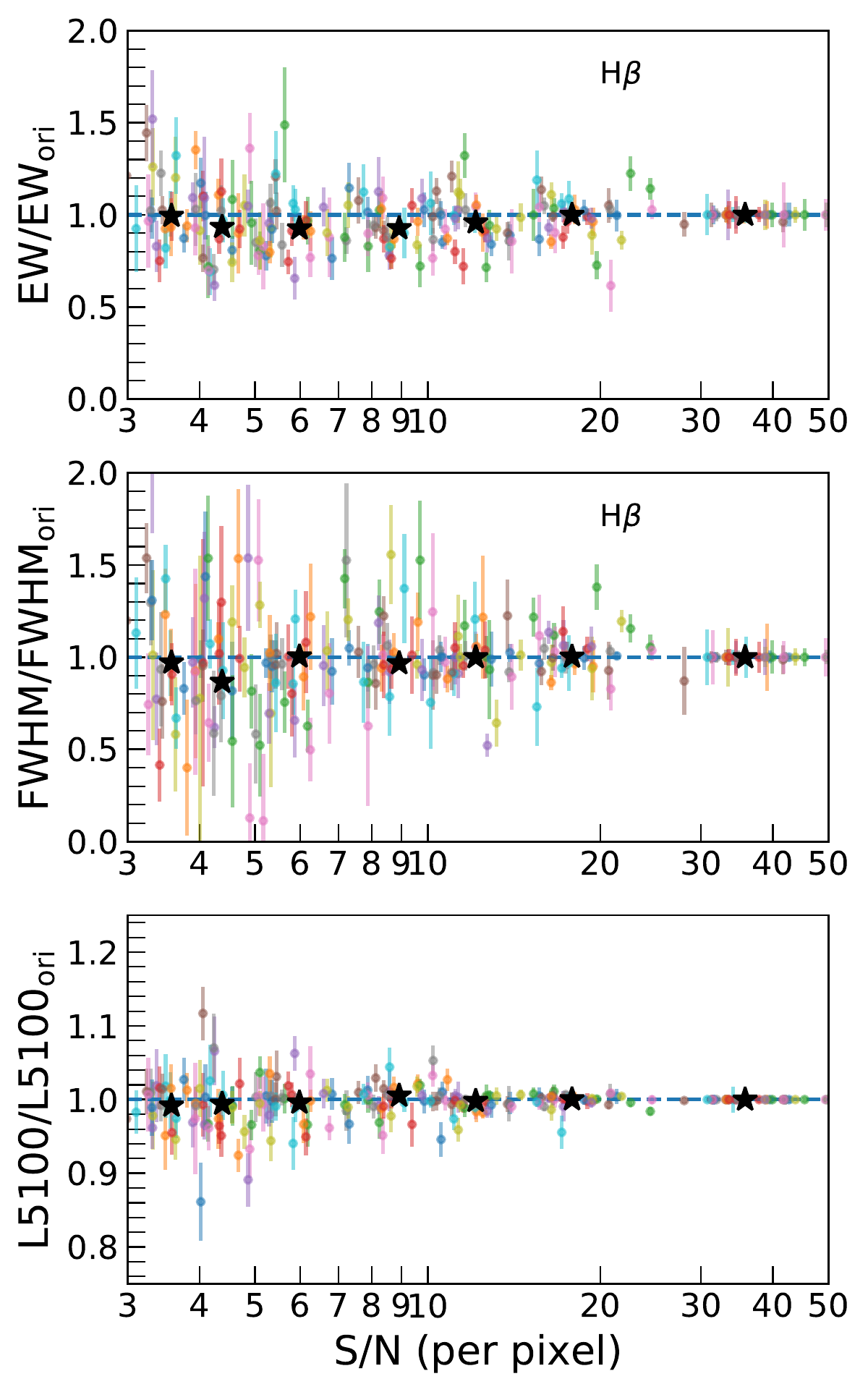}}
\resizebox{5cm}{9cm}{\includegraphics{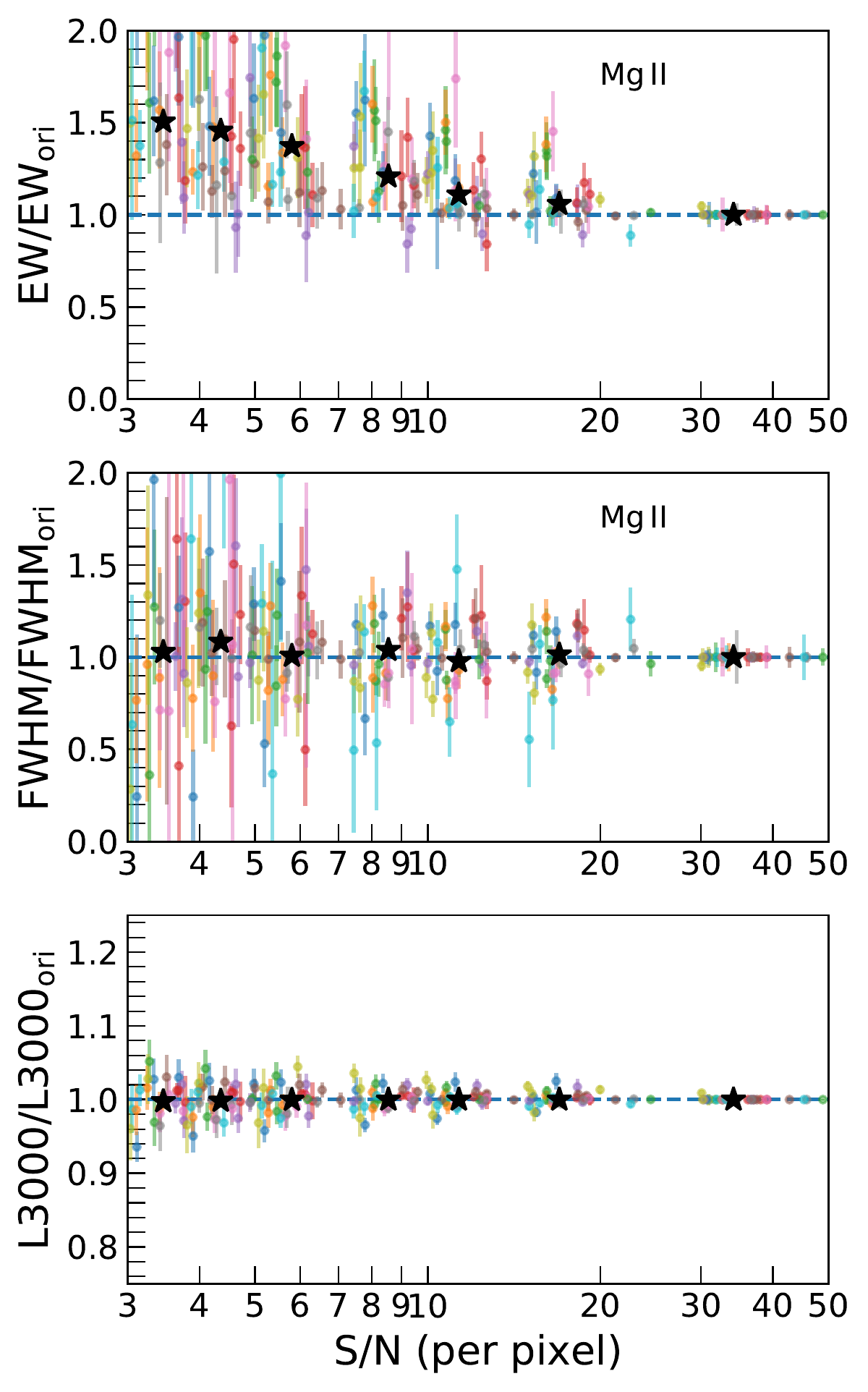}}
\resizebox{5cm}{9cm}{\includegraphics{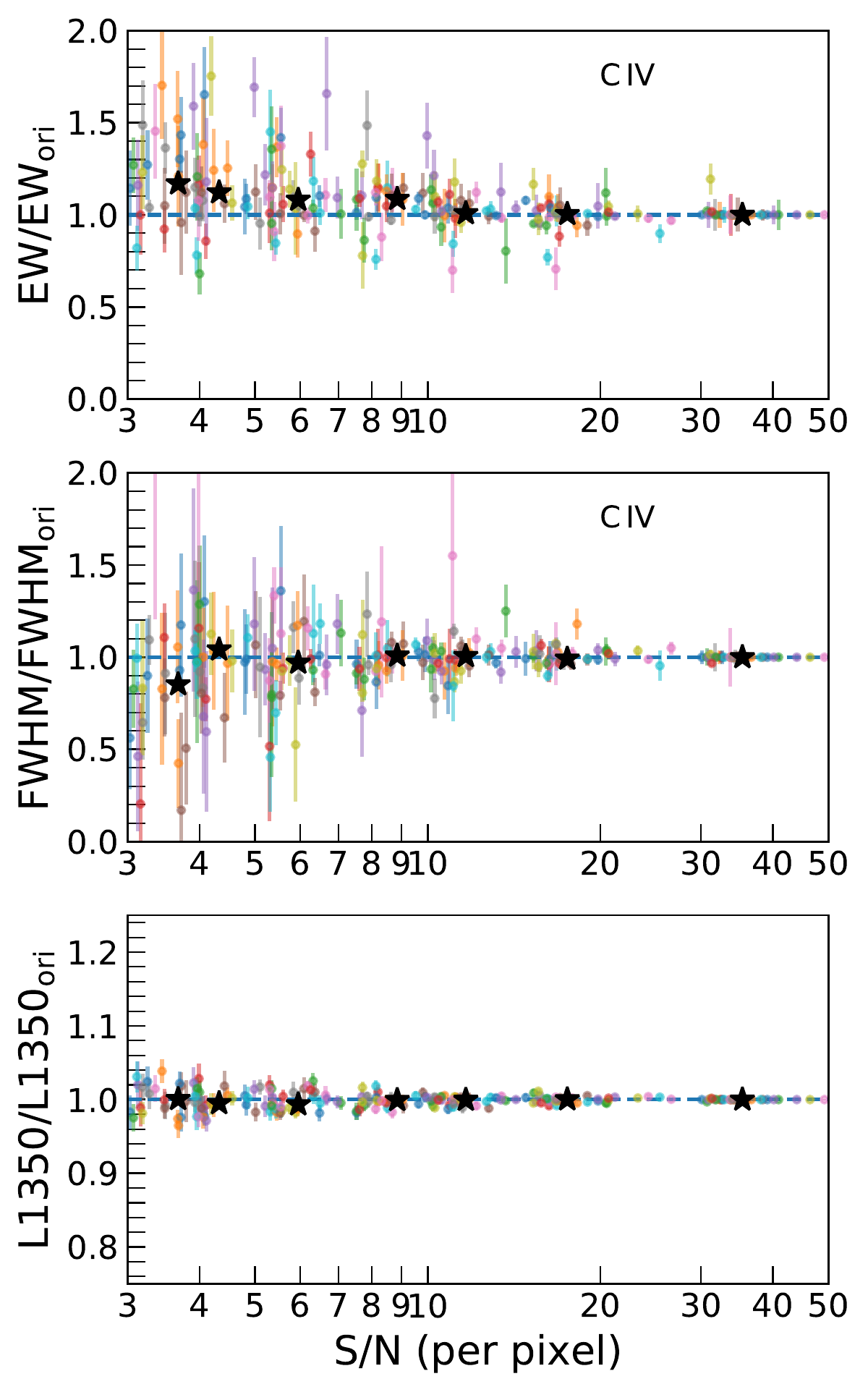}}
\caption{The ratio of EW (top), FWHM (middle) and continuum luminosity (bottom) of the S/N degraded spectra to the original spectra. A sample of 30 high-S/N spectra (measurement from individual spectrum is represented by an unique color) was chosen for each line and their S/N is degraded by adding noise by a factor of 2, 3, 4, 6, 8 and 10 to their original flux errors. The sample median is also shown (star-marker). Although on average the measurements are consistent up to a very low S/N $\sim 3$, any individual spectrum can deviate by 50\% or more.}\label{Fig:sn_impact} 
\end{figure*}

\section{Impact of S/N on spectral quantities}\label{sec:SNR}
  Although spectral decomposition of high S/N spectra can be reliable, the decomposition of low S/N spectra is usually difficult. In Figure \ref{Fig:cont_snr}, we plot the density contours of median continuum S/N with redshift in the upper panel and the distribution of median continuum S/N at different redshift range in the lower panel. The tail of the S/N distribution decreases rapidly at higher redshift. At low redshift $z<0.8$, the fraction of sources with S/N$>3$ pixel$^{-1}$ is 84\%, while for high-redshift $z>2.2$,  the fraction of sources with S/N$>3$ pixel$^{-1}$ is 54\%. The total number of sources with S/N$>3$ pixel$^{-1}$ in our catalog is 332,204 i.e., about 63\% of the total sample.

  Several authors \citep[e.g., ][]{2011ApJS..194...45S,2016ApJS..224...14D,2019ApJS..241...34S} have investigated the impact of S/N on the spectral decomposition method. They found that for high equivalent width (EW) objects, FWHMs and EWs are biased by less than $\pm 20$\% if line S/N reduced to as low as about 3, while for low-EW objects, the FWHMs and EWs are biased by $>20$\% for $S/N <5$. However, in all cases even at very low S/N, continuum luminosity measurements are unbiased. To investigate the impact of S/N on the measurement of our spectral quantities, we followed an approach similar to the previous studies. First, we selected a sample of thirty high-S/N original spectra independently for H$\beta$ in the redshift of  $z<0.8$, Mg II in the redshift range of $z=0.8-1.8$ and CIV in the redshift range of $z=1.8-3.2$. Then for each spectrum, we multiplied a constant factor of 2, 3, 4, 6, 8 and 10 to their original flux errors and added to the original spectrum a Gaussian random deviate of zero mean and standard deviation given by the new flux errors. We then repeated our spectral decomposition method as used in the decomposition of high-S/N original spectra, re-measured all the spectral quantities from the de-graded spectra, and finally compared them with the high-S/N original spectra. In Figure \ref{Fig:sn_impact}, we plot the ratio of the measurements from degraded spectra to the original high-S/N spectra as a function of the median continuum S/N for all thirty objects for each line (measurement from individual spectrum is represented by an unique color). With decreasing S/N, measurement uncertainties increase as per expectation. For example, when the sample median S/N reduced by a factor of 10 from 36.5 to 3.6, uncertainty in EW increased from 4.8\AA\, to 10.2\AA, and uncertainty in FWHM increased from 221 km s$^{-1}$ to 1246 km s$^{-1}$. The offsets represented by the sample median (star-marker) are negligible even at very low S/N suggesting the measurements are unbiased. However, measurements of any individual spectrum can have 50\% or more deviation. The Mg II EW (top-middle) shows a systematic offsets with decreasing S/N. However, this is not present in the case of H$\beta$ (top-left) and CIV (top-right) EW. The reason of this offset in Mg II could be due to the blending UV Fe II and Mg II line. The continuum luminosity is unbiased even at very low S/N.      
  
  The above investigation suggests that on average our spectral decomposition method recovers the measurements of the high S/N ratio spectra although individual measurements can deviate by 50\% or more. For peculiar sources, e.g., ones with double peak emission line our decomposition may fail badly. For this purpose, we provide various quality flags for each object, as described in detail in Appendix \ref{sec:flag}, based on several criteria. These quality flags give the reliability of our measurements.

\begin{figure}
\centering
\resizebox{7cm}{11cm}{\includegraphics{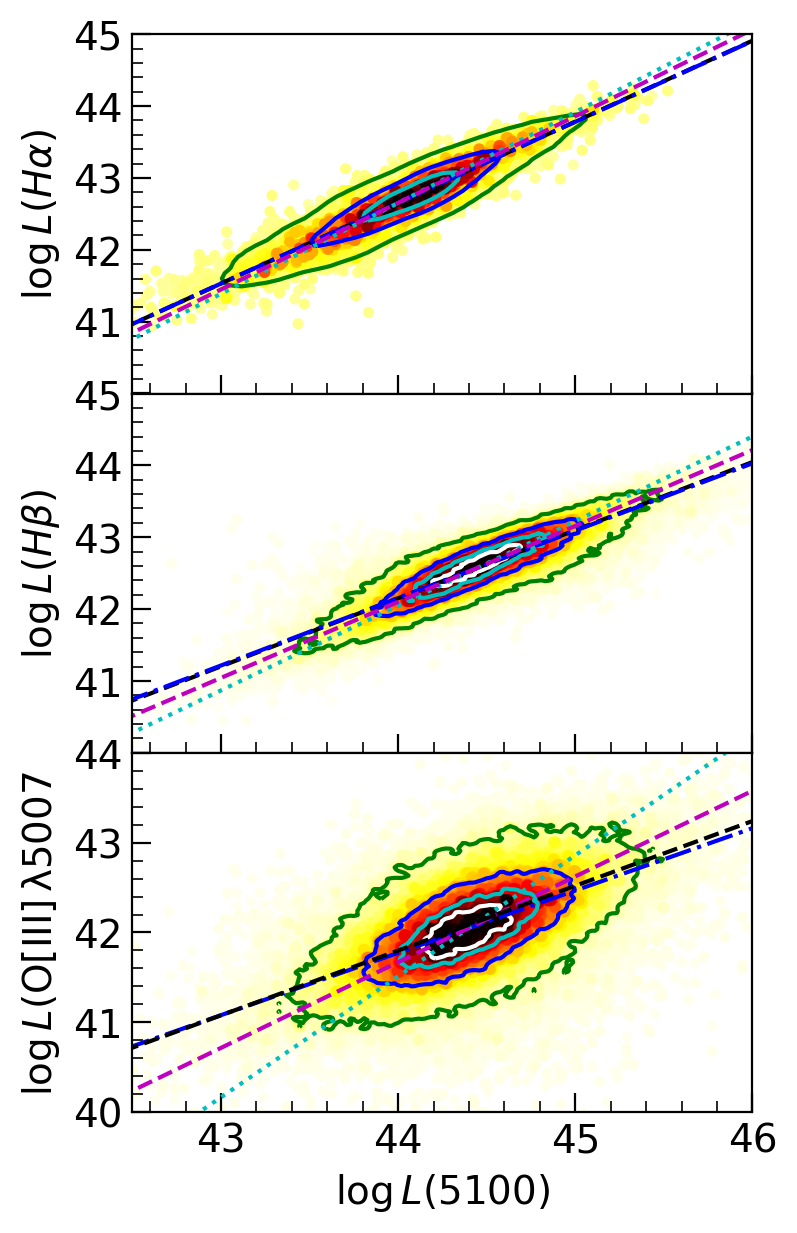}}
\caption{The H$\alpha$ (top), H$\beta$ (middle) and [O III] ($\lambda$5007) line luminosities is plotted against 5100\AA \,continuum luminosity. The best linear fit using \textsc{linmix} is shown by the black-dashed line while the fits obtained by \textsc{bces} considering x-axis as independent variable (blue dashed-dot line), y-axis as independent variable (cyan dotted line) and orthogonal least squares fit (magenta-dashed line) are shown. The 20 (white), 40 (cyan), 68 (blue) and 95 (green) percentile density contours along with the density map are shown.}\label{Fig:OIII_L5100} 
\end{figure}

\begin{figure*}
\centering
\resizebox{18cm}{22cm}{\includegraphics{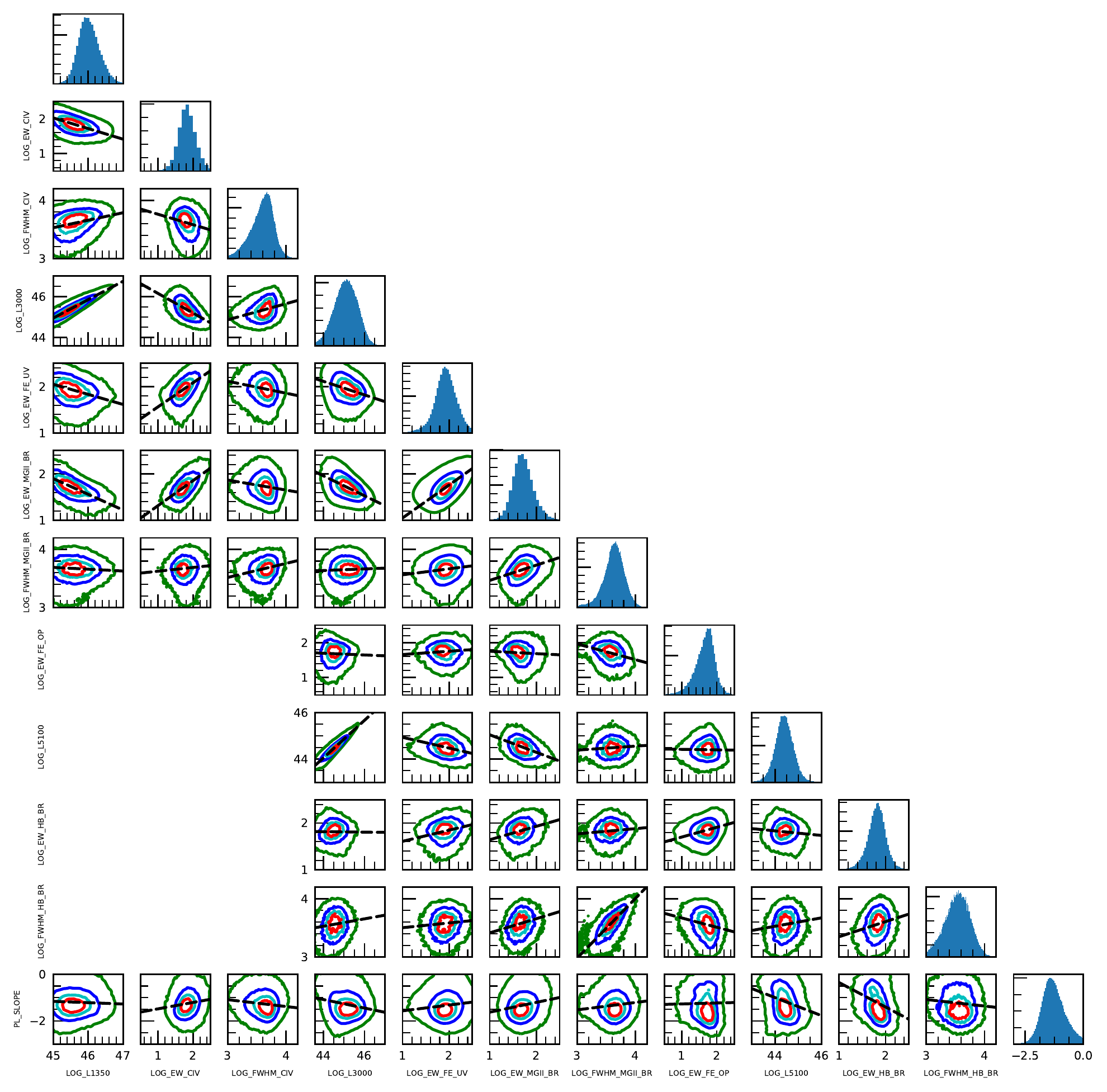}}
\caption{Correlation between various quasar parameters for objects with quality flag $=0$. Linear fits to the diagram is shown by a straight line in each plot using \textsc{linmix}. The corresponding best fit coefficients are mentioned in Table \ref{Table:corr}. The 20 (red), 40 (cyan), 68 (blue) and 95 (green) percentile density contours are also shown. The histograms are the projected one-dimensional distributions.}\label{Fig:corr} 
\end{figure*}

\begin{figure}
\centering
\resizebox{9cm}{9cm}{\includegraphics{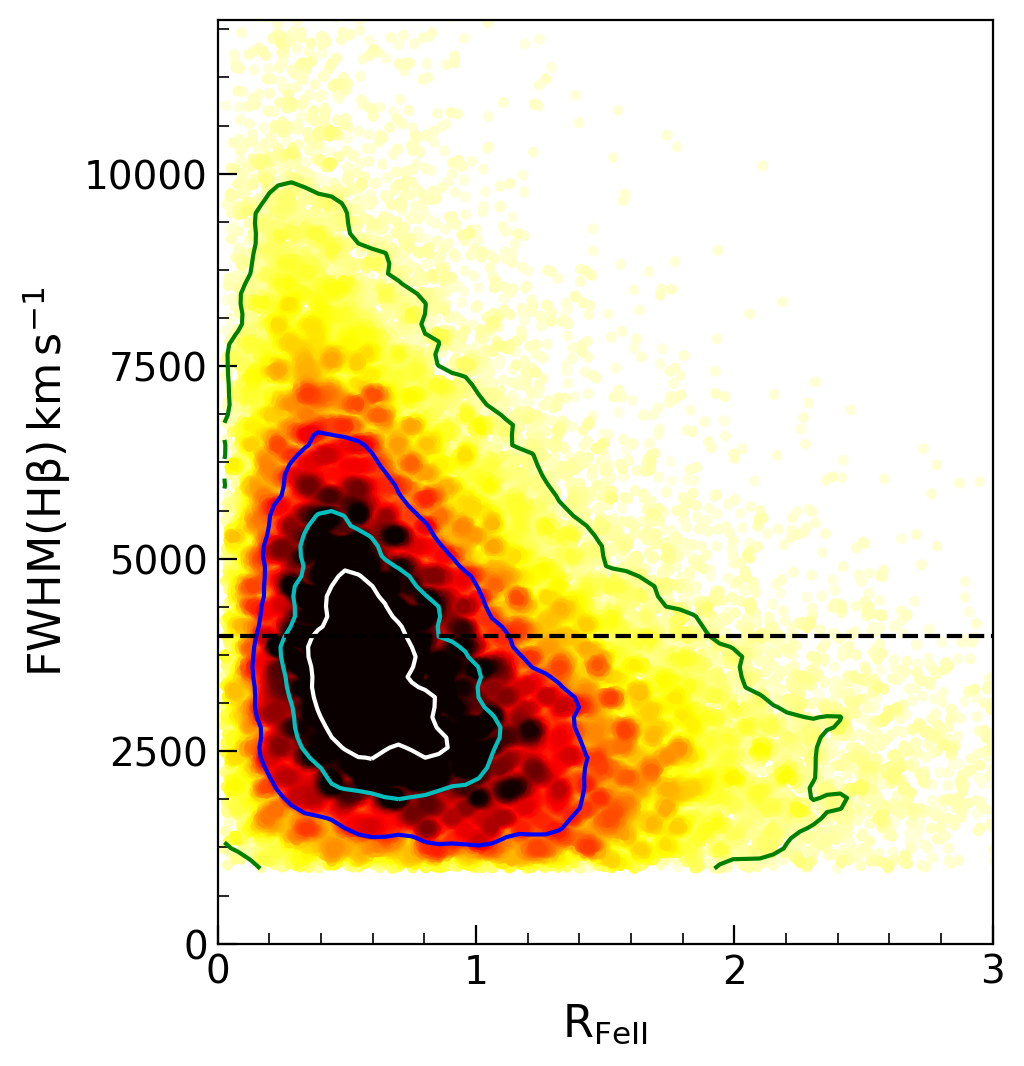}}
\caption{H$\beta$ line width vs. optical Fe \textsc{ii} strength. The 20, 40, 68 and 95 percentile density contours along with the density map are shown. The horizontal line at 4000 km s$^{-1}$ is also shown.}\label{Fig:R4570_FWHM_HB} 
\end{figure}

\begin{figure}
\centering
\resizebox{9cm}{5cm}{\includegraphics{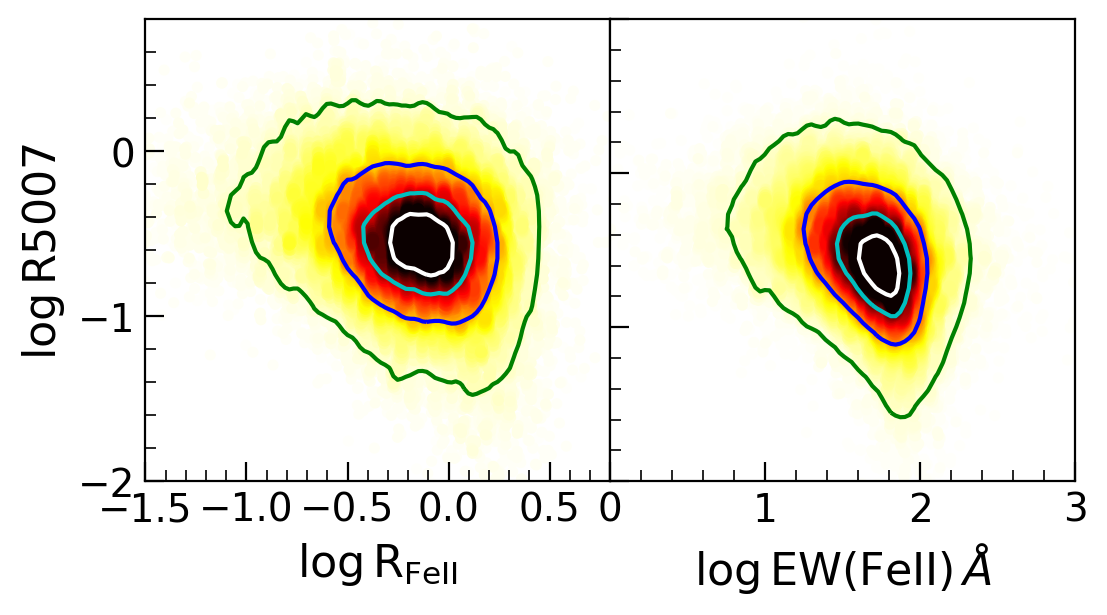}}
\caption{The R5007 (ratio of [O \textsc{iii}] to H$\beta$ equivalent width) against R$_{\mathrm{Fe \, \textsc{ii}}}$ (left) and Fe \textsc{ii} equivalent width (right).  The 20, 40, 68 and 95 percentile density contours along with the density map are shown.}\label{Fig:OIII_feII} 
\end{figure}


\section{Applications}\label{sec:discussion}

\subsection{Correlation analysis}

The spectral catalog generated in this work for a large number of quasars
can be used to investigate the correlation between various line and continuum 
properties in detail. Here, we studied some of the correlations. For this, we considered measurements with a quality flag of zero. The luminosity 
of Balmer lines show strong correlation with the monochromatic continuum luminosity at 5100 \AA\ over a wide range of redshift and luminosity suggesting that the physical 
mechanisms behind the correlation are the same in different AGN across all redshift 
and luminosity range \citep[e.g.,][]{2005ApJ...630..122G,2015ApJ...806..109J,2017ApJS..229...39R}. 
In Figure \ref{Fig:OIII_L5100}, we plot the luminosity of H$\alpha$ (top panel) and 
H$\beta$ (middle panel) against $L_{5100}$. In both cases, a strong correlation is 
found with $r_s$ of 0.93 and 0.86, respectively. We performed linear regression 
analysis with measurement errors on both axes using Bayesian 
code \textsc{linmix}\footnote{\url{https://github.com/jmeyers314/linmix}} \citep{2007ApJ...665.1489K} and obtained

\begin{equation}
\log L(H\alpha)   = (1.126 \pm 0.004)\log L(5100) + (-6.89\pm 0.20)
\label{eq:ha}
\end{equation}  

\begin{equation}
\log L(H\beta)   = (0.947 \pm 0.002)\log L(5100) + (-0.45\pm 0.09)
\label{eq:hb}
\end{equation} 

with an intrinsic scatter of $0.025\pm0.001$ and $0.035\pm0.001$, respectively. These correlations are shown by the black dashed line in Figure \ref{Fig:OIII_L5100}. For equation \ref{eq:ha}, a linear regression using \textsc{bces}\footnote{\url{https://github.com/rsnemmen/BCES}} \citep{1996ApJ...470..706A,2012Sci...338.1445N} gives a slope (m) of $1.125\pm0.005$  and intercept (c) of $-6.86\pm0.22$ considering $\log L(5100)$ as the independent variable (blue dashed-dot line), m$=1.264\pm0.006$ and c$=-12.97\pm0.27$ considering $\log L(H\alpha)$ as the independent variable (cyan dotted line) and m$=1.120\pm0.005$ and c$=-10.37\pm0.25$ for orthogonal least squares (magenta-dashed line). The same for equation \ref{eq:hb} is found to be $0.937\pm0.002$ and $0.89\pm0.11$ (m, c) for $\log L(5100)$ as independent variable, m$=1.177\pm0.002$ and c$=-9.75\pm0.11$ considering $\log L(H\beta)$ as the independent variable and m$=1.057\pm0.002$ and c$=-4.41\pm0.10$ for orthogonal least squares. The slopes of the $\log L(H\alpha)- \log L(5100)$ and $\log L(H\beta)- \log L(5100)$ correlations agree with the previous studies. For example, using a sample of low-z ($z<0.35$) SDSS quasars, \citet{2005ApJ...630..122G} found a slope of $1.157 \pm 0.005$ for equation \ref{eq:ha} and $1.133\pm0.005$ for equation \ref{eq:hb}. For high redshift ($z=1.5-2.2$) and high-luminosity ($\log L_{5100}=45.4-47.0$) quasars, \citet{2012ApJ...753..125S} obtained a slope of $1.010\pm0.042$ and $1.251\pm0.067$ for equations \ref{eq:ha} and \ref{eq:hb}, respectively. \citet{2015ApJ...806..109J} investigated $\log L(H\alpha)- \log L(5100)$ correlation using quasars of $z=0.0-6.2$ and luminosity of $\log L_{5100}=41.7-47.2$ and obtained a slope of $1.044\pm0.008$.

The correlation between $L\mathrm{[O III]}-L(5100)$ has been widely used to 
estimate bolometric luminosity for Type 2 AGN since their host galaxy 
contamination prevents reliable estimation of $L_{5100}$ 
\citep[see][]{2003MNRAS.346.1055K,Heckman_2004}. However, this relation has a 
large scatter \citep{Heckman_2004,2011ApJS..194...45S}. 
The L([O III]$\lambda5007$) as a function of $L_{5100}$ is plotted in the bottom 
panel ($r_s=0.58$) of Figure \ref{Fig:OIII_L5100}. The best-fit linear regression using \textsc{linmix} gives  

\begin{equation}
\log L\mathrm{[O III]} = (0.720 \pm 0.003)\log L(5100) + (10.08\pm 0.13)
\label{eq:oiii}
\end{equation}
with an intrinsic scatter of $0.102\pm0.001$. But the same using \textsc{bces} is found to be m$=0.693\pm0.003$ and c$=11.24\pm0.15$ when $\log L(5100)$ is the independent variable, m$=1.347\pm0.031$ and c$=-17.74\pm1.41$ when $\log L\mathrm{[O III]}$ is the independent variable and m$=0.953\pm0.015$ and c$=-0.30\pm0.69$ for orthogonal least squares. Depending on the treatment of the independent variable, the $L\mathrm{[O III]}-L(5100)$ relation shows a range of slopes of $0.7$ to $1.3$ due to large scatter. This agrees with \citet{2011ApJS..194...45S}, who noted a scatter of $\sim$0.4 dex around the best-fit $L\mathrm{[O III]}-L(5100)$ relation and a slope of 0.77 when $L_{5100}$ is considered as the independent variable, and 1.34 for bisector linear regression fit.

In Figure \ref{Fig:corr}, we plot various such line and continuum quantities 
and performed correlation analysis between them. 
The fits to the data  are shown by 
dashed lines in Figure \ref{Fig:corr} and the results of the fitting are given 
in Table \ref{Table:corr}. Most of the correlation agrees with the previous works based on smaller samples. For example, the continuum luminosity at 1350\AA, 3000\AA, and 5100\AA \, are strongly correlated with each other \citep[e.g.,][]{2015ApJ...806..109J}. The line width of H$\beta$ and Mg II shows a strong correlation \citep[e.g.,][]{2009ApJ...707.1334W}, however, the correlation is very weak between Mg II and CIV line FWHM. All the parameters are uncorrelated with spectral index except for a weak anti-correlation with $L_{5100}$ and H$\beta$ EW \citep{2011ApJS..194...45S}. Emission line FWHM is weakly correlated with line EW both for H$\beta$ and Mg II lines similar to what has been noted for Mg II line by \citet{2009ApJ...703L...1D}. The well-known anti-correlation between continuum luminosity and line EW is found both for Mg II and CIV \citep{1977ApJ...214..679B} but not for H$\beta$ EW. Although no correlation between EW of optical Fe II (measured from the best-fit optical Fe II template in the wavelength range of 4435-4685$\AA$) and EW of H$\beta$ is found, EW of UV Fe II (measured from the best-fit UV Fe II template in the wavelength range of 2200-3090$\AA$) is strongly correlated with the EW of Mg II line and C IV \citep[e.g.,][]{2015ApJS..221...35K}.

 \begin{table*}
 \caption{Linear regression analysis to the y vs. x correlation for objects with quality flag =0 using \textsc{linmix} having slope (m), intercept (c) and intrinsic scatter ($\sigma_{\mathrm{int}}$). The Spearman rank correlation coefficient ($r_s$) is also noted.}
	\begin{center}
	\hspace*{-1.1cm}
 	\resizebox{0.7\linewidth}{!}{%
     \begin{tabular}{ l l r r r r}\toprule
      		y				    &  x                    &    m       			&  c                         & $\sigma_{\mathrm{int}}$ & $r_s$ \\ \midrule
      				   		
	PL\_SLOPE	 		& LOG\_FWHM\_HB\_BR	 	&$-$0.280	$\pm$  0.018 & $-$0.239	    $\pm$ 0.064	& 0.463	$\pm$ 0.003	& $-$0.09 \\
	PL\_SLOPE	 		& LOG\_EW\_HB\_BR	 	&$-$1.051	$\pm$  0.017 & 0.702	 	$\pm$ 0.030	& 0.445	$\pm$ 0.003	& $-$0.30 \\
	PL\_SLOPE	 		& LOG\_L5100	 	 	&$-$0.389	$\pm$  0.006 & 16.100	 	$\pm$ 0.256	& 0.508	$\pm$ 0.003	& $-$0.24 \\
	PL\_SLOPE	 		& LOG\_EW\_FE\_OP	  	&   0.040	$\pm$  0.013 & $-$1.308	    $\pm$ 0.021	& 0.492	$\pm$ 0.003	& $-$0.05 \\
	PL\_SLOPE	 		& LOG\_FWHM\_MGII\_BR	&   0.317	$\pm$  0.008 & $-$2.476	    $\pm$ 0.030	& 0.311	$\pm$ 0.001	& 0.03 \\
	PL\_SLOPE	 		& LOG\_EW\_MGII\_BR	 	&   0.420	$\pm$  0.005 & $-$2.042	    $\pm$ 0.009	& 0.313	$\pm$ 0.001	& 0.15 \\
	PL\_SLOPE	 		& LOG\_EW\_FE\_UV	 	&   0.250	$\pm$  0.005 & $-$1.816	    $\pm$ 0.010	& 0.297	$\pm$ 0.001	& 0.07 \\
	PL\_SLOPE	 		& LOG\_L3000	 		&$-$0.183	$\pm$  0.002 & 6.972	    $\pm$ 0.073	& 0.341	$\pm$ 0.001	& $-$0.12 \\
	PL\_SLOPE	 		& LOG\_FWHM\_CIV	 	&$-$0.293	$\pm$  0.007 & $-$0.208	    $\pm$ 0.024	& 0.258	$\pm$ 0.001	& $-$0.08 \\
	PL\_SLOPE	 		& LOG\_EW\_CIV	 		&   0.274	$\pm$  0.004 & $-$1.754	    $\pm$ 0.008	& 0.254	$\pm$ 0.001	& 0.15 \\
	PL\_SLOPE	 		& LOG\_L1350	        &$-$0.057	$\pm$  0.002 & 1.407	    $\pm$ 0.100	& 0.317	$\pm$ 0.001	& $-$0.03 \\
	LOG\_FWHM\_HB\_BR	& LOG\_EW\_HB\_BR	    &   0.260	$\pm$  0.005 & 3.087	    $\pm$ 0.008	& 0.031	$\pm$ 0.001	& 0.22 \\
	LOG\_FWHM\_HB\_BR	& LOG\_L5100	        &   0.072	$\pm$  0.002 & 0.338	    $\pm$ 0.086	& 0.030	$\pm$ 0.001	& 0.16 \\
	LOG\_FWHM\_HB\_BR	& LOG\_EW\_FE\_OP	    &$-$0.159	$\pm$  0.004 & 3.829	    $\pm$ 0.006	& 0.030	$\pm$ 0.001	& $-$0.21 \\
	LOG\_FWHM\_HB\_BR	& LOG\_FWHM\_MGII\_BR	&   1.023	$\pm$  0.004 & $-$0.089	    $\pm$ 0.014	& 0.005	$\pm$ 0.001	& 0.67 \\
	LOG\_FWHM\_HB\_BR	& LOG\_EW\_MGII\_BR	    &   0.243	$\pm$  0.004 & 3.171	    $\pm$ 0.007	& 0.024	$\pm$ 0.001	& 0.21 \\
	LOG\_FWHM\_HB\_BR	& LOG\_EW\_FE\_UV	    &   0.082	$\pm$  0.005 & 3.420	    $\pm$ 0.010	& 0.024	$\pm$ 0.001	& 0.09 \\
	LOG\_FWHM\_HB\_BR	& LOG\_L3000	        &   0.064	$\pm$  0.002 & 0.730	    $\pm$ 0.075	& 0.030	$\pm$ 0.001	& 0.17 \\
	LOG\_EW\_HB\_BR	 	& LOG\_L5100	        &$-$0.049	$\pm$  0.002 & 3.987	    $\pm$ 0.091	& 0.034	$\pm$ 0.001	& $-$0.04 \\
	LOG\_EW\_HB\_BR	 	& LOG\_EW\_FE\_OP	    &   0.209	$\pm$  0.003 & 1.491	    $\pm$ 0.005	& 0.027	$\pm$ 0.001	& 0.24 \\
	LOG\_EW\_HB\_BR	 	& LOG\_FWHM\_MGII\_BR	&   0.112	$\pm$  0.007 & 1.433	    $\pm$ 0.027	& 0.037	$\pm$ 0.001	& 0.09 \\
	LOG\_EW\_HB\_BR	 	& LOG\_EW\_MGII\_BR	 	&   0.288	$\pm$  0.005 & 1.354	    $\pm$ 0.008	& 0.035	$\pm$ 0.001	& 0.24 \\
	LOG\_EW\_HB\_BR	 	& LOG\_EW\_FE\_UV	    &   0.240	$\pm$  0.006 & 1.367	    $\pm$ 0.012	& 0.041	$\pm$ 0.001	& 0.21 \\
	LOG\_EW\_HB\_BR	 	& LOG\_L3000	        &$-$0.005	$\pm$  0.002 & 2.029	    $\pm$ 0.083	& 0.034	$\pm$ 0.001	& 0.04 \\
	LOG\_L5100	 		& LOG\_EW\_FE\_OP	    &$-$0.022	$\pm$  0.006 & 44.438	    $\pm$ 0.011	& 0.150	$\pm$ 0.001	& 0.02 \\
	LOG\_L5100	 		& LOG\_FWHM\_MGII\_BR	&   0.169	$\pm$  0.009 & 43.874	    $\pm$ 0.034	& 0.135	$\pm$ 0.001	& 0.10 \\
	LOG\_L5100	 		& LOG\_EW\_MGII\_BR	    &$-$0.752	$\pm$  0.006 & 45.792	    $\pm$ 0.010	& 0.107	$\pm$ 0.001	& $-$0.42 \\
	LOG\_L5100	 		& LOG\_EW\_FE\_UV	    &$-$0.463	$\pm$  0.007 & 45.396	    $\pm$ 0.014	& 0.118	$\pm$ 0.001	& $-$0.24 \\
	LOG\_L5100	 		& LOG\_L3000	        &   0.806	$\pm$  0.001 & 8.576	    $\pm$ 0.044	& 0.019	$\pm$ 0.001	& 0.93 \\
	LOG\_EW\_FE\_OP	 	& LOG\_FWHM\_MGII\_BR	&$-$0.450	$\pm$  0.009 & 3.312	    $\pm$ 0.033	& 0.053	$\pm$ 0.001	& $-$0.24 \\
	LOG\_EW\_FE\_OP	 	& LOG\_EW\_MGII\_BR	    &$-$0.068	$\pm$  0.007 & 1.820	    $\pm$ 0.012	& 0.057	$\pm$ 0.001	& $-$0.13 \\
	LOG\_EW\_FE\_OP	 	& LOG\_EW\_FE\_UV	    &   0.094	$\pm$  0.008 & 1.564	    $\pm$ 0.015	& 0.047	$\pm$ 0.001	& 0.00 \\
	LOG\_EW\_FE\_OP	 	& LOG\_L3000	        &$-$0.025	$\pm$  0.003 & 2.792	    $\pm$ 0.119	& 0.062	$\pm$ 0.001	& 0.00 \\
	LOG\_FWHM\_MGII\_BR	& LOG\_EW\_MGII\_BR	    &   0.262	$\pm$  0.001 & 3.207	    $\pm$ 0.002	& 0.017	$\pm$ 0.001	& 0.32 \\
	LOG\_FWHM\_MGII\_BR	& LOG\_EW\_FE\_UV	    &   0.103	$\pm$  0.001 & 3.461	    $\pm$ 0.003	& 0.018	$\pm$ 0.001	& 0.12 \\
	LOG\_FWHM\_MGII\_BR	& LOG\_L3000	        &   0.014	$\pm$  0.001 & 3.041	    $\pm$ 0.025	& 0.019	$\pm$ 0.001	& 0.06 \\
	LOG\_FWHM\_MGII\_BR	& LOG\_FWHM\_CIV	    &   0.241	$\pm$  0.003 & 2.797	    $\pm$ 0.013	& 0.013	$\pm$ 0.001	& 0.17 \\
	LOG\_FWHM\_MGII\_BR	& LOG\_EW\_CIV	        &   0.064	$\pm$  0.002 & 3.563	    $\pm$ 0.003	& 0.014	$\pm$ 0.001	& 0.06 \\
	LOG\_FWHM\_MGII\_BR	& LOG\_L1350	        &$-$0.031	$\pm$  0.001 & 5.100	    $\pm$ 0.045	& 0.015	$\pm$ 0.001	& $-$0.02 \\
	LOG\_EW\_MGII\_BR	& LOG\_EW\_FE\_UV	    &   0.708	$\pm$  0.002 & 0.342	    $\pm$ 0.003	& 0.021	$\pm$ 0.001	& 0.58 \\
	LOG\_EW\_MGII\_BR	& LOG\_L3000	        &$-$0.214	$\pm$  0.001 & 11.388	    $\pm$ 0.029	& 0.033	$\pm$ 0.001	& $-$0.49 \\
	LOG\_EW\_MGII\_BR	& LOG\_FWHM\_CIV	    &$-$0.215	$\pm$  0.006 & 2.511	    $\pm$ 0.020	& 0.047	$\pm$ 0.001	& $-$0.12 \\
	LOG\_EW\_MGII\_BR	& LOG\_EW\_CIV	        &   0.536	$\pm$  0.002 & 0.781	    $\pm$ 0.004	& 0.026	$\pm$ 0.001	& 0.62 \\
	LOG\_EW\_MGII\_BR	& LOG\_L1350	        &$-$0.338	$\pm$  0.001 & 17.098	    $\pm$ 0.064	& 0.028	$\pm$ 0.001	& $-$0.65 \\
	LOG\_EW\_FE\_UV	 	& LOG\_L3000	        &$-$0.146	$\pm$  0.001 & 8.516	    $\pm$ 0.039	& 0.043	$\pm$ 0.001	& $-$0.27 \\
	LOG\_EW\_FE\_UV	 	& LOG\_FWHM\_CIV	    &$-$0.257	$\pm$  0.006 & 2.882	    $\pm$ 0.022	& 0.047	$\pm$ 0.001	& $-$0.10 \\
	LOG\_EW\_FE\_UV	 	& LOG\_EW\_CIV	        &   0.515	$\pm$  0.003 & 1.048	    $\pm$ 0.005	& 0.029	$\pm$ 0.001	& 0.54 \\
	LOG\_EW\_FE\_UV	 	& LOG\_L1350	        &$-$0.217	$\pm$  0.002 & 11.805	    $\pm$ 0.084	& 0.043	$\pm$ 0.001	& $-$0.31 \\
	LOG\_L3000	 		& LOG\_FWHM\_CIV	    &   0.784	$\pm$  0.007 & 42.513	    $\pm$ 0.027	& 0.173	$\pm$ 0.001	& 0.28 \\
	LOG\_L3000	 		& LOG\_EW\_CIV	        &$-$0.952	$\pm$  0.003 & 47.099	    $\pm$ 0.006	& 0.120	$\pm$ 0.001	& $-$0.57 \\
	LOG\_L3000	 		& LOG\_L1350	        &   0.896	$\pm$  0.001 & 4.624	    $\pm$ 0.047	& 0.033	$\pm$ 0.001	& 0.91 \\
	LOG\_FWHM\_CIV	 	& LOG\_EW\_CIV	        &$-$0.179	$\pm$  0.002 & 3.938	    $\pm$ 0.003	& 0.025	$\pm$ 0.001	& $-$0.21 \\
	LOG\_FWHM\_CIV	 	& LOG\_L1350	        &   0.129	$\pm$  0.001 &$-$2.278	    $\pm$ 0.038	& 0.024	$\pm$ 0.001	& 0.33 \\
	LOG\_EW\_CIV	 	& LOG\_L1350	        &$-$0.305	$\pm$  0.001 & 15.749	 	$\pm$ 0.042	& 0.038	$\pm$ 0.001	& $-$0.53 \\
	
      		\bottomrule
        \end{tabular} } 
        \label{Table:corr}
        \end{center}
    \end{table*}

\begin{figure*}
\centering
\resizebox{8cm}{12cm}{\includegraphics{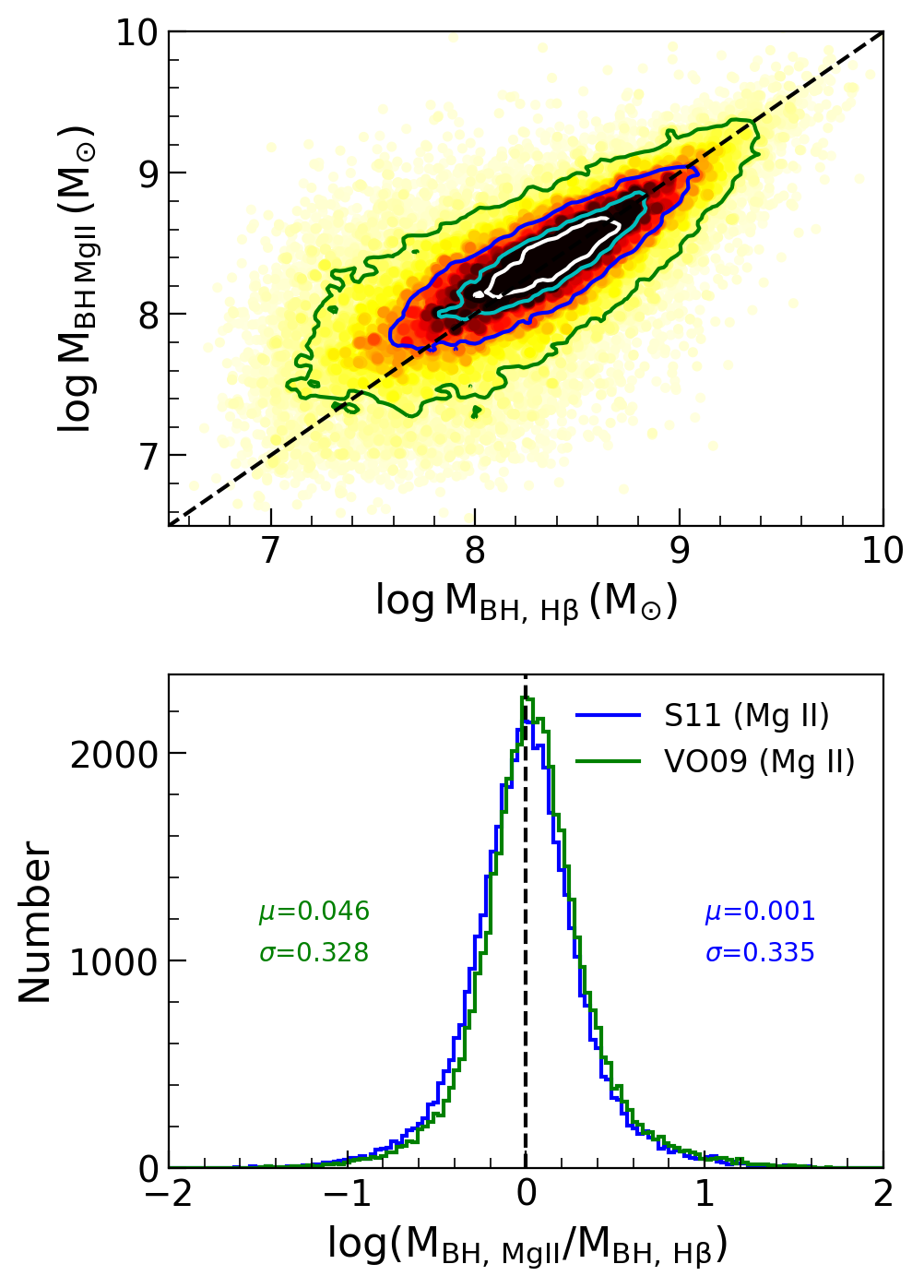}}
\resizebox{8cm}{12cm}{\includegraphics{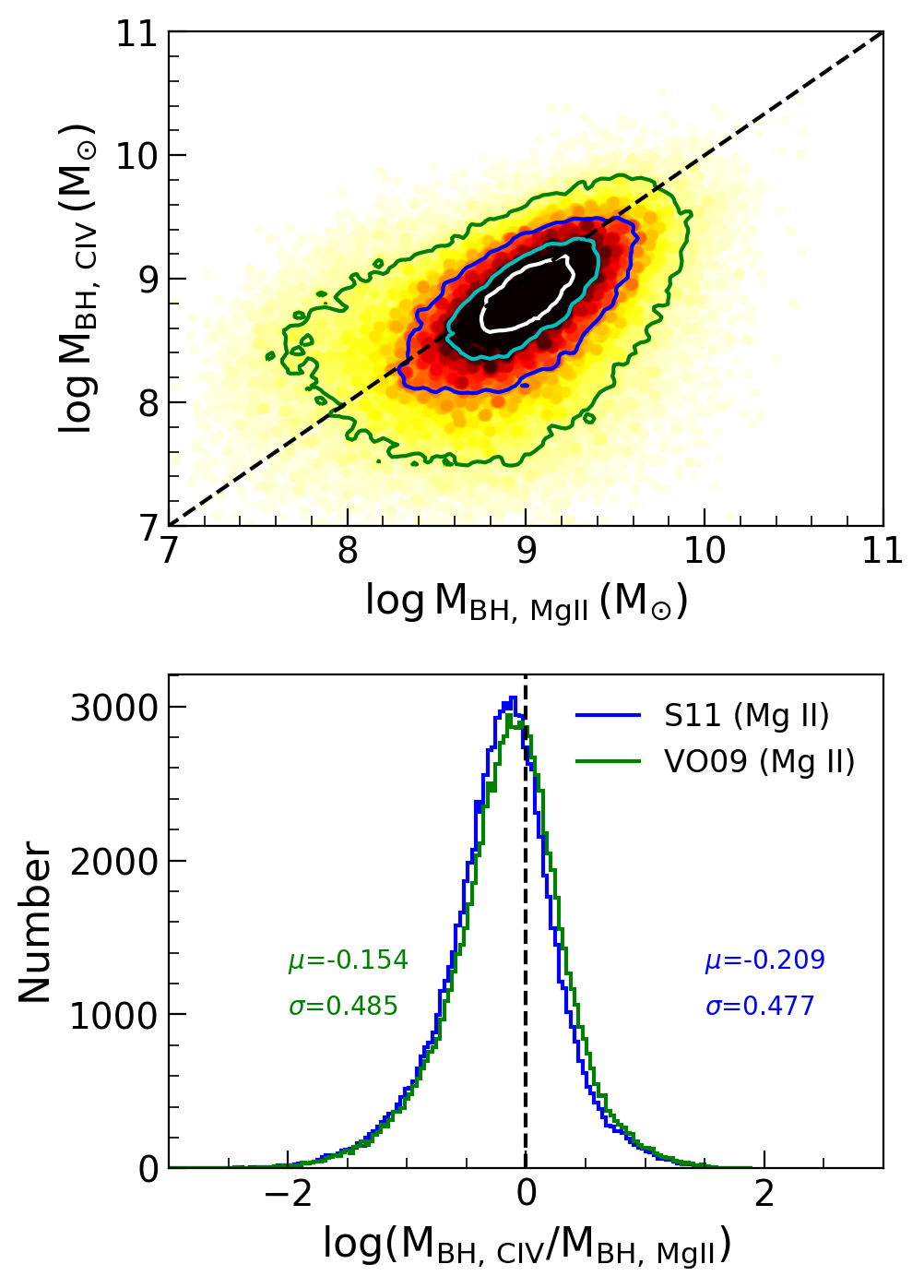}}
\caption{Left: Comparison between black holes masses estimated based on Mg \textsc{ii} and H$\beta$ line. The distribution of the mass ratios is shown in the lower panel along with the number of objects for which both the masses were estimated, sample mean ($\mu$) and dispersion ($\sigma$) are noted. Right: Comparison between C \textsc{iv} and Mg \textsc{ii} based mass measurements for objects having both the lines. The offset and dispersion of M$_{\mathrm{BH, C \textsc{iv}}}$/M$_{\mathrm{BH, MgII}}$ is larger than that of M$_{\mathrm{BH, MgII}}$/M$_{\mathrm{BH, H\beta}}$. The 20, 40, 68 and 95 percentile density contours along with the density map are shown in the upper panels. Only sources with quality flag =0 are included.}\label{Fig:Mbh} 
\end{figure*}

\citet{1992ApJS...80..109B} performed PCA using a sample of 87 PG quasars ($z<0.5$) and found various correlations involving  optical Fe II, [O III]5007 and H$\beta$ broad component, radio to optical flux ratio and the optical to X-ray spectral index. The first PCA eigenvector (Eigenvector 1; hereafter E1) strongly anti-correlates with $R_{\mathrm{Fe \, II}}$ (defined by the ratio of 
the EW of Fe II (4435$-$4685\AA\,) to H$\beta$ broad line) and luminosity of [O III]$\lambda 4959,5007$. The main parameters of the well-known 4DE1 
\citep{2000ApJ...536L...5S,2002ApJ...566L..71S}, which can account for the diverse nature of broad line AGN, are 
the FWHM of broad H$\beta$ line and $R_{\mathrm{Fe \, \textsc{ii}}}$.  These two quantities are 
plotted in Figure \ref{Fig:R4570_FWHM_HB} for objects with quality flag =0 in the catalog.  Firstly, quasars with a wide range of Fe \textsc{ii} strength can 
be found at a given FWHM(H$\beta$). Similarly, at a given Fe \textsc{ii} strength, the 
H$\beta$ can have a large range. The $R_{\mathrm{Fe \, II}}$ distribution peaks at $\sim0.7$ which can be occupied by quasars with FWHM(H$\beta$) of $\sim1000-10,000$ km s$^{-1}$. This dispersion is suggested to be due to an orientation effect \citep[see][]{2014Natur.513..210S,2015ApJ...804L..15S}. Secondly, the well-known trend of decreasing FWHM with increasing $R_{\mathrm{FeII}}$ is noticed as also shown in previous studies \citep[e.g.,][]{2011ApJS..194...45S,2019A&A...625A.123C}. Thus, quasars with a very broad H$\beta$ line and strong Fe \textsc{ii} strength is rare, especially in the low redshift SDSS sample. However, IR spectroscopic study of high-z quasars shows a systematically larger FWHM(H$\beta$) compared to the low-z sources at high $R_{\mathrm{Fe \,  II}}$ \citep{2016ApJ...817...55S}. The dashed line represents the separation of quasars into two populations; the population A (FWHM(H$\beta$, broad) $\le 4000$ km s$^{-1}$) sources with strong Fe \textsc{ii} and soft X-ray excess, and population B (FWHM(H$\beta$, broad) $> 4000$ km s$^{-1}$) sources with weak Fe \textsc{ii} and a lack of soft X-ray excess \citep{2000ARA&A..38..521S}.   

 Previous studies found an anti-correlation between [O \textsc{iii}] and Fe \textsc{ii} emission, i.e., objects with strong [O \textsc{iii}] are found to be weak Fe \textsc{ii} emitters and vice versa \citep[e.g.,][]{1992ApJS...80..109B,1999A&A...350..805G,1999ApJ...514...40M,2017ApJS..229...39R} although \citet{2001A&A...372..730V} found such anti-correlation to be very weak. In Figure \ref{Fig:OIII_feII}, we plot R5007, i.e. the ratio of EW of [O III]5007 to the EW of total H$\beta$ (sum of the EW of broad and narrow H$\beta$ components) against R$_{\mathrm{Fe \, \textsc{ii}}}$ (left) and EW of Fe \textsc{ii} (right). We found that the anti-correlation, although weak, is present with correlation coefficient $r_s=-0.23$ between R5007 and R$_{\mathrm{Fe \, \textsc{ii}}}$, while the R5007 vs. EW (Fe \textsc{ii}) anti-correlation is moderately strong with $r_s=-0.32$. The anti-correlations become stronger with $r_s=-0.37$ for R5007 vs. R$_{\mathrm{Fe \, II}}$ and $r_s=-0.45$ for R5007 vs. EW (Fe II) when sources with continuum S/N$>10$ pixel$^{-1}$ is considered. The correlation found here is consistent with previous studies. For example, \citet{1992ApJS...80..109B} found $r_s=-0.36$ for R5007 vs. R$_{\mathrm{Fe \, II}}$ relation and $-0.52$ for R5007 vs. EW (Fe II) relation. Similarly, \citet{1999ApJ...514...40M} found $r_s=-0.43$ for R5007 vs. R$_{\mathrm{Fe \, II}}$ relation and $r_s= -0.54$ for R5007 vs. EW (Fe II) relation while \citet{1999A&A...350..805G} found $r_s = -0.36$ for R5007 vs. EW (Fe II) relation. Therefore, the strong Fe \textsc{ii} sources may have weak [O \textsc{iii}] emission or vice versa.

\begin{figure}
\centering
\resizebox{9cm}{8.5cm}{\includegraphics{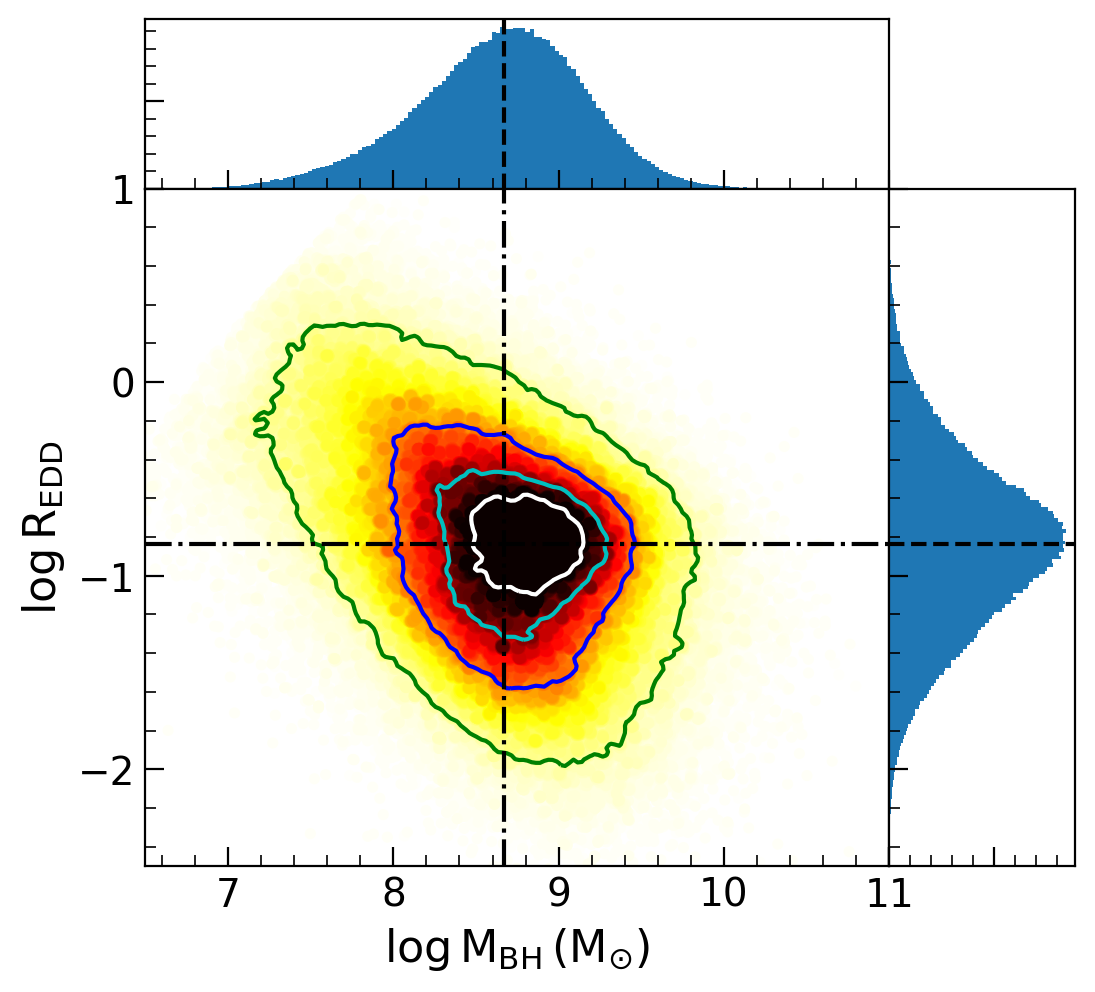}}
\caption{Eddington ratio vs. black hole masses for all quasars in the catalog. The dashed lines represent the median of the distribution. The 20, 40, 68 and 95 percentile density contours along with the density map are shown. Only sources with quality flag =0 are included.}\label{Fig:REDD_Mbh} 
\end{figure}

\subsection{Black hole mass measurement}

The value of $M_{\mathrm{BH}}$ in an AGN can be estimated using virial relation 
from single-epoch spectrum for which continuum luminosity and line width 
measurements are available as follows 
 \begin{equation}
 \log \left(\frac{M_{\mathrm{BH}}}{M_{\odot}} \right) = A + B \log \left(\frac{\lambda L_{\lambda}}{10^{44} \mathrm{erg\, s^{-1}}} \right) + 2\log \left(\frac{\mathrm{FWHM}}{\mathrm{km\, s^{-1}}} \right)
 \label{eq:mass}
 \end{equation}
 where $A$ and $B$ are the constants empirically calibrated by various authors using 
reverberation mapping $R-L$ relations of local AGN \citep{2000ApJ...533..631K,
2005ApJ...629...61K,2013ApJ...767..149B}  as well as internally calibrated based 
on the availability of multiple strong emission lines such as Mg \textsc{ii} with 
H$\beta$ \citep[e.g.,][]{2002MNRAS.337..109M,2004MNRAS.352.1390M,2006ApJ...641..689V,
2011ApJS..194...45S,2018ApJ...859..138W} and C \textsc{iv} with Mg \textsc{ii} or 
H$\beta$ \citep[e.g.,][]{2006ApJ...641..689V,2011ApJ...742...93A,2017ApJ...839...93P}. 
In this work, we used the black hole mass calibrations from \citet[][thereafter VP06]{2006ApJ...641..689V}, \citet[][thereafter A11]{2011ApJ...742...93A}, \citet[][thereafter VO09]{2009ApJ...699..800V} and S11.  
   \[
        A, B= 
     \begin{cases}
         (0.910, 0.50), & \text{for H$\beta$; VP06} \\
         (0.895, 0.52), & \text{for H$\beta$; A11} \\
         (0.860, 0.50), & \text{for Mg \textsc{ii}; VO09} \\
         (0.740, 0.62), & \text{for Mg \textsc{ii}; S11}  \\       
         (0.660, 0.53), & \text{for C \textsc{iv};  VP06}
     \end{cases}
     \]

We caution that the derived virial $M_{\mathrm{BH}}$ values could have uncertainty $>0.4$ 
dex due to the different systematics \citep[e.g., different line width measures, geometry and kinematics of BLR, see][]{2006A&A...456...75C,2013BASI...41...61S} involved in the calibrations used, which have 
not been taken in to account. The uncertainties in the virial $M_{\mathrm{BH}}$ values that 
are provided in the catalog are only the measurement uncertainties calculated via 
error propagation of Equation \ref{eq:mass}. In Figure \ref{Fig:Mbh} we compare 
$M_{\mathrm{BH}}$ calculated using different lines. We plot Mg II based black hole masses against H$\beta$ (upper left) and the ratio of the two masses (lower panels). Both S11 and VO09 provide consistent Mg II based masses with negligible offsets between Mg II and H$\beta$ based masses but with a dispersion of $\sim0.3$. In the right panel, we compare CIV based masses against Mg II based masses. Here, we notice a larger offset and dispersion in the ratio of CIV to Mg II based masses compared to Mg II to H$\beta$ based mass ratio. In the catalog, we provide `fiducial' virial $M_{\mathrm{BH}}$ values calculated based on  
(a) H$\beta$ line (for $z< 0.8$) using the calibration of VP06, (b) Mg \textsc{ii} 
line (for $ 0.8\leq z<1.9$) using the calibration provided
by VO09, and (c) C \textsc{iv} line (for $z\geq 1.9$) using VP06 calibration.

We estimated Eddington ratio ($R_{\mathrm{Edd}}$), which is the ratio of bolometric (see section \ref{sec:param}) to Eddington luminosity ($L_{\mathrm{Edd}}$). The values of $M_{\mathrm{BH}}$  and Eddington ratios for all quasars are also provided 
in the catalog. In Figure \ref{Fig:REDD_Mbh}, we plot $R_{\mathrm{Edd}}$ against 
$M_{\mathrm{BH}}$ for sources with quality flag =0. Firstly, the accretion rate decreases with increasing $M_{\mathrm{BH}}$ 
and highly accreting quasars with massive black holes are rare which is expected considering that $R_{\mathrm{Edd}}$ is inversely proportional to $M_{\mathrm{BH}}$ that increases with line width. Secondly, low 
accreting and low mass black holes are also rare due to the flux limit of SDSS. 
The distribution of $R_{\mathrm{Edd}}$ and $M_{\mathrm{BH}}$ is also plotted. 
The median of the $\log M_{\mathrm{BH}}$ distribution is $8.67^{+0.45}_{-0.53} \, M_{\odot}$ 
having a range of $7.1-9.9 \, M_{\odot}$ ($3\sigma$ around the median) and the  
$\log R_{\mathrm{Edd}}$ distribution has a median of $-0.83^{+0.42}_{-0.44}$ with a  range of $-2.06$ to $0.43$.

\section{Summary}\label{sec:summary}
We have carried out detailed spectral decompositions, that include host galaxy 
subtraction, AGN continuum, and emission line modeling for more than 500,000 quasars 
spectra from SDSS DR14 quasar catalog and estimated spectral properties such as 
line flux, FWHM, wavelength shift, etc. We estimated virial $M_{\mathrm{BH}}$ 
and Eddington ratios for the quasars. We performed various correlation analysis 
to show the applicability of the measurements presented in this work 
in a larger context. The strong Fe \textsc{ii} emitters with larger line FWHM, and highly 
accreting high mass sources are found to be rare in this large sample of quasars. 
In particular, we found the well-known inverse correlation between EW and continuum flux in C \textsc{iv} and Mg \textsc{ii}, and the strong correlation between 
Balmer line and continuum luminosity. We provide all the measurements in the form 
of a catalog, which is the largest catalog containing the spectral properties of quasars till 
date. This catalog will be of immense use to the community to study various properties of quasars.

\acknowledgments

We thank the referee for useful suggestions, which significantly improved the clarity of the manuscript. We thank Hengxiao Guo for making the spectral fitting code \textsc{PyQSOFit} publicly available and for useful discussions. JK acknowledges financial support from the Academy of Finland, grant 311438. SR thanks Neha Sharma (KHU, South Korea) for carefully reading the manuscript. Throughout this work, we have used FGCI-cluster Titan. This work has made use of SDSS spectroscopic data. Funding for SDSS-III has been provided by the Alfred P. Sloan Foundation, the Participating Institutions, the National Science Foundation, and the U.S. Department of Energy Office of Science. The SDSS-III web site is http://www.sdss3.org/. SDSS-III is managed by the Astrophysical Research Consortium for the Participating Institutions of the SDSS-III Collaboration including the University of Arizona, the Brazilian Participation Group, Brookhaven National Laboratory, Carnegie Mellon University, University of Florida, the French Participation Group, the German Participation Group, Harvard University, the Instituto de Astrofisica de Canarias, the Michigan State/Notre Dame/JINA Participation Group, Johns Hopkins University, Lawrence Berkeley National Laboratory, Max Planck Institute for Astrophysics, Max Planck Institute for Extraterrestrial Physics, New Mexico State University, New York University, Ohio State University, Pennsylvania State University, University of Portsmouth, Princeton University, the Spanish Participation Group, University of Tokyo, University of Utah, Vanderbilt University, University of Virginia, University of Washington, and Yale University.

\software{PyQSOFit \citep{2018ascl.soft09008G}, TOPCAT \citep{2005ASPC..347...29T}, NumPy \citep{numpy}, SciPy  \citep{Virtanen_2020}, Matplotlib \citep{Matplotlib}, Astropy \citep{astropy}, linmix \citep{2007ApJ...665.1489K}, BCES \citep{1996ApJ...470..706A}, Kapteyn \citep{KapteynPackage}}

 \bibliographystyle{apj}
 \bibliography{ref}
 
\clearpage
\newpage
 \appendix
\section{Quality flag}\label{sec:flag}
We provide quality flags on the various spectral quantities in the catalog to access the reliability of the measurements following \citet{2017MNRAS.472.4051C}. The quality flags are an integer number which is calculated as 2$^{\mathrm{Bit}\_0}$+ 2$^{\mathrm{Bit} \_1}$+ $2^{\mathrm{Bit} \_2}$+...+$2^{\mathrm{Bit} \_n}$ where `Bits' can have values of 0 (no flag raised) or 1 (flag raised). Therefore, a quality flag of zero means all Bits are zero and the associated quantity is reliable while a flag $>$0 means the associated quantity should be used with caution. Below we tabulate the quality flag statistics and mention the criteria used to set those quality flags for continuum and line quantities as footnotes in Tables \ref{Table:PCA_flag}, \ref{Table:cont_flag} and \ref{Table:line_flag}. 
Here we summarize the criteria used to define the flag. 

The quality flags are assigned based on the measured quantities and their uncertainties. For emission line quantities, if line PEAK\_FLUX $<$3$\times$ PEAK\_FLUX\_ERR i.e. relative uncertainty in PEAK\_FLUX\_ERR/PEAK\_FLUX $>$ 1/3, a bit is assigned to have a value of 1 and the quantity for a given line is unreliable. Moreover, if the line luminosity, FWHM and velocity offsets and their uncertainties $=0$ or infinite, the associated bits have a value of 1. Also, relative uncertainty (i.e. the ratio of the uncertainty to the reported value) in line luminosity $>$1.5 and FWHM$>$2, and uncertainties in the velocity offsets $>1000$ km s$^{-1}$ are considered for the associated bits to have a value of 1. Similarly, for PCA decomposition, if the host fraction at 4200$\AA$ or 5100$\AA$ is $>100$, the PCA decomposition is considered to be unreliable and assigned to have a value of 1. Moreover, if the reduced-$\chi^2>15$ and host fraction is $>0.3$ a bit equal to 1 is assigned. For AGN continuum luminosity, if luminosity or continuum slope, and their uncertainty is zero or infinite a bit is assigned to a value of 1. We also considered sources with relative uncertainty in continuum luminosity $>1.5$ and slope $>0.3$, and reduced-$\chi^2>50$ in the continuum fit to have a value of 1.

\begin{ThreePartTable}
\begin{TableNotes}
\footnotesize
\item [+] decomposition is not applied.
\item [++] decomposition is applied.
\item [*] all bits set to zero.
\item [a] fraction of host to the total flux is higher than 100\% at 4200\AA\ or 5100\AA.
\item [b] reduced $\chi^2$ of host-galaxy decomposition $>15$ and fraction of host$>0.3$.  
\end{TableNotes}

\begin{longtable}{ll}\\
\caption{Quality flag on host galaxy decomposition using PCA. The criteria used to set the quality flags are also given in the footnote.} \\

  \toprule
Quality  flag  & Number of sources with host decomposition \\
 \midrule
			 \endfirsthead
No PCA\tnote{+}          & 513458    \\
PCA\tnote{++}            & 12807      \\
Good quality\tnote{*}    & 12674 ( 99.0\%)  \\
Bit\_0\tnote{a}          &  126 ( 1.0\%)   \\
Bit\_1\tnote{b}          &   7 ( 0.1\%)   \\

 \bottomrule
 \insertTableNotes  
 \endlastfoot
 \label{Table:PCA_flag}
 \end{longtable}
\end{ThreePartTable}

\begin{ThreePartTable}
\begin{TableNotes}
\footnotesize
\item [+] wavelength is outside the observed range.
\item [++] wavelength is inside the observed range.
\item [*] all bits set to zero. If PCA flag is $>0$ then a value of 1000 is added to the final continuum flag.  
\item [a] luminosity or its uncertainty is zero or NaN.   
\item [b] relative uncertainty of luminosity $>$ 1.5.
\item [c] slope or its uncertainty is zero or NaN. 
\item [d] slope hits a limit in the fit.
\item [e] slope uncertainty $>$ 0.3.
\item [f] reduced $\chi^2$ of the continuum fit $>$ 50. 
\end{TableNotes}

\begin{longtable}{lllll}
  \caption{Quality flag statistics for continuum luminosity. The criteria used to set the quality flags are also given in the footnote.} \\ 

  \toprule
  Quality  flag  &    &  continuum luminosity & &  \\ 
  \cmidrule(l){2-5}
			 & L5100 			& L4400          &  L3000              	& L1350   \\
			 
			 \midrule
			 \endfirsthead
			 
No Cont\tnote{+}             & 436243       	& 379399          &  118583  			& 234287   \\
Cont\tnote{++}               & 90022       		& 146866          &  407682  			& 291978   \\ 
Good quality\tnote{*}        & 88821 (98.7\%)   & 145586 (99.1\%) & 405085 (99.4\%)  	& 290334 (99.4\%) \\
Bit\_0\tnote{a}              & 103 (0.1\%)      & 109 (0.1\%)     &  281 (0.1\%)  		& 211 (0.1\%) \\
Bit\_1\tnote{b}		         & 6 (0.0\%)        & 7 (0.0\%)       &  234  (0.1\%)  		& 285 (0.1\%) \\ 
Bit\_2\tnote{c} 	         &  149 (0.2\%)     & 169 (0.1\%)     &  272  (0.1\%)  		& 171 (0.\%) \\
Bit\_3\tnote{d}              &  388 (0.4\%)     & 425 (0.3\%)     &   694 (0.2\%)       & 378 (0.1\%) \\
Bit\_4\tnote{e}              & 909 (1.0\%)      & 947 (0.6\%)     &  1915 (0.5\%)  	    & 1095 (0.4\%) \\
Bit\_5\tnote{f}              &    7 (0.0\%)     & 14 (0.0\%)      &    39 (0.0\%)  		& 75 (0.0\%)  \\

 \bottomrule
 \insertTableNotes  
 \endlastfoot
 \label{Table:cont_flag}
 \end{longtable}
\end{ThreePartTable}

\begin{ThreePartTable}

\begin{TableNotes}
\footnotesize
\item [+] line fitting window is outside the observed range.
\item [++] line fitting window is inside the observed range.
\item [*] all bits set to zero.
\item [a] relative uncertainty of peak flux $>$ 1/3.   
\item [b] luminosity or its uncertainty is zero or NaN.
\item [c] relative uncertainty of luminosity $>$ 1.5.
\item [d] FWHM or its uncertainty is zero or NaN. 
\item [e] FWHM value hits lower or upper limit in the fit.
\item [f] relative uncertainty of FWHM $>$ 2.
\item [g] Velocity offset or its uncertainty is zero or NaN.
\item [h] Velocity offset value hits lower or upper limit in the fit.
\item [i] uncertainty in velocity offset $>$1000kms${-1}$.
\end{TableNotes}

 \begin{longtable} 
 {
    *{8}{p{\dimexpr.11\linewidth-0.5\tabcolsep}}
  }
  \caption{Quality flag statistics for different emission lines. The criteria used to set the quality flags are also given in the footnote.} \\ 

  \toprule
Quality  flag  &    & & &Emission lines & &  \\ 
\cmidrule(l){2-8}
               & H$\alpha$              & H$\beta$           		 & H$\gamma$ 				 & MgII              & CIII 			   	& CIV                & Ly$\alpha$ \\ 
\midrule
\endfirsthead

 No line\tnote{+}       & 515758             & 436306                 & 383649            & 107258   		 & 89887  				& 175854  			& 318884\\
 Line\tnote{++}         & 10507       	     & 89959                  & 142616            & 419007   		 & 436378  				& 350411  			& 207381\\
 Good quality\tnote{*}  & 8115 (77.2\%) & 63312 (70.4\%)   & 51588 (36.2\%)    & 309166 (73.8\%)  & 312896 (71.7\%)   & 290053 (82.8\%) & 150779 (72.7\%)\\
 Bit$\_$0\tnote{a}      & 1112 (10.6\%) & 18179 (20.2\%)   & 84024 (58.9\%)    & 60226 (14.4\%)   & 85667 (19.6\%)  	& 20586 (5.9\%)   & 9829 (4.7\%)\\
 Bit$\_$1\tnote{b}      & 1378 (13.1\%) & 3204 (3.6\%)    & 22431 (15.7\%)    & 46903 (11.2\%)   & 16234 (3.7\%)  		& 25474 (7.3\%)     & 18884 (9.1\%)\\
 Bit$\_$2\tnote{c}      & 23 (0.2\%)          & 521 (0.6\%)           & 5801 (4.1\%)      & 637 (0.2\%)   	 & 134 (0.1\%)  		& 410 (0.1\%)       & 451 (0.2\%)\\
 Bit$\_$3\tnote{d}      & 1378 (13.1\%) & 3204 (3.6\%)    & 22431 (15.7\%)    & 46905 (11.2\%)   & 16234 (3.7\%)  		& 25474 (7.3\%)     & 18884 (9.1\%)\\
 Bit$\_$4\tnote{e}      & 1406 (13.4\%)    & 3826 (4.3\%)       & 35091 (24.6\%)    & 47192 (11.3\%)   & 36912 (8.5\%)  		& 31504 (9.0\%)     & 36125 (17.4\%)\\
 Bit$\_$5\tnote{f}      & 43 (0.4\%)       & 1639 (1.8\%)             & 9175 (6.4\%)      & 15157 (3.6\%)    & 27241 (6.2\%)  		& 6533 (1.9\%)      & 5005 (2.4\%)\\
 Bit$\_$6\tnote{g}      & 22 (0.2\%)       & 741 (0.8\%)              & 17627 (12.4\%)    & 291 (0.1\%)   	 & 146 (0.1\%)  		& 525 (0.2\%)       & 275 (0.1\%)\\
 Bit$\_$7\tnote{h}      & 62 (0.6\%)       & 2138 (2.4\%)             & 30612 (21.5\%)    & 11230 (2.7\%)    & 6846 (1.6\%)  		& 4202 (1.2\%)      & 8548 (4.1\%)\\
 Bit$\_$8\tnote{i}      & 409 (3.9\%)      & 18538 (20.6\%)     & 76873 (53.9\%)    & 74823 (17.9\%)   & 45939 (10.5\%)  	    & 23637 (6.7\%)     & 12964 (6.3\%)\\

 \bottomrule
 \insertTableNotes  
 \endlastfoot
 \label{Table:line_flag}
 \end{longtable} 

\end{ThreePartTable}

 \section{Catalog format and column information}\label{sec:catalog}

  We provide two catalogs\footnote{\url{https://www.utu.fi/sdssdr14/}}:
  \begin{enumerate}
  \item The main catalog (``dr14q\_spec\_prop.fits'') is based on the spectral information from this study consisting of 274 columns, which are described in Table \ref{Table:catalog}. 
  \item An extended catalog (``dr14q\_spec\_prop\_ext.fits'') where all the columns of DR14Q \citep{2018A&A...613A..51P} is appended after the main catalog (i.e., after column \# 274). The extended catalog has a total of 380 columns.
  \end{enumerate}

   \begin{longtable}{lllll} 
   \caption{FITS catalog description and column information of the spectral catalog: (1) FITS column number, (2) name of columns, (3) format (4) unit, and (5) description. The unmeasurable values are indicated with $-999$. All the measured continuum and line spectral quantities are from the model and their uncertainties are from the Monte Carlo simulation as mentioned in the text.} \\  \toprule
  
   Number & Column Name              & Format    & Unit  & Description \\ 
   (1)    &  (2)                     & (3)       & (4)   &   (5)       \\ \midrule
   \endfirsthead
  
1	&	SDSS\_NAME					&	String	&	     		& \pbox{40cm}{The DR14 object designation as given in\\ DR14 quasars catalog} 		\\
2	&	RA							&	Double	& Degree 		& Right Ascension (J2000)		\\
3	&	DEC							&	Double	& Degree 		& Declination (J2000)	\\
4	&	SDSS\_ID					&	String	&				& PLATE-MJD-FIBER	\\
5	&	PLATE						&	Long	&				& SDSS plate number	\\
6	&	MJD							&	Long	&				& MJD when spectrum was observed	\\
7	&	FIBERID						&	Long	&				& SDSS fiber	\\
8	&	REDSHIFT					&	Double	&				& Redshift	\\
9	&	SN\_RATIO\_CONT				&	Double	&				& \pbox{40cm}{Continuum median S/N per pixel estimated at wavelength \\ around 1350, 2245, 3000, 4210, 5100 \AA\, depending on the \\ spectral coverage}	\\
10	&	MIN\_WAVE					&	Double	& \AA			& Minimum wavelength of the rest frame spectrum 	\\
11	&	MAX\_WAVE					&	Double	& \AA			& Maximum wavelength of the rest frame spectrum	\\
12	&	PL\_NORM					&	Double	& erg s$^{-1}$ cm$^{-2}$ $\AA^{-1}$				& Normalization parameter-AGN power-law\\
13	&	PL\_NORM\_ERR				&	Double	& erg s$^{-1}$ cm$^{-2}$ $\AA^{-1}$				& Measurement error in PL\_NORM\\
14	&	PL\_SLOPE					&	Double	&				& Slope of AGN power-law\\
15	&	PL\_SLOPE\_ERR				&	Double	&				& Measurement error in PL\_SLOPE\\
16	&   CONT\_RED\_CHI2				& 	Double	&            	& Reduced $\chi^2$ of the continuum fitting \\
17	&   HOST\_FR\_4200				&	Double	&				& \pbox{40cm}{Fraction of host galaxy flux with respect to the total flux \\ at 4200\AA} \\
18	&	HOST\_FR\_5100				&	Double	&				& same as Col. 17 but at 5100\AA \\
19	&	PCA\_RED\_CHI2				&	Double	&				& Reduced $\chi^2$ of the PCA decomposition \\
20	&	QUALITY\_PCA				& 	Double	&				& Quality flag of PCA decomposition\\ 
21	&   LOG\_L1350					&	Double	& erg s$^{-1}$	& Logarithmic continuum luminosity at rest-frame 1350 \AA	\\
22	&   LOG\_L1350\_ERR				&	Double	& erg s$^{-1}$	& Measurement error in LOG\_L1350	\\
23	&   QUALITY\_L1350				& 	Double	&				& Quality flag of L1350 measurement \\
24	&   LOG\_L3000					&	Double	& erg s$^{-1}$	& Logarithmic continuum luminosity at rest-frame 3000 \AA	\\
25	&   LOG\_L3000\_ERR				&	Double	& erg s$^{-1}$	& Measurement error in LOG\_L3000	\\
26	&   QUALITY\_L3000				&   Double	&				& Quality flag of L3000 measurement\\
27	&   LOG\_L4400					&	Double	& erg s$^{-1}$	& Logarithmic continuum luminosity at rest-frame 4400 \AA	\\
28	&   LOG\_L4400\_ERR				&	Double	& erg s$^{-1}$	& Measurement error in LOG\_L4400	\\
29	&   QUALITY\_L4400				&   Double	&				& Quality flag of L4400 measurement\\
30	&   LOG\_L5100					&	Double	& erg s$^{-1}$	& Logarithmic continuum luminosity at rest-frame 5100 \AA	\\
31	&   LOG\_L5100\_ERR				&	Double	& erg s$^{-1}$	& Measurement error in LOG\_L5100	\\
32	&   QUALITY\_L5100				&	Double	&				& Quality flag of L5100 measurement\\
33	&   FBC\_FR\_3000				&	Double  &				& Fraction of Balmer continuum to total continuum at 3000 \AA\\
34	&   LOGL\_FE\_UV				&	Double	& erg s$^{-1}$	& \pbox{40cm}{Logarithmic luminosity of the UV Fe II complex \\ within the 2200-3090 \AA}	\\
35	&   LOGL\_FE\_UV\_ERR			&	Double	& erg s$^{-1}$	& Measurement error in LOGL\_FE\_UV	\\
36	&   LOGL\_FE\_OP				&	Double	& erg s$^{-1}$	& \pbox{40cm}{Logarithmic luminosity of the optical Fe II complex \\ within the 4435-4685 \AA}	\\
37	&   LOGL\_FE\_OP\_ERR			&	Double	& erg s$^{-1}$	& Measurement error in LOGL\_FE\_OP\\
38	&   EW\_FE\_UV					&	Double	& \AA			& \pbox{40cm}{Rest-frame equivalent width of UV Fe II complex \\ within the 2200-3090 \AA}	\\
39	&   EW\_FE\_UV\_ERR				&	Double	& \AA			& Measurement error in EW\_FE\_UV	\\
40	&   EW\_FE\_OP					&	Double	& \AA			& \pbox{40cm}{Rest-frame equivalent width of optical Fe II complex \\ within the 4435-4685 \AA} \\
41	&   EW\_FE\_OP\_ERR				&	Double	& \AA			& Measurement error in	EW\_FE\_OP\\
42	&   LINE\_NPIX\_HA				&	Double	&				& Number of good pixels for the rest-frame 6400-6765 \AA	\\
43	&   LINE\_MED\_SN\_HA			&	Double	&				& Median S/N per pixel for the rest-frame  6400-6765 \AA	\\
44	&   LINE\_NPIX\_HB				&	Double	&				& Number of good pixels for the rest-frame 4750-4950 \AA	\\
45	&   LINE\_MED\_SN\_HB			&	Double	&				& Median S/N per pixel for the rest-frame  4750-4950 \AA	\\
46	&   LINE\_NPIX\_HG				&	Double	&				& Number of good pixels for the rest-frame 4280-4400 \AA	\\
47	&   LINE\_MED\_SN\_HG			&	Double	&				& Median S/N per pixel for the rest-frame  4280-4400 \AA	\\
48	&   LINE\_NPIX\_MGII			&	Double	&				& Number of good pixels for the rest-frame 2700-2900 \AA	\\
49	&   LINE\_MED\_SN\_MGII			&	Double	&				& Median S/N per pixel for the rest-frame  2700-2900 \AA	\\
50	&   LINE\_NPIX\_CIII			&	Double	&				& Number of good pixels for the rest-frame 1850-1970 \AA	\\
51	&   LINE\_MED\_SN\_CIII			&	Double	&				& Median S/N per pixel for the rest-frame  1850-1970 \AA    \\
52	&   LINE\_NPIX\_CIV				&	Double	&				& Number of good pixels for the rest-frame 1500-1600 \AA	\\
53	&   LINE\_MED\_SN\_CIV			&	Double	&				& Median S/N per pixel for the rest-frame  1500-1600 \AA	\\
54	&   LINE\_NPIX\_LYA				&	Double	&				& Number of good pixels for the rest-frame 1150-1290 \AA	\\
55	&   LINE\_MED\_SN\_LYA			&	Double	&				& Median S/N per pixel for the rest-frame  1150-1290 \AA	\\
56	&   LYA\_LINE\_STATUS			&	Long	&				& Line fitting status\footnote{An integer number returned by KMPFIT code \citep{KapteynPackage} which is used in PyQSOfit to perform the non-linear least-squares fitting. Values larger than zero can represent success (however STATUS=5 may indicate failure to converge). More information about fitting status can be found in \url{https://idlastro.gsfc.nasa.gov/ftp/pro/markwardt/mpfit.pro}} in Ly$\alpha$ fitting		\\
57	&   LYA\_LINE\_CHI2				&	Double	&				& $\chi^2$ in Ly$\alpha$ fitting	\\
58	&   LYA\_LINE\_RED\_CHI2		&	Double	&				& Reduced $\chi^2$ in Ly$\alpha$ fitting	\\
59	&   LYA\_NDOF					&	Long	&				& Degrees of freedom in L$\alpha$ fitting \\
60	&   CIV\_LINE\_STATUS			&	Long	&				& Line fitting status in CIV fitting	\\
61	&   CIV\_LINE\_CHI2				&	Double	&				& $\chi^2$ in CIV fitting	\\
62	&   CIV\_LINE\_RED\_CHI2		&	Double	&				& Reduced $\chi^2$ in CIV fitting	\\
63	&   CIV\_NDOF					&	Long	&				& Degrees of freedom in CIV fitting	\\
64	&   CIII\_LINE\_STATUS			&	Long	&				& Line fitting status in CIII fitting	\\
65	&   CIII\_LINE\_CHI2			&	Double	&				& $\chi^2$ in CIII fitting	\\
66	&   CIII\_LINE\_RED\_CHI2		&	Double	&				& Reduced $\chi^2$ in CIII fitting	\\
67	&   CIII\_NDOF					&	Long	&				& Degrees of freedom in CIII fitting	\\
68	&   MGII\_LINE\_STATUS			&	Long	&				& Line fitting status in Mg II fitting	\\
69	&   MGII\_LINE\_CHI2			&	Double	&				& $\chi^2$ in Mg II fitting	\\
70	&   MGII\_LINE\_RED\_CHI2		&	Double	&				& Reduced $\chi^2$ in Mg II fitting	\\
71	&   MGII\_NDOF					&	Long	&				& Degrees of freedom in Mg II fitting	\\
72	&   HG\_LINE\_STATUS			&	Long	&				& Line fitting status in H$\gamma$ fitting	\\
73	&   HG\_LINE\_CHI2				&	Double	&				& $\chi^2$ in H$\gamma$  fitting	\\
74	&   HG\_LINE\_RED\_CHI2			&	Double	&				& Reduced $\chi^2$ in H$\gamma$  fitting	\\
75	&   HG\_NDOF					&	Long	&				& Degrees of freedom in H$\gamma$  fitting	\\
76	&   HB\_LINE\_STATUS			&	Long	&				& Line fitting status in H$\beta$ fitting	\\
77	&   HB\_LINE\_CHI2				&	Double	&				& $\chi^2$ in H$\beta$  fitting	\\
78	&   HB\_LINE\_RED\_CHI2			&	Double	&				& Reduced $\chi^2$ in H$\beta$  fitting	\\
79	&   HB\_NDOF					&	Long	&				& Degrees of freedom in H$\beta$  fitting \\
80	&   HA\_LINE\_STATUS			&	Long	&				& Line fitting status in H$\alpha$ fitting	\\
81	&   HA\_LINE\_CHI2				&	Double	&				& $\chi^2$ in H$\alpha$  fitting	\\
82	&   HA\_LINE\_RED\_CHI2			&	Double	&				& Reduced $\chi^2$ in H$\alpha$  fitting	\\
83	&   HA\_NDOF					&	Long	&				& Degrees of freedom in H$\alpha$  fitting \\
84	&   LOGL\_HA\_NA				&	Double	& erg s$^{-1}$	& Logarithmic line luminosity of the H$\alpha$ narrow component	\\
85	&   LOGL\_HA\_NA\_ERR			&	Double	& erg s$^{-1}$	& Measurement error in	LOGL\_HA\_NA \\
86	&   EW\_HA\_NA					&	Double	& \AA	        & Rest-frame equivalent width of H$\alpha$ narrow component	\\
87	&   EW\_HA\_NA\_ERR				&	Double	& \AA	        & Measurement error in EW\_HA\_NA	\\
88	&   FWHM\_HA\_NA				&	Double	& km s$^{-1}$	& FWHM of H$\alpha$ narrow component	\\
89	&   FWHM\_HA\_NA\_ERR			&	Double	& km s$^{-1}$	& Measurement error in FWHM\_HA\_NA	\\
90	&   LOGL\_NII6549				&	Double	& erg s$^{-1}$	& Logarithmic line luminosity of NII6549	\\
91	&   LOGL\_NII6549\_ERR			&	Double	& erg s$^{-1}$	& Measurement error in LOGL\_NII6549	\\
92	&   EW\_NII6549					&	Double	& \AA	        & Rest-frame equivalent width of NII6549	\\
93	&   EW\_NII6549\_ERR			&	Double	& \AA	        & Measurement error in EW\_NII6549	\\
94	&   LOGL\_NII6585				&	Double	& erg s$^{-1}$	& Logarithmic line luminosity of NII6585	\\
95	&   LOGL\_NII6585\_ERR			&	Double	& erg s$^{-1}$	& Measurement error in LOGL\_NII6585	\\
96	&   EW\_NII6585					&	Double	& \AA	        & Rest-frame equivalent width of NII6585	\\
97	&   EW\_NII6585\_ERR			&	Double	& \AA	        & Measurement error in EW\_NII6585	\\
98	&   LOGL\_SII6718				&	Double	& erg s$^{-1}$	& Logarithmic line luminosity of SII6718	\\
99	&   LOGL\_SII6718\_ERR			&	Double	& erg s$^{-1}$	& Measurement error in LOGL\_SII6718	\\
100	&   EW\_SII6718					&	Double	& \AA	        & Rest-frame equivalent width of SII6718	\\
101	&   EW\_SII6718\_ERR			&	Double	& \AA	        & Measurement error in EW\_SII6718	\\
102	&   LOGL\_SII6732				&	Double	& erg s$^{-1}$	& Logarithmic line luminosity of SII6732 \\
103	&   LOGL\_SII6732\_ERR			&	Double	& erg s$^{-1}$	& Measurement error in LOGL\_SII6732	\\
104	&   EW\_SII6732					&	Double	& \AA			& Rest-frame equivalent width of SII6732	\\
105	&   EW\_SII6732\_ERR			&	Double	& \AA			& Measurement error in EW\_SII6732	\\
106	&   FWHM\_HA\_BR				&	Double	& km s$^{-1}$	& FWHM of H$\alpha$ broad component 	\\
107	&   FWHM\_HA\_BR\_ERR			&	Double	& km s$^{-1}$	& Measurement error in FWHM\_HA\_BR	\\
108	&   SIGMA\_HA\_BR				&	Double	& km s$^{-1}$   & Line dispersion (second moment) of H$\alpha$ broad component 	\\
109	&   SIGMA\_HA\_BR\_ERR			&	Double	& km s$^{-1}$   & Measurement error in SIGMA\_HA\_BR	\\
110	&   EW\_HA\_BR					&	Double	& \AA			& Rest-frame equivalent width of H$\alpha$ broad component	\\
111	&   EW\_HA\_BR\_ERR				&	Double	& \AA			& Measurement error in EW\_HA\_BR	\\
112	&   PEAK\_HA\_BR				&	Double	& \AA			& Peak wavelength of H$\alpha$ broad component	\\
113	&   PEAK\_HA\_BR\_ERR			&	Double	& \AA			& Measurement error in PEAK\_HA\_BR	\\
114	&   PEAK\_FLUX\_HA\_BR			&	Double	& erg s$^{-1}$ cm$^{-2}$ $\AA^{-1}$   & Peak flux of H$\alpha$ broad component	\\
115	&   PEAK\_FLUX\_HA\_BR\_ERR		&	Double	& erg s$^{-1}$ cm$^{-2}$ $\AA^{-1}$   &  Measurement error in PEAK\_FLUX\_HA\_BR \\
116	&   LOGL\_HA\_BR				&	Double	& erg s$^{-1}$	& Logarithmic line luminosity of H$\alpha$ broad component	\\
117	&   LOGL\_HA\_BR\_ERR			&	Double	& erg s$^{-1}$	& Measurement error in LOGL\_HA\_BR	\\
118	&   QUALITY\_HA					&	Double	&				& Quality of H$\alpha$ line fitting \\
119	&   LOGL\_HB\_NA				&	Double	& erg s$^{-1}$	& Logarithmic line luminosity of H$\beta$ narrow component	\\
120	&   LOGL\_HB\_NA\_ERR			&	Double	& erg s$^{-1}$	& Measurement error in LOGL\_HB\_NA	\\
121	&   EW\_HB\_NA					&	Double	& \AA			& Rest-frame equivalent width of H$\beta$ narrow component\\
122	&   EW\_HB\_NA\_ERR				&	Double	& \AA			& Measurement error in EW\_HB\_NA	\\
123	&   FWHM\_HB\_NA				&	Double	& km  s$^{-1}$	& FWHM of H$\beta$ narrow component	\\
124	&   FWHM\_HB\_NA\_ERR			&	Double	& km  s$^{-1}$	& Measurement error in FWHM\_HB\_NA	\\
125	&   LOGL\_OIII4959C				&	Double	& erg s$^{-1}$	& Logarithmic line luminosity of OIII4959 core component	\\
126	&   LOGL\_OIII4959C\_ERR		&	Double	& erg s$^{-1}$	& Measurement error in LOGL\_OIII4959C	\\
127	&   EW\_OIII4959C				&	Double	& \AA 			& Rest-frame equivalent width of OIII4959 core component	\\
128	&   EW\_OIII4959C\_ERR			&	Double	& \AA 			& Measurement error in EW\_OIII4959C	\\
129	&   LOGL\_OIII5007C				&	Double	& erg s$^{-1}$	& Logarithmic line luminosity of OIII5007 core component	\\
130	&   LOGL\_OIII5007C\_ERR		&	Double	& erg s$^{-1}$	& Measurement error in LOGL\_OIII5007C	\\
131	&   EW\_OIII5007C				&	Double	& \AA 			& Rest-frame equivalent width of OIII5007 core component 	\\
132	&   EW\_OIII5007C\_ERR			&	Double	& \AA 			& Measurement error in EW\_OIII5007C	\\
133	&   LOGL\_OIII4959W				&	Double	& erg s$^{-1}$	& Logarithmic line luminosity of OIII4959 wing component	\\
134	&   LOGL\_OIII4959W\_ERR		&	Double	& erg s$^{-1}$	& Measurement error in LOGL\_OIII4959W	\\
135	&   EW\_OIII4959W				&	Double	& \AA 			& Rest-frame equivalent width of OIII4959 wing component	\\
136	&   EW\_OIII4959W\_ERR			&	Double	& \AA 			& Measurement error in EW\_OIII4959W	\\
137	&   LOGL\_OIII5007W				&	Double	& erg s$^{-1}$	& Logarithmic line luminosity of OIII5007 wing component	\\
138	&   LOGL\_OIII5007W\_ERR		&	Double	& erg s$^{-1}$	& Measurement error in LOGL\_OIII5007W		\\
139	&   EW\_OIII5007W				&	Double	& \AA 			& Rest-frame equivalent width of OIII5007 wing component	\\
140	&   EW\_OIII5007W\_ERR			&	Double	& \AA 			& Measurement error in EW\_OIII5007W	\\
141	&   LOGL\_OIII4959				&	Double	& erg s$^{-1}$	& Logarithmic line luminosity of entire OIII4959 	\\
142	&   LOGL\_OIII4959\_ERR			&	Double	& erg s$^{-1}$	& Measurement error in LOGL\_OIII4959	\\
143	&   EW\_OIII4959				&	Double	& \AA 			& Rest-frame equivalent width of entire OIII4959	\\
144	&   EW\_OIII4959\_ERR			&	Double	& \AA  			& Measurement error in EW\_OIII4959	\\
145	&   LOGL\_OIII5007				&	Double	& erg s$^{-1}$	& Logarithmic line luminosity of entire OIII5007	\\
146	&   LOGL\_OIII5007\_ERR			&	Double	& erg s$^{-1}$	& Measurement error in LOGL\_OIII5007	\\
147	&   EW\_OIII5007				&	Double	& \AA 			& Rest-frame equivalent width of entire OIII5007	\\
148	&   EW\_OIII5007\_ERR			&	Double	& \AA 			& Measurement error in EW\_OIII5007	\\
149	&   LOGL\_HEII4687\_BR			&	Double	& erg s$^{-1}$	& Logarithmic line luminosity of HeII4687 broad component	\\
150	&   LOGL\_HEII4687\_BR\_ERR		&	Double	& erg s$^{-1}$	& Measurement error in LOGL\_HEII4687\_BR	\\
151	&   EW\_HEII4687\_BR			&	Double	& \AA 			& Rest-frame equivalent width of HeII4687 broad component	\\
152	&   EW\_HEII4687\_BR\_ERR		&	Double	& \AA 			& Measurement error in EW\_HEII4687\_BR	\\
153	&   LOGL\_HEII4687\_NA			&	Double	& erg s$^{-1}$	& Logarithmic line luminosity of HeII4687 narrow component	\\
154	&   LOGL\_HEII4687\_NA\_ERR		&	Double	& erg s$^{-1}$	& Measurement error in LOGL\_HEII4687\_NA	\\
155	&   EW\_HEII4687\_NA			&	Double	& \AA 			& Rest-frame equivalent width of HeII4687 narrow component	\\
156	&   EW\_HEII4687\_NA\_ERR		&	Double	& \AA 			& Measurement error in EW\_HEII4687\_NA	\\
157	&   FWHM\_HB\_BR				&	Double	& km s$^{-1}$	& FWHM of H$\beta$ broad component	\\
158	&   FWHM\_HB\_BR\_ERR			&	Double	& km s$^{-1}$	& Measurement error in FWHM\_HB\_BR	\\
159	&   SIGMA\_HB\_BR				&	Double	& km s$^{-1}$   & Line dispersion (second moment) of H$\beta$ broad component 	\\
160	&   SIGMA\_HB\_BR\_ERR			&	Double	& km s$^{-1}$   & Measurement error in SIGMA\_HB\_BR	\\
161	&   EW\_HB\_BR					&	Double	& \AA 			& Rest-frame equivalent width of H$\beta$ broad component	\\
162	&   EW\_HB\_BR\_ERR				&	Double	& \AA 			& Measurement error in EW\_HB\_BR	\\
163	&   PEAK\_HB\_BR				&	Double	& \AA 			& Peak wavelength of H$\beta$ broad component	\\
164	&   PEAK\_HB\_BR\_ERR			&	Double	& \AA 			& Measurement error in PEAK\_HB\_BR	\\
165	&   PEAK\_FLUX\_HB\_BR			&	Double	& erg s$^{-1}$ cm$^{-2}$ $\AA^{-1}$    & Peak flux of H$\beta$ broad component	\\
166	&   PEAK\_FLUX\_HB\_BR\_ERR		&	Double	& erg s$^{-1}$ cm$^{-2}$ $\AA^{-1}$    &  Measurement error in PEAK\_FLUX\_HB\_BR \\
167	&   LOGL\_HB\_BR				&	Double	& erg s$^{-1}$	& Logarithmic line luminosity of H$\beta$ broad component	\\
168	&   LOGL\_HB\_BR\_ERR			&	Double	& erg s$^{-1}$ 	& Measurement error in LOGL\_HB\_BR	\\
169	&   QUALITY\_HB					&	Double	&				& Quality of H$\beta$ line fitting \\
170	&   LOGL\_HG\_NA				&	Double	& erg s$^{-1}$	& Logarithmic line luminosity of H$\gamma$ narrow component	\\
171	&   LOGL\_HG\_NA\_ERR			&	Double	& erg s$^{-1}$	& Measurement error in LOGL\_HG\_NA	\\
172	&   EW\_HG\_NA					&	Double	& \AA 			& Rest-frame equivalent width of H$\gamma$ narrow component	\\
173	&   EW\_HG\_NA\_ERR				&	Double	& \AA 			& Measurement error in EW\_HG\_NA	\\
174	&   LOGL\_OIII4364				&	Double	& erg s$^{-1}$	& Logarithmic line luminosity of OIII4364	\\
175	&   LOGL\_OIII4364\_ERR			&	Double	& erg s$^{-1}$	& Measurement error in LOGL\_OIII4364	\\
176	&   EW\_OIII4364				&	Double	& \AA 			& Rest-frame equivalent width of OIII4364	\\
177	&   EW\_OIII4364\_ERR			&	Double	& \AA 			& Measurement error in EW\_OIII4364	\\
178	&   FWHM\_HG\_BR				&	Double	& km s$^{-1}$	& FWHM of H$\gamma$ broad component	\\
179	&   FWHM\_HG\_BR\_ERR			&	Double	& km s$^{-1}$	& Measurement error in FWHM\_HG\_BR	\\
180	&   SIGMA\_HG\_BR				&	Double	& km s$^{-1}$   & Line dispersion (second moment) of H$\gamma$ broad component 	\\
181	&   SIGMA\_HG\_BR\_ERR			&	Double	& km s$^{-1}$   & Measurement error in SIGMA\_HG\_BR	\\
182	&   EW\_HG\_BR					&	Double	& \AA 			& Rest-frame equivalent width of H$\gamma$ broad component	\\
183	&   EW\_HG\_BR\_ERR				&	Double	& \AA 			& Measurement error in EW\_HG\_BR	\\
184	&   PEAK\_HG\_BR				&	Double	& \AA 			& Peak wavelength of H$\gamma$ broad component	\\
185	&   PEAK\_HG\_BR\_ERR			&	Double	& \AA 			& Measurement error in PEAK\_HG\_BR	\\
186	&   PEAK\_FLUX\_HG\_BR			&	Double	& erg s$^{-1}$ cm$^{-2}$ $\AA^{-1}$    & Peak flux of H$\gamma$ broad component	\\
187	&   PEAK\_FLUX\_HG\_BR\_ERR		&	Double	& erg s$^{-1}$ cm$^{-2}$ $\AA^{-1}$    &  Measurement error in PEAK\_FLUX\_HG\_BR \\
188	&   LOGL\_HG\_BR				&	Double	& erg s$^{-1}$	& Logarithmic line luminosity of H$\gamma$ broad component	\\
189	&   LOGL\_HG\_BR\_ERR			&	Double	& erg s$^{-1}$	& Measurement error in LOGL\_HG\_BR	\\
190	&   QUALITY\_HG					&	Double	&				& Quality of H$\gamma$ line fitting \\
191	&   LOGL\_MGII\_NA				&	Double	& erg s$^{-1}$	& Logarithmic line luminosity of Mg II narrow component	\\
192	&   LOGL\_MGII\_NA\_ERR			&	Double	& erg s$^{-1}$	& Measurement error in LOGL\_MGII\_NA	\\
193	&   EW\_MGII\_NA				&	Double	& \AA 			& Rest-frame equivalent width of Mg II narrow component	\\
194	&   EW\_MGII\_NA\_ERR			&	Double	& \AA 			& Measurement error in EW\_MGII\_NA	\\
195	&   FWHM\_MGII\_NA				&	Double	& km s$^{-1}$ 	& FWHM of Mg II narrow component  	\\
196	&   FWHM\_MGII\_NA\_ERR			&	Double	& km s$^{-1}$	& Measurement error in FWHM\_MGII\_NA	\\
197	&   FWHM\_MGII\_BR				&	Double	& km s$^{-1}$	& FWHM of Mg II broad component 	\\
198	&   FWHM\_MGII\_BR\_ERR			&	Double	& km s$^{-1}$	& Measurement error in FWHM\_MGII\_BR	\\
199	&   SIGMA\_MGII\_BR				&	Double	& km s$^{-1}$   & Line dispersion (second moment) of MGII broad component 	\\
200	&   SIGMA\_MGII\_BR\_ERR		&	Double	& km s$^{-1}$   & Measurement error in SIGMA\_MGII\_BR	\\
201	&   EW\_MGII\_BR				&	Double	& \AA 			& Rest-frame equivalent width of Mg II broad component	\\
202	&   EW\_MGII\_BR\_ERR			&	Double	& \AA 			& Measurement error in EW\_MGII\_BR	\\
203	&   PEAK\_MGII\_BR				&	Double	& \AA 			& Peak wavelength of Mg II broad component	\\
204	&   PEAK\_MGII\_BR\_ERR			&	Double	& \AA 			& Measurement error in PEAK\_MGII\_BR	\\
205	&   PEAK\_FLUX\_MGII\_BR		&	Double	& erg s$^{-1}$ cm$^{-2}$ $\AA^{-1}$   & Peak flux of MGII broad component	\\
206	&   PEAK\_FLUX\_MGII\_BR\_ERR	&	Double	& erg s$^{-1}$ cm$^{-2}$ $\AA^{-1}$   &  Measurement error in PEAK\_FLUX\_MGII\_BR \\
207	&   LOGL\_MGII\_BR				&	Double	& erg s$^{-1}$	& Logarithmic line luminosity of Mg II broad component	\\
208	&   LOGL\_MGII\_BR\_ERR			&	Double	& erg s$^{-1}$	& Measurement error in LOGL\_MGII\_BR	\\
209	&   QUALITY\_MGII				&	Double	&				& Quality of MGII line fitting \\
210	&   FWHM\_CIII					&	Double	& km  s$^{-1}$	& FWHM of entire CIII	\\
211	&   FWHM\_CIII\_ERR				&	Double	& km  s$^{-1}$	& Measurement error in FWHM\_CIII	\\
212	&   SIGMA\_CIII					&	Double	& km s$^{-1}$   & Line dispersion (second moment) of entire CIII 	\\
213	&   SIGMA\_CIII\_ERR			&	Double	& km s$^{-1}$   & Measurement error in SIGMA\_CIII	\\
214	&   EW\_CIII					&	Double	& \AA 			& Rest-frame equivalent width of entire CIII	\\
215	&   EW\_CIII\_ERR				&	Double	& \AA 			& Measurement error in EW\_CIII	\\
216	&   PEAK\_CIII					&	Double	& \AA 			& Peak wavelength of entire CIII	\\
217	&   PEAK\_CIII\_ERR				&	Double	& \AA 			& Measurement error in PEAK\_CIII	\\
218	&   PEAK\_FLUX\_CIII			&	Double	& erg s$^{-1}$ cm$^{-2}$ $\AA^{-1}$   &  Peak flux of CIII 	\\
219	&   PEAK\_FLUX\_CIII\_ERR		&	Double	& erg s$^{-1}$ cm$^{-2}$ $\AA^{-1}$   &  Measurement error in PEAK\_FLUX\_CIII \\
220	&   LOGL\_CIII					&	Double	& erg s$^{-1}$	& Logarithmic line luminosity of entire CIII	\\
221	&   LOGL\_CIII\_ERR				&	Double	& erg s$^{-1}$	& Measurement error in LOGL\_CIII	\\
222	&   QUALITY\_CIII				&	Double	&				& Quality of CIII line fitting \\
223	&   FWHM\_CIV					&	Double	& km  s$^{-1}$	& FWHM of entire CIV	\\
224	&   FWHM\_CIV\_ERR				&	Double	& km  s$^{-1}$	& Measurement error in FWHM\_CIV	\\
225	&   SIGMA\_CIV					&	Double	& km s$^{-1}$   & Line dispersion (second moment) of entire CIV 	\\
226	&   SIGMA\_CIV\_ERR				&	Double	& km s$^{-1}$   & Measurement error in SIGMA\_CIV	\\
227	&   EW\_CIV						&	Double	& \AA 			& Rest-frame equivalent width of entire CIV	\\
228	&   EW\_CIV\_ERR				&	Double	& \AA  			& Measurement error in EW\_CIV		\\
229	&   PEAK\_CIV					&	Double	& \AA 			& Peak wavelength of entire CIV	\\
230	&   PEAK\_CIV\_ERR				&	Double	& \AA 			& Measurement error in PEAK\_CIV	\\
231	&   PEAK\_FLUX\_CIV				&	Double	& erg s$^{-1}$ cm$^{-2}$ $\AA^{-1}$   &  Peak flux of CIV \\
232	&   PEAK\_FLUX\_CIV\_ERR		&	Double	& erg s$^{-1}$ cm$^{-2}$ $\AA^{-1}$   &  Measurement error in PEAK\_FLUX\_CIV \\
233	&   LOGL\_CIV					&	Double	& erg s$^{-1}$	& Logarithmic line luminosity of entire CIV	\\
234	&   LOGL\_CIV\_ERR				&	Double	& erg s$^{-1}$	& Measurement error in LOGL\_CIV	\\
235	&   QUALITY\_CIV				&	Double	&				& Quality of CIV line fitting \\
236	&   FWHM\_LYA					&	Double	& km  s$^{-1}$	& FWHM of entire Ly$\alpha$	\\
237	&   FWHM\_LYA\_ERR				&	Double	& km  s$^{-1}$	& Measurement error in FWHM\_LYA	\\
238	&   SIGMA\_LYA					&	Double	& km s$^{-1}$   & Line dispersion (second moment) of entire LYA 	\\
239	&   SIGMA\_LYA\_ERR				&	Double	& km s$^{-1}$   & Measurement error in SIGMA\_LYA	\\
240	&   EW\_LYA						&	Double	& \AA 			& Rest-frame equivalent width of entire Ly$\alpha$	\\
241	&   EW\_LYA\_ERR				&	Double	& \AA  			& Measurement error in EW\_LYA	\\
242	&   PEAK\_LYA					&	Double	& \AA  			& Peak wavelength of entire Ly$\alpha$	\\
243	&   PEAK\_LYA\_ERR				&	Double	& \AA 			& Measurement error in PEAK\_LYA	\\
244	&   PEAK\_FLUX\_LYA				&	Double	& erg s$^{-1}$ cm$^{-2}$ $\AA^{-1}$   &  Peak flux of LYA \\
245	&   PEAK\_FLUX\_LYA\_ERR		&	Double	& erg s$^{-1}$ cm$^{-2}$ $\AA^{-1}$   &  Measurement error in PEAK\_FLUX\_LYA \\
246	&   LOGL\_LYA					&	Double	& erg s$^{-1}$	& Logarithmic line luminosity of entire Ly$\alpha$	\\
247	&   LOGL\_LYA\_ERR				&	Double	& erg s$^{-1}$	& Measurement error in LOGL\_LYA	\\
248	&   QUALITY\_LYA				&	Double	&				& Quality of LYA line fitting \\
249	&   LOGL\_NV					&	Double	& erg s$^{-1}$	& Logarithmic line luminosity of Nv1240 	\\
250	&   LOGL\_NV\_ERR				&	Double	& erg s$^{-1}$	& Measurement error in LOGL\_NV	\\
251	&   EW\_NV						&	Double	& \AA 			& Rest-frame equivalent width of Nv1240 	\\
252	&   EW\_NV\_ERR					&	Double	& \AA 			& Measurement error in EW\_NV	\\
253	&   FWHM\_NV					&	Double	& km s$^{-1}$	& FWHM of Nv1240	\\
254	&   FWHM\_NV\_ERR				&	Double	& km s$^{-1}$	& Measurement error in FWHM\_NV	\\
255	&   LOG\_MBH\_HB\_VP06			&	Double	& $M_{\odot}$	& \pbox{40cm}{Logarithmic single-epoch BH mass estimate based \\on H$\beta$ (VP06)}\\
256	&   LOG\_MBH\_HB\_VP06\_ERR		&	Double	& $M_{\odot}$	& Measurement error in LOG\_MBH\_HB\_VP06	\\
257	&   LOG\_MBH\_HB\_A11			&	Double	& $M_{\odot}$	& \pbox{40cm}{Logarithmic single-epoch BH mass estimate based \\on H$\beta$ (A11)}	\\
258	&   LOG\_MBH\_HB\_A11\_ERR		&	Double	& $M_{\odot}$	& Measurement error in 	LOG\_MBH\_HB\_A11	\\
259	&   LOG\_MBH\_MGII\_VO09		&	Double	& $M_{\odot}$	& \pbox{40cm}{Logarithmic single-epoch BH mass estimate based \\on Mg II (VO09)}	\\
260	&   LOG\_MBH\_MGII\_VO09\_ERR	&	Double	& $M_{\odot}$	& Measurement error in LOG\_MBH\_MGII\_VO09	\\
261	&   LOG\_MBH\_MGII\_S11			&	Double	& $M_{\odot}$	& \pbox{40cm}{Logarithmic single-epoch BH mass estimate based \\on Mg II (S11)}	\\
262	&   LOG\_MBH\_MGII\_S11\_ERR	&	Double	& $M_{\odot}$	& Measurement error in LOG\_MBH\_MGII\_S11	\\
263	&   LOG\_MBH\_CIV\_VP06			&	Double	& $M_{\odot}$	& \pbox{40cm}{Logarithmic single-epoch BH mass estimate based \\on CIV (VP06)}	\\
264	&   LOG\_MBH\_CIV\_VP06\_ERR	&	Double	& $M_{\odot}$	& Measurement error in LOG\_MBH\_CIV\_VP06	\\
265	&   LOG\_MBH					&	Double	& $M_{\odot}$	& Logarithmic fiducial single-epoch BH mass	\\
266	&   LOG\_MBH\_ERR				&	Double	& $M_{\odot}$	& Measurement error in LOG\_MBH	\\
267	&   QUALITY\_MBH				&	Double	&				& \pbox{40cm}{Quality of MBH estimation (A sum of quality \\of continuum luminosity and quality of line FWHM)} \\
268	&   LOG\_LBOL					&	Double	& erg s$^{-1}$	& Logarithmic fiducial bolometric luminosity	\\
269	&   QUALITY\_LBOL				&	Double	&				& \pbox{40cm}{Quality of LBOL estimation (=quality \\of continuum luminosity)} \\
270	&   LOG\_REDD					&	Double	&				& \pbox{40cm}{Logarithmic Eddington ratio based on \\fiducial single-epoch BH mass}	\\
271	&   QUALITY\_REDD				&	Double	&				& \pbox{40cm}{Quality of REDD estimation (A sum of quality of \\continuum luminosity and quality of line MBH)} \\
272	&   BI\_CIV						&	Double	& km s$^{-1}$	& \pbox{40cm}{BALnicity Index of C IV absorption trough \\from Paris et al. (2018)}  \\
273	&   ERR\_BI\_CIV				&	Double	& km s$^{-1}$	& Measurement error in BI\_CIV from Paris et al. (2018)	 \\
274	&   BAL\_FLAG					&			&				& BAL flag from Shen et al. (2011) \\
 \bottomrule
         \label{Table:catalog}
     \end{longtable}    
    
\end{document}